%% file: MPSPaperII_final.tex
\documentclass[a4paper,10pt]{article}
\usepackage{graphicx}
\usepackage{amsmath}
\usepackage{citesort}
\usepackage[sort&compress,numbers]{natbib}
\usepackage{amsfonts}
\usepackage{amssymb}
\usepackage{latexsym}
\usepackage{color}
\usepackage{ntheorem}
\usepackage{braket}
\usepackage[colorlinks=true]{hyperref}
\usepackage{authblk}
\usepackage{psfrag}
\usepackage{multirow}

\input{LeeNewcommandv3}

\newcommand{\Dconst}{C_{\mathrm{el}}}
\newcommand{\Cconst}{C_{\mathrm{mag}}}

\def\gamSthree{\gamma^{\mathbb{S}^3}}

\def\c{\widehat{c}}
\def\k{\widehat{k}}
\def\etah{\widehat{\eta}}

\def\F{\mathcal F}
\def\Fsq{\F^2}
\def\I{\mathcal I}
\def\C{\mathcal C}
\def\y{x}
\def\v{\beta}
\DeclareFontFamily{OT1}{rsfs}{}
\DeclareFontShape{OT1}{rsfs}{m}{n}{ <-7> rsfs5 <7-10> rsfs7 <10-> rsfs10}{}
\DeclareMathAlphabet{\mycal}{OT1}{rsfs}{m}{n}

{\catcode `\@=11 \global\let\AddToReset=\@addtoreset}
\AddToReset{equation}{section}

{\catcode `\@=11 \global\let\AddToReset=\@addtoreset}
\AddToReset{figure}{section}

{\catcode `\@=11 \global\let\AddToReset=\@addtoreset}
\AddToReset{table}{section}

\begin{document}

\title{Classification of Kerr-de Sitter-like spacetimes with conformally flat $\scri$
\thanks{Preprint UWThPh-2016-20. }
}
\author[1]{Marc Mars}
\author[2]{Tim-Torben Paetz}
\author[3]{Jos\'e M. M. Senovilla}
\affil[1]{Instituto de F\'isica Fundamental y Matem\'aticas, Universidad de Salamanca, Plaza de la Merced s/n, 37008 Salamanca, Spain}
\affil[2]{Gravitational Physics, University of Vienna, Boltzmanngasse 5, 1090 Vienna, Austria}
\affil[3]{F\'isica Te\'orica, Universidad del Pa\'is Vasco, Apartado 644, 48080 Bilbao, Spain}

\maketitle

\vspace{-0.2em}

\begin{abstract}
We provide a classification of $\Lambda>0$-vacuum spacetimes which admit a Killing vector field with respect to which the associated ``Mars-Simon tensor'' (MST) vanishes
and having a conformally flat $\scri^-$ (or $\scri^+$).
To that end, we also give a complete classification of conformal Killing vector fields on the $3$-sphere $\mathbb{S}^3$ up to M\"obius transformations shedding light on the two fundamental constants that characterize the family of Kerr-de Sitter-like spacetimes, which turn out to be well-defined geometrical invariants.
The topology of $\scri$ is determined in every case, and a characterization result at $\scri$ of the Kerr-de Sitter family presented.
\end{abstract}


\tableofcontents



\section{Introduction}
Understanding the asymptotic properties of spacetimes with a positive cosmological constant  $\Lambda$ is indispensable but at the same time has revealed as more involved than in the absence of $\Lambda$ ---no matter how small this constant may be. It is necessary because there is little doubt at present that the cosmological constant is strictly positive, e.g. \cite{Planck}; but also more complex because, even though the conformal completion \cite{p2} of spacetimes can be equally built, the positivity of $\Lambda$ implies the existence of particle horizons \cite{Wald} or, put simply, that null infinity ---denoted by $\scri$--- is {\em spacelike}, leading to a series of notable difficulties, see e.g. \cite{ashtekar}.

An advantageous fact, though, is the existence of the conformal field equations \cite{F2,F2B} which allow for the definition of a well-posed ``asymptotic Cauchy problem'' by giving appropriate asymptotic initial data at infinity. This was used in the companion paper \cite{mpss} where we analyzed a class of solutions of Einstein's vacuum field equations with positive $\Lambda$ which admit a Killing vector field (KVF) as well as a smooth conformal completion at infinity \`a la Penrose.
The Kerr-de Sitter family is the most prominent example of an explicitly known solution belonging to this class, and it happens to have the additional property that the so-called ``Mars-Simon tensor'' (MST) associated to 
a particular KVF
vanishes.
Thus, in \cite{mpss} we characterized spacetimes with vanishing MST in terms of their asymptotic Cauchy data prescribed at past ---or future--- null infinity.

The aim of this paper is to provide a classification of  $\Lambda>0$-vacuum spacetimes with vanishing MST in terms of those asymptotic Cauchy data. The classification is partial for we will impose the additional hypothesis that null infinity is locally conformally flat.
In such a case asymptotic Cauchy data comprise a locally conformally flat Riemannian 3-manifold
(the conformal gauge freedom allows one to take 
a domain of the round 3-sphere $\mathbb{S}^3$)
together with a ``mass''-parameter and a conformal Killing vector field (CKVF). 
Taking the remaining conformal gauge freedom into account, the classification is reduced to the classification
of CKVFs on $\mathbb{S}^3$ up to M\"obius transformations.
Apart from a certain degenerate case, such CKVFs are completely determined by two invariants $\widehat c$ and $\widehat k$, so that
asymptotic Cauchy data for the class of spacetimes described above consists of three parameters in the generic cases, plus another invariant $\bar r\in \{1,2\}$ which distinguishes two possibilities in the degenerate case.

These asymptotic data need to be related to the $\Lambda>0$-vacuum spacetimes  they generate.
For this  one simply guesses all the spacetimes in this class and shows that all parameter ranges are covered.
Interestingly, proceeding this way we come upon a  $\Lambda>0$-vacuum spacetime which does not seem to have been discussed 
in the literature hitherto, despite the fact that it can be shown to belong to the well-known Pleba\'{n}ski family. Actually, this solution can be obtained as a particular limit of the Kerr-de Sitter family, has one free parameter and seems to describe a inhomogeneization of de Sitter spacetime, facts indicating that it might be of physical relevance. The analysis of this unanticipated solution will be given elsewhere.

The paper is organized as follows:
In Section~\ref{section_background}
we will review the notion of a smooth conformal completion at infinity and the definition of the Mars-Simon tensor.
We will further recall  some results concerning the asymptotic Cauchy problem, crucial for our analysis, and discuss a conformal gauge freedom
contained in asymptotic Cauchy data sets.

In Section~\ref{section_invariants}
we introduce the invariants which are used to accomplish the classification of CKVFs and provide a summary of the most relevant results,
which are derived in detail in the Appendices~\ref{app_SHM}-\ref{app_conf_Eucl}.
Specifically, in Appendix~\ref{app_SHM} we relate the class of CKVFs 
on the $n$-dimensional unit sphere $\mathbb{S}^n$  to the set of
two-forms at the origin in the Minkowski spacetime $\mathbb{M}^{1,n+1}$ by exploiting 
the relation between the conformal group of 
$\mathbb{S}^n$ and the isometry group of hyperbolic space $\mathbb{H}^{n+1}$ viewed
as a spacelike hyperboloid in flat spacetime. We then study the relation between 
the action of the conformal group of the $n$-sphere on CKVFs with 
the action of the orthochronous Lorentz group on two-forms at the origin.
In Appendix~\ref{App:2-forms} we classify algebraically equivalence classes
of two-forms related by orthochronous Lorentzian transformations on the tangent
space of any point in a Lorentzian $d$-dimensional manifold. The classification is achieved
in terms of a set of $[d/2]$ polynomial invariants $I_a$ 
constructed from (a representative) of the two-form and a rank parameter $\bar r$.
 In Appendix~\ref{app_conf_Eucl}
we particularize to the case of $\mathbb{S}^3$, or equivalently
$\mathbb{E}^3$, and obtain the relation between the invariants $I_1, I_2$ ---as now $d=5$--- and the constants $\k(Y)$ and $\c(Y)$ associated to the CKVF $Y$ on $\mathbb{S}^3$ which were shown \cite{mpss} to be
the traces on $\scri$ of geometrically defined scalar quantities associated to
KVFs in $\Lambda >0$-vacuum spacetimes.  The domain of variation for $\k(Y)$ and $\c(Y)$ is
determined explicitly, and we show that except for the values $\k =0$, $\c <0$, these two
constants suffice to label univocally the equivalence classes 
of CKVFs under the action of the conformal group of $\mathbb{S}^3$. 
When $\k=0$, $\c <0$ there are two equivalence classes  which are distinguised by
the value of the rank parameter $\bar r =1$ or $\bar r =2$. For each equivalence
class, we give a canonical representative 
in the $\mathbb{E}^3$ conformal representation of $\mathbb{S}^3 \setminus \{p\}$.

We also include a further Appendix~\ref{sec_hyp_orth}
which  contains a characterization of asymptotic KIDs ensuring that the KVF of the emerging $\Lambda>0$-vacuum spacetime
is hypersurface orthogonal. This shows e.g.\ that the Schwarzschild-de Sitter spacetime admits a hypersurface orthogonal KVF just by checking
the data given at $\scri$.

The classification of $\Lambda>0$-vacuum spacetimes with vanishing MST and locally conformally flat $\scri$ is completed
in Section~\ref{section_alternative}. 
Our main result is formulated in Section~\ref{sec_main_result}, Theorem~\ref{second_main_thm}, where we also discuss the topology
of the $\scri$'s of those spacetimes which arise in our classification.

In \cite{mars-senovilla, mars-senovilla-null} a complete classification (from a spacetime perspective) of $\Lambda$-vacuum spacetimes with vanishing MST has been provided in terms of explicit solutions.
Given a solution of Einstein's field equations it is usually a non-trivial issue to analyze whether or not it admits a smooth conformal completion at infinity. 
Our approach straightforwardly distinguishes those spacetimes in \cite{mars-senovilla} with $\Lambda >0$ which admit a (locally conformally flat) null infinity.
This is the contents of Section~\ref{sec_comparison}.

While most of our characterization is of a local nature, we provide in Section~\ref{section_global} a global characterization of the Kerr-de Sitter family at null infinity, or, more precisely, of the domain of dependence of a connected component $\scri^-\cong\mathbb{S}^3\setminus\{p_1,p_2\}$  of $\scri$.
In addition to the requirement that $\scri^-$ be locally conformally flat, one merely needs to impose certain conditions near one of the poles $p_1$ or $p_2$.
This is achieved by employing the ``Killing initial data'' (KID) equation \cite{ttp2} that the CKVF needs to satisfy on $\scri$.

\section{Background 
}
\label{section_background}
We consider a 4-dimensional connected, oriented and time-oriented spacetime $(\mcM, g)$,
where the metric tensor field $g$ is a smooth solution of Einstein's vacuum field equations
\begin{equation}
R_{\mu\nu}\,=\,\Lambda g_{\mu\nu} \label{efe}
\end{equation}
with a positive cosmological constant $\Lambda >0$. Here $R_{\mu\nu}$ is the Ricci tensor of $(\mcM, g)$, and the corresponding Weyl tensor will be denoted by $C_{\mu\nu\alpha}{}^\beta$  (our sign conventions follow e.g. \cite{Wald}). We will use both abstract index and index-free notations. Boldface letters denote differential forms. Greek indices are spacetime indices while Latin indices will be used for spatial indices corresponding to $\scri$, to be defined presently.

\subsection{Conformal completion, $\scri$ and reduced KID equations}
\label{sec_prelim}
A convenient and useful tool to analyze asymptotic properties of spacetimes, and even properties ``at infinity'', is Penrose's conformal technique \cite{p2}:
\begin{definition}[Conformal completion and $\scri$]
A spacetime $(\mcM, g)$ admits a smooth conformal compactification at infinity if
there exists a spacetime $(\widetilde{\mcM\enspace}\hspace{-0.5em} ,\widetilde g)$ ---called ``unphysical"---and a conformal embedding
\begin{equation}
\mcM  \overset{\phi}{\hookrightarrow} \widetilde{\mcM\enspace} \, , \quad \phi^*(\Theta^{-2}\widetilde g) =  g\;, \quad \Theta\in C^\infty(\widetilde{\mcM\enspace},\mathbb{R}),
\quad \Theta|_{\phi(\mcM)}>0\;, \label{completion}
\end{equation}
such that $\scri=\partial\phi(\mcM)\cap \{ \Theta=0\;, \enspace \mathrm{d}\Theta \ne 0\}$ is a smooth hypersurface called {\em null infinity}.
\end{definition}
We will implicitly identify $\mcM$ with its image $\phi(\mcM)\subset\widetilde{\mcM\enspace}$ and sometimes simply write $\widetilde g =\Theta^2 g$. Indices of physical and unphysical fields are raised and
lowered with $g$ and $\widetilde g$, respectively.

Null infinity $\scri$ is not necessarily connected. One defines {\em past} null infinity as the subset of $\scri$ that contains no endpoint of a future-directed causal curve defined within $(\mcM, g)$, and analogously for {\em future} null infinity. These sets, themselves, are not connected in general either. If the Einstein equations (\ref{efe}) hold with $\Lambda >0$ then $\scri$ is spacelike. This leads to the idea of defining a spacetime by providing an appropriate initial data set at a given connected component of past null infinity, say $\scri^-$ (everything remains valid by defining a {\em final} data set given at a connected component $\scri^+$ of future null infinity). Such an ``asymptotic Cauchy problem'' was proven to be well posed by Friedrich \cite{F_lambda,F2}. More precisely, the Einstein field equations (\ref{efe}) were shown to be equivalent to a set of hyperbolic differential equations on $(\widetilde{\mcM\enspace}\hspace{-0.5em} ,\widetilde g)$, called the Conformal Field Equations, with the important property that they remain regular across $\Theta =0$. The required initial data are just the conformal properties of the hypersurface, that is a 3-manifold with a conformal class of Riemannian metrics ---which plays the role of $\scri^-$ once the equations are solved and the spacetime is built---, supplemented with a 2-covariant, symmetric, traceless and divergence-free tensor field $D$ on that hypersurface.

%

In this paper, in addition to (\ref{efe}) and the existence of a conformal completion in the above sense, we will assume the existence of a Killing vector field $X$. Obviously, $\phi_* X$ is a conformal Killing vector field (CKVF) of $(\mcM,\widetilde g)$ and it happens to extend smoothly as a tangential vector to $\scri$. We will denote by $Y$ the restriction of $\phi_* X$ to $\scri$.
 Then, $Y$ is a CKVF of $\scri$ and furthermore the data $D$ satisfy a transport equation along the flow lines of $Y$. As shown by Paetz \cite{ttp2}, these conditions are also sufficient.

We collect and combine these results in the following fundamental theorem, which is one of the basis for our analysis.

\begin{theorem}[\cite{F_lambda,ttp2}]
\label{thm_well-posedness}
Let $(\Sigma, h)$ be a connected Riemannian 3-dimensional manifold endowed with a symmetric 2-covariant tensor field $D$ and let $\Lambda>0$ be a positive constant. 
\begin{enumerate}
\item[(i)]
The triple $(\Sigma, h, D)$ defines a unique ---up to isometries--- maximal globally hyperbolic development $(\mcM, g)$
of the field equations (\ref{efe}) admitting a conformal completion (\ref{completion}) with isometric embedding $\iota : (\Sigma, h) \hookrightarrow (\widetilde{\mcM\enspace}\hspace{-0.5em} ,\widetilde g)$ such that $\iota(\Sigma) = \scri^{-}$ and $\iota^{\star} ( \Theta^{-1} \widetilde C(n,\cdot,n,\cdot)) = D$
($n$ is the unit normal to $\scri^{-}$ and 
$\widetilde C$ is the Weyl
tensor of $(\widetilde{\mcM\enspace}, \widetilde g)$)
if and only if $D$ is traceless with vanishing divergence.
\item[(ii)] The unphysical spacetime $(\widetilde{\mcM\enspace}\hspace{-0.5em} ,\widetilde g)$ admits a CKVF $X$ which is tangential to $\scri^-$ as well as a Killing vector field of $(\mcM,g)$
if and only if, in addition, the vector field $Y$ defined by $X|_\scri := \iota_\star Y$ is a CKVF of $(\Sigma, h)$
which satisfies the ``reduced KID equations":
\begin{equation}
   \mcL_Y D  + \frac{1}{3} (\mbox{div}_h Y) D =0
\; .
\label{reduced_KID}
\end{equation}
\end{enumerate}
\end{theorem}

A tensor with the above properties of $D$
is said to be a ``traceless and transversal" tensor, or just a ``TT tensor'' for short. A triple $(\Sigma, h, D)$ with the properties of the previous theorem is called {\em asymptotic Cauchy data}. The quadruple $(\Sigma,h,D,Y)$ where $Y$ is a CKVF of $(\Sigma,h)$ satisfying (\ref{reduced_KID}) is called ``asymptotic Killing initial data'', or simply {\em asymptotic KID}.

\begin{remark}
\label{rem_gauge_freedom}
{\rm
Some remarks concerning gauges freedom are in order:
\begin{enumerate}
\item[a)]
Given any diffeomorphism $\Psi: \Sigma \mapsto \Sigma$,
the covariance of the field equations imply that the asymptotic Cauchy data
$(\Sigma, \Psi^{\star}(h), \Psi^{\star}(D))$ are geometrically equivalent
to $(\Sigma, h, D)$.
Standard properties of local groups of transformations imply that
$\Psi^{-1}_{\star}(Y)$ is a CKVF of
$(\Sigma, \Psi^{\star}(h))$. Therefore, it is clear that
$(\Sigma,\Psi^{\star}(h), \Psi^{\star}(D), \Psi^{-1}_{\star}(Y))$ defines asymptotic KID geometrically
equivalent to $(\Sigma,h,D,Y)$. We call this {\bf diffeomorphism invariance} of the data.
\item[b)]
Given the conformal invariance of the conformal field equations, only the conformal class of the initial data
$(\Sigma, h, D)$ matters geometrically \cite{F_lambda}, i.e.\
replacing $h$ by $\Omega^{-2} h$
and $D$ by $\Omega D$, with $\Omega$ any positive smooth function, yields the same physical spacetime. We call this gauge freedom the {\bf conformal rescaling freedom}.
It will play an important role in our classification below.
\item[c)] Applying the gauge transformation  $\Theta\mapsto -\Theta$ would imply
$h\mapsto h$ and $D\mapsto -D$, so that the sign of $D$ is a matter of gauge. However, we assume the conformal factor $\Theta$ to be positive on the physical spacetime $(\mcM,g)$ 
whence the freedom to choose $\mathrm{sign}(D)$ is lost. This will have implications on the allowed range of the physical parameters of the resulting spacetimes.
\end{enumerate}
}
\end{remark}

We will implicitly identify $\iota (\Sigma)$ and $\scri^-$ when there is no risk of confusion. As mentioned above, tensors in $(\scri^-,h)$ carry Latin indices $i,j,k, \cdots$. Norms in $(\scri^-,h)$ will be denoted by
$| \cdot |^2$ and $\nablah$ will refer to the Levi-Civita covariant
derivative of $h$.  We will also need the Cotton-York tensor of the three-dimensional metric $h$, defined by
\begin{align*}
\Cot_{ij} = - \frac{1}{2} \etah_{i}^{\phantom{i}kl}
\Big ( \nablah_l \widehat{L}_{jk} - \nablah_k \widehat{L}_{j l} \Big),
\quad \quad \widehat{L}_{ij} := \widehat{R}_{ij} - \frac{1}{4} \widehat{R}
h_{ij}
\end{align*}
where $\etah$ is the volume form of $h$ and
$\widehat{R}_{ij}$, $\widehat{R}$ are the Ricci tensor and
scalar curvature of $h$ respectively.  Hats will be placed on objects defined on $(\Sigma,h)$ ---except for $h$, $D$ and $Y$.

\subsection{Alignment condition: Mars-Simon tensors}
The non-trivial Killing vector field $X$ of $(\mcM, g)$ satisfies by definition
\begin{equation*}
(\mcL_X g)_{\mu\nu} \,\equiv\, 2\nabla_{(\mu}X_{\nu)} \,=\, 0
\end{equation*}
and thus $F_{\mu\nu}:= \nabla_{\mu}X_{\nu}$ is a two-form  $F_{(\mu\nu)}\,=\,0$. Using  the volume 4-form
$\eta_{\mu\nu\sigma\rho}$ of $g$ one defines the Hodge dual operation $\star$ in the standard way, in particular $2F^{\star}_{\mu\nu}:=\eta_{\mu\nu\sigma\rho}F^{\sigma\rho}$ and the {\em Killing self-dual 2-form} is defined as
$$
\mathcal{F}_{\mu\nu} := F_{\mu\nu} +i F^{\star}_{\mu\nu}
$$
and satisfies $\mathcal{F}^{\star}{}_{\mu\nu} = - i \mathcal{F}_{\mu\nu}$. We set by definition $\mathcal{F}^2 := \mathcal{F}_{\mu\nu} \mathcal{F}^{\mu\nu}$.

The space of self-dual two-forms can be naturally endowed with a metric, described by the symmetric double two-form
$$
\mathcal{I}_{\mu\nu\sigma\rho} := \frac{1}{4} (g_{\mu\sigma}g_{\nu\rho} -g_{\mu\rho}g_{\nu\sigma} + i\eta_{\mu\nu\sigma\rho} ),
$$
in the sense that ${\mathcal{I}}_{\mu\nu\sigma\rho} {\mathcal W}^{\sigma\rho} = {\mathcal W}_{\mu\nu}$ for any self-dual two-form ${\mathcal W}_{\mu\nu}$. Then, using only the Killing vector field $X$ (and $g$) we can define
$$
\mathcal{U}_{\mu\nu\sigma\rho}  := -  \mathcal{F}_{\mu\nu}\mathcal{F}_{\sigma\rho} + \frac{1}{3}\mathcal{F}^2\mathcal{I}_{\mu\nu\sigma\rho}
$$
which is a symmetric double two-form with all the properties \cite{mars,IK} of the Weyl tensor ($\mathcal{U}_{\mu[\nu\sigma\rho]} =0, \, \,  \mathcal{U}^\sigma{}_{\nu\sigma\rho} =0$). It is also self-dual in the sense that $\mathcal{U}^\star_{\mu\nu\sigma\rho} =-i \mathcal{U}_{\mu\nu\sigma\rho}$ (recall that for {\em traceless} double symmetric two-forms the left and right Hodge duals coincide, see e.g. \cite{penrose,S}) . Hence, it seems natural to compare $\mathcal{U}_{\mu\nu\sigma\rho}$ with the self-dual Weyl tensor
\begin{align*}
 \mathcal{C}_{\mu\nu\sigma\rho} :=& C_{\mu\nu\sigma\rho} +i C^{\star}_{\mu\nu\sigma\rho} ,
\end{align*}
the simplest case arising when $\mathcal{U}_{\mu\nu\sigma\rho}$ and $\mathcal{C}_{\mu\nu\sigma\rho}$ are proportional to each other. This was the crucial condition used in \cite{mars,mars1} to characterize the Kerr-NUT metrics among vacuum solutions with a Killing vector, and has since been used successfully  in a number of interesting programs, e.g. \cite{IK,gase,gava,mpss}. The very same ``alignment" condition permits to characterize an important number of distinguished spacetimes in the general case with a non-zero $\Lambda$ \cite{mars-senovilla,mars-senovilla-null}. A summary with all possible $\Lambda$-vacuum (i.e. satisfying (\ref{efe}))
metrics subject to this alignment condition 
can be found in the recent \cite{mars-senovilla-null}.

An equivalent way of stating that the two tensors $\mathcal{U}_{\mu\nu\sigma\rho}$ and $\mathcal{C}_{\mu\nu\sigma\rho}$ are proportional to each other
is the existence of a function $Q$ such that the self-dual double symmetric two-form
\begin{equation}
\mathcal{S}_{\mu\nu\sigma\rho}\,:=\, \mathcal{C}_{\mu\nu\sigma\rho}  + Q\, \mathcal{U}_{\mu\nu\sigma\rho}
\;,
\label{dfn_mars-simon}
\end{equation}
vanishes, i.e. 
\begin{align}
\C_{\mu\nu\rho\sigma} = Q \left (\F_{\mu\nu} \F_{\rho\sigma} 
- \frac{1}{3} \F^2 \I_{\mu\nu\rho\sigma}  \right ).
\label{alig}
\end{align}
The tensors (\ref{dfn_mars-simon}) have come to be known in the literature as \emph{Mars-Simon tensors (MST)} (cf.\ e.g.\ \cite{IK,gase,mpss}). Observe that the function $Q$ has so far been left undefined. We want, however, to analyze the situation where $\mathcal{S}_{\mu\nu\sigma\rho}$ vanishes for some choice of $Q$.
In that case we will necessarily have
\begin{equation}
\mathcal{S}_{\mu\nu\sigma\rho}\mathcal{F}^{\mu\nu}\mathcal{F}^{\sigma\rho} \,=\,0 \quad \Longleftrightarrow \quad  Q\mathcal{F}^{4}\,=\, \frac{3}{2} \mathcal{F}^{\mu\nu}\mathcal{F}^{\sigma\rho} \mathcal{C}_{\mu\nu\sigma\rho}
\;,
\label{definition_Q}
\end{equation}
which determines $Q$ as long as $\mathcal{F}^2$ has no zeros. But the results in \cite{mars-senovilla, mars-senovilla-null} inform us that $\Lambda$-vacuum spacetimes with $\mathcal{F}^2$ vanishing somewhere (and vanishing MST $\mathcal{S}$) necessarily satisfy $\mathcal{F}^2 =0$ {\em everywhere} and furthermore $\Lambda \leq 0$.

Therefore, we can assume without loss generality that $\mathcal{F}^2$ has no zeros and that \eq{definition_Q} determines $Q$.
This choice of $Q$ is denoted by $Q_0$ \cite{mpss},
\begin{equation}
 Q_0\,:=\, \frac{3}{2} \mathcal{F}^{-4}\mathcal{F}^{\mu\nu}\mathcal{F}^{\sigma\rho} \mathcal{C}_{\mu\nu\sigma\rho}
\label{definition_Q0}
\end{equation}
and, from now on, MST will mean $\mathcal{S}$ as in (\ref{dfn_mars-simon}) with $Q=Q_0$.
%

(In some situations, in particular if evolution equations for the MST are needed, it  is convenient to take a different  choice of the function $Q$. We refer the reader to \cite{mpss} for a discussion of this issue).

In \cite{mpss} we have introduced the notion of \emph{Kerr-de Sitter-like spacetimes}:
\begin{definition}
\label{KdS_lile}
Let  $(\mcM,g)$ be a $\Lambda>0$-vacuum spacetime
admitting a smooth  conformal compactification and a corresponding null
infinity $\scri$.  $(\mcM,g)$ is called
``Kerr-de Sitter-like'' at a connected component
$\scri^{-}$ of $\scri$ if it admits a KVF $X$ in the domain of dependence (DoD) of $\scri^-$ such that the associated MST vanishes.
\end{definition}
We will actually say that the spacetime is ``Kerr-de Sitter like" in the entire DoD of $\scri^-$.
In \cite{mpss} we have proven the following result:
%
\begin{theorem}
\label{first_main_thm2}
Let $(\Sigma,h)$ be a Riemannian 3-manifold which admits a CKVF $Y$ with $|Y|^2>0$,
 complemented by
a TT-tensor $D$ to asymptotic KID.
Then there exists a
Kerr-de Sitter-like  spacetime $(\mcM,g)$
such that $\Sigma$ can be identified with $\scri^-$ if and only if, with
$\bm{Y} := h(Y,\cdot)$
\begin{enumerate}
\item[(i)] The Cotton-York tensor of $h$ takes the form $\widehat C = \Cconst |Y|^{-5}( \bm{Y}
\otimes \bm{Y} -\frac{1}{3}|Y|^2 h)$ for some constant $\Cconst\in\mathbb{R}$, and
\item[(ii)]    $D =\Dconst  |Y|^{-5} ( \bm{Y} \otimes \bm{Y}  -
\frac{1}{3}|Y|^2 h)$ for some constant $\Dconst\in\mathbb{R}$.
\end{enumerate}
\end{theorem}
From point c) in Remark \ref{rem_gauge_freedom} we will have to keep the whole real line as the range of the constant $\Dconst$. The reader should keep in mind that the range of the mass-parameter of e.g.\ the Kerr-de Sitter spacetime which will be generated by  asymptotic Cauchy data in our choice of gauge will then happen to be the whole real line as well.

\subsection{Conformal group and the conformal rescaling freedom}
Our aim is to provide a classification of the Kerr-de Sitter like spacetimes according to Theorem \ref{first_main_thm2}, and in this paper this will be done under the additional hypothesis $\Cconst =0$. To that end, and {\em independently} of any assumption on $\widehat C$, our starting point is an analysis of the consequences of the conformal rescaling freedom introduced in point b) of Remark \ref{rem_gauge_freedom}.


Assume therefore that $\Psi$ is an element of the conformal
group of $(\Sigma,h)$, i.e.\ a diffeomorphism $\Psi: \Sigma
\mapsto \Sigma$ satisfying $\Psi^{\star} (h) = \Omega^2 h$
for a smooth positive function $\Omega$. Combining the
diffeomorphism invariance (point a) in Remark \ref{rem_gauge_freedom} and the conformal rescaling freedom we find
(we denote by $\approx$ the geometric equivalence of the data, i.e. the fact
that both sets of data generate the same spacetime up to diffeomorphism):
\begin{align}
(\Sigma,h,D, \Psi_{\star} (Y)) & \approx
(\Sigma,\Psi^{\star} (h), \Psi^{\star}(D), Y)
= ( \Sigma, \Omega^2 h, \Psi^{\star} (D), Y) \nonumber \\
& \approx
( \Sigma, h , \Omega \Psi^{\star} (D), Y).
\label{confgroup}
\end{align}

Thus, under any element of the conformal group we can replace the
CKVF $Y$ by the CKVF $\Psi_{\star}(Y)$,
provided we also transform $D$ as dictated  in (\ref{confgroup}).
We are going to apply this to the case when $D = \Dconst D_{Y}$, with
\be
D_Y := |Y|^{-5} ( \bm{Y} \otimes \bm{Y}  - \frac{1}{3}|Y|^2 h). \label{DY}
\ee

\begin{Lemma}
\label{conflemma}
Let $(\widehat{\Sigma},h)$ be Riemannian, three-dimensional
and admitting a CKVF $Y$ with no zeros. Then the
asymptotic KID $(\Sigma,h, \Dconst D_{Y},Y)$ is geometrically
equivalent to $(\Sigma,h, \Dconst D_{\Psi_{\star}(Y)}, \Psi_{\star}(Y))$,
where $\Psi$ is any element of the conformal group of $(\Sigma,h)$.
\end{Lemma}

\begin{remark}
{\rm Observe that, as remarked in \cite{mpss}, given any three-dimensional Riemannian space $(\Sigma,h)$ admitting a non-trivial CKVF $Y$ and letting
$\widehat{\Sigma}$ be any connected component of
$\Sigma \setminus \{ p \in \Sigma, Y(p)=0\}$,
the quadruple $(\widehat{\Sigma}, h, \Dconst D_{Y},Y )$
defines asymptotic KID.}
\end{remark}

\begin{proof}
The lemma will follow from (\ref{confgroup}) applied to
$ D = \Dconst D_{\Psi_{\star}(Y)}$ provided we can show
\begin{align}
\Omega \Psi^{\star} (D_{\Psi_{\star}(Y)}) = D_Y.
\label{tobeproven}
\end{align}
Notice that $D_{\Psi_{\star}(Y)}$ is defined with respect
to the metric $h$ (cf. (\ref{confgroup})). Since $\Sigma$ now has
two metrics, we distinguish norms by a subscript.
\begin{align*}
|\Psi_{\star}(Y)|^2_h = h( \Psi_{\star}(Y),\Psi_{\star}(Y)) =
\Psi^{\star}(h) (Y,Y) = |Y|^2_{\Psi^{\star}(h)}.
\end{align*}
Let $V_1, V_2$ be arbitrary vector fields on $\Sigma$. The definition of
$D_{\Psi_{\star}(Y)}$ gives
\begin{align*}
\Psi^{\star}&  (D_{\Psi_{\star}(Y)}) (V_1, V_2) =
D_{\Psi_{\star}(Y)}(\Psi_{\star}(V_1),\Psi_{\star}(V_2)) \\
& =  \frac{1}{|\Psi_{\star}(Y)|_h^5}
\Big [ h(\Psi_{\star}(V_1), \Psi_{\star}(Y))
h(\Psi_{\star}(V_2), \Psi_{\star}(Y))
- \frac{1}{3} |\Psi_{\star}(Y)|^{2}_{h}
h(\Psi_{\star}(V_1),\Psi_{\star}(V_2)) \Big ] \\
& =
\frac{1}{|Y|^5_{\Psi^{\star}(h)}}
\Big ( \Psi^{\star}(h) (V_1,Y)
\Psi^{\star}(h) (V_2,Y)  - \frac{1}{3} |Y|^2_{\Psi^{\star}(h)}
\Psi^{\star}(h) (V_1,V_2) \Big ) \\
& = \frac{1}{\Omega |Y|^2_h}
\Big ( h(V_1,Y) h(V_2,Y) - \frac{1}{3} |Y|^2_h
h (V_1,V_2 ) \Big ) \\
& = \frac{1}{\Omega} D_{Y} (V_1,V_2),
\end{align*}
where in the third equality we used $\Psi^{\star}(h) = \Omega^2 h$.
Thus (\ref{tobeproven}) is established and the lemma follows.
\qed
\end{proof}

On any Riemannian space $(\Sigma,h)$, let $\Conf(\Sigma)$
be the conformal group. We introduce an equivalence relation
between CKVFs of $(\Sigma,h)$
defined by $Y_1 \approx Y_2$ iff there exists
an element $\Psi \in \Conf(\Sigma)$ such that
$\Psi_{\star}(Y_1) = Y_2$. The corresponding equivalence classes
are denoted by $[Y]$.
Lemma \ref{conflemma} states that
the geometric KID $(\Sigma,h,\Dconst D_Y,Y)$ depends only on the
equivalence class and not on the representative. Obviously, this
property is interesting as long as the conformal group of
$\Sigma$ is non-trivial, and it is more relevant the larger
the conformal group is. In this paper we are
interested in $(\Sigma,h)$ being locally conformally flat, so that
$(\Sigma,h)$ is locally conformally isometric to an open subset
of Euclidean space $(\mathbb{R}^3, g_E)$, or also
conformally isometric to an open subset of the standard
3-sphere $(\mathbb{S}^3,\gamSthree)$.
The conformal group of $(\mathbb{S}^3,\gamSthree)$
is large (in fact of maximal dimension)
 so in order to classify the KID data
$(\Sigma,h,\Dconst D_Y, Y)$ we need to understand the
equivalence classes $[Y]$ in
$(\mathbb{S}^3,\gamSthree)$ ---and/or in $(\mathbb{R}^3, g_E)$.

\section{Geometrically invariant properties of asymptotic KID at conformally flat $\scri$}
\label{section_invariants}
In this paper we are going to provide a classification of all  spacetimes which can be constructed by ``Kerr-de Sitter like'' asymptotic KID, that is, according to Theorem \ref{first_main_thm2},
\begin{align}
 \Cot_{ij} &= \Cconst|Y|^{-5}\Big(Y_iY_j -\frac{1}{3}|Y|^{2} h_{ij}\Big)\;,
\label{asympt_Cauchy_data_1}\\
\quad  D_{ij} &
=\Dconst  |Y|^{-5} \Big(Y_iY_j -\frac{1}{3}|Y|^{2} h_{ij}\Big)
\;,
\label{asympt_Cauchy_data_2}
\end{align}
where $Y$ is a CKVF of $(\scri^-,h)$, by
making the \emph{additional hypothesis} that
\begin{equation*}
\text{$(\scri^-,h)$ is locally conformally flat.}
\end{equation*}
%
In particular, we are interested in a characterization of the Kerr-de Sitter metric which is  well-known (cf.\ e.g.\ \cite{ashtekar})  to  admit a  locally conformally flat $\scri^-$.

A 3-dimensional Riemannian manifold is locally conformally flat  if and only if its Cotton-York tensor $\Cot_{ij}$
vanishes.
In our setting this will be the case if and only if
%
\begin{equation}
\Cconst\,=\,0
\;,
\label{conf_flat_scri}
\end{equation}
which we assume from now on. Decisive advantages of the assumption \eq{conf_flat_scri} are
that the CKVF $Y$ and other quantities can be computed fully explicitly on a
locally conformally flat $\scri$, and that we can exploit the conformal rescaling freedom as explained in the previous section.

Assume, for the time being, that the constant $\Dconst$ vanished as well. Then $D_{ij}=0$, and the reduced KID equations (\ref{reduced_KID}) would be automatically satisfied, whence any CKVF of $(\scri^-,h)$ would extend to a Killing vector field of $(\mcM,g)$. If $(\scri^-,h)$ is locally conformally flat it admits 10 independent CKVFs, so
the emerging spacetime would be maximally symmetric and
we recover the well-known fact \cite{F_lambda} 
that a locally conformally flat $\scri$ with $D=0$ generates a spacetime 
which is locally the de Sitter spacetime.
Let us therefore assume henceforth that
\begin{equation}
\Dconst\,\ne \, 0
\;.
\end{equation}

We stress that our classification herein is completely independent of the one given in \cite{mars-senovilla} as our analysis will be carried out completely at $\scri^-$. It is an open issue to do a corresponding analysis without the assumption \eq{conf_flat_scri} of a locally conformally flat $\scri$.

\subsection{The constants $\widehat c$ and $\widehat k$
and the equivalence classes $[Y]$ in $(\mathbb{S}^3,
\gamSthree)$}
\label{section_constant}

Two functions $\c(Y)$ and $\k(Y)$ ---which end up, in our setting, being constants--- defined in~\cite{mpss} and associated to $[Y]$ on $(\scri,h)$ turn out to be a key ingredient in our analysis. They are the remnants of two other functions introduced in  \cite{mars-senovilla} for spacetimes $(\mcM,g)$ satisfying (\ref{efe}) and with a Killing vector field $X$ with $\mathcal{F}^2 \neq 0$ and $Q_0 \mathcal{F}^2 - 4 \Lambda \neq 0$.

More precisely, define four real functions $b_1$, $b_2$, $c$ and $k$ by \cite{mars-senovilla, mpss}
\begin{eqnarray}
 b_2-ib_1 &=& - \frac{ 36 Q_0 (\mathcal{F}^2)^{5/2} }{(Q_0\mathcal{F}^2-4\Lambda)^3}
\label{equation_b1b2}
\;,
\\
c &=& - g(X,X)-  \mathrm{Re}\Big(  \frac{6\mathcal{F}^2(Q_0\mathcal{F}^2+2\Lambda)}{(Q_0\mathcal{F}^2-4\Lambda)^2} \Big)
\;,
\label{equation_c}
\\
k  &=&  \Big|\frac{36\mathcal{F}^2}{(Q_0\mathcal{F}^2-4\Lambda)^2}\Big| \nabla_{\mu}Z\nabla^{\mu}Z -  b_2Z +cZ^2 +\frac{\Lambda}{3} Z^4
\;,
\label{equation_k}
\end{eqnarray}
where
\begin{equation}
 Z = 6\,\mathrm{Re} \Big( \frac{\sqrt{\mathcal{F}^2}}{Q_0\mathcal{F}^2-4\Lambda}\Big)
\;.
\end{equation}
For spacetimes with vanishing MST they turn out to be  constant
\cite[Theorem 4 \& 6]{mars-senovilla}. In \cite{mpss} we computed their traces left over on $\scri$ when $\Lambda >0$:
\begin{eqnarray}
(  b_2-ib_1)|_{\scri} &=&        6 \Cconst \Lambda^{-2}
 + 2i \Dconst \Lambda^{-1} \sqrt{ \frac{3}{\Lambda}}
\;,
\label{expression_b1}
\\
\c(Y)\,:=\, \frac{\Lambda}{3} c|_{\scri}
&=& - \frac{1}{4}  \Big( |N|^2  - \frac{4}{9}f^2
+ \frac{8}{3} Y^i\widehat  \nabla_{i}f
+ 8 Y^iY^j \widehat L_{ij}
\Big)
\;,
\label{expression_c}
\\
\k(Y)\,:=\, \Big(\frac{\Lambda}{3}\Big)^3 k |_{\scri}
&=& \frac{1}{18}|Y|^{-2}(Y_kN^k )^2 \Big(f^2
- 3Y^i\widehat  \nabla_{i}f
- 9 Y^iY^j \widehat L_{ij}
\Big)
\nonumber
\\
&& + \frac{1}{2} \Cot_{ij} Y^i Y^j
(Y_kN^k) -\frac{1}{8}\widehat\nabla_i\log |Y|^2 \widehat\nabla^i(Y_kN^k)^2
\nonumber
\\
&&
 +\frac{1}{4} \widehat\nabla_i(Y_jN^j) \widehat\nabla^i(Y_kN^k)
\;,
\label{simplified_k}
\end{eqnarray}
where we have set
\begin{eqnarray}
 N &:=& \mathrm{curl}_h\, Y \;,
\\
f &:=& \mathrm{div}_h\, Y
\;.
\end{eqnarray}
Hence, the constants $b_1$ and $b_2$ are essentially $\Dconst$ and $\Cconst$, respectively. In particular, in our case $b_2$ vanishes.

On the other hand, the quantities $\c(Y)$, $\k(Y)$ can be defined on any
Riemannian $3$-manifold $(\Sigma,h)$ and for any vector
field $Y$ 
($\k(Y)$ only away from the zero set of $Y$).  As a consequence of Theorem
\ref{first_main_thm2}
and the fact that $c$ and $k$ are spacetime constants in the case
of vanishing MST, it follows that $\c(Y)$ and $\k(Y)$ are constant
on $(\Sigma,h)$ satisfying (\ref{asympt_Cauchy_data_1})
provided $Y$ is a CKVF.
This holds, in particular, in the case of $(\Sigma,h)$ being locally
conformally flat we study in this paper.

Since any such $(\Sigma,h)$ can always be viewed (locally) as a subset
of the standard three-sphere, it becomes relevant to study $\c(Y)$ and
$\k(Y)$ for $Y$ a CKVF on $(\mathbb{S}^3,\gamSthree)$.
It turns out that understanding these constants is closely related to
classifying the equivalence classes $\{ [Y] \}$ on the
standard three-sphere. We devote Appendices
\ref{app_SHM}, \ref{App:2-forms} and
\ref{app_conf_Eucl}
to study the classification of CKVFs
of $(\mathbb{S}^3,\gamSthree)$ up to conformal diffeomorphisms, and to
discuss the relation between $[Y]$ and
the constants $\c(Y)$ and $\k(Y)$. A (partial) summary
of the results in
Appendices \ref{app_SHM}, \ref{App:2-forms} and
\ref{app_conf_Eucl} is given in the following
theorem.

\begin{theorem}
\label{Classification}
Let $Y$ denote a CKVF in $(\mathbb{S}^3,\gamSthree)$. The following
properties hold:
\begin{enumerate}
\item The constants $\c(Y)$ and $\k(Y)$ depend only on the
conformal class $[Y]$.
\item The range of $(\c(Y),\k(Y))$ is
\begin{align*}
{\mathcal A}:=\{ \k \geq 0, \c \geq 0\} \cup
\{ \c <0, 4 \k + \c^2 \geq 0 \} \subset \mathbb{R}^2.
\end{align*}
\item Given constants $(\c_1,\k_1) \in {\mathcal A} $
there exists precisely one equivalence class $[Y]$ with
$\c([Y]) = \c_1 ,\k([Y]) = \k_1$ unless $\c_1 <0$ and $\k_1=0$.
\item If $\c_1 <0$ and $\k_1=0$ there are precisely two equivalence classes $[Y]_{\bar r}$
distinguished by a constant $\bar r\in\{1,2\}$ (cf.\ Appendix~\ref{App:2-forms} for its definition).
\item Let $\Psi := \mathbb{S}^3 \setminus \{ p \} \mapsto \mathbb{R}^3$
be the stereographic projection centered at $p$. A canonical representative
$Y$ for each class $[Y]$ exists and, in Cartesian coordinates
$\{ x,y,z \}$ of $\mathbb{R}^3$, is given by:
\begin{itemize}
\item[] Case 1 (a): $\k < 0, \c  \leq - 2 |\k|^{\frac{1}{2}}$. With
$\mu_0 \in (0,1]$ defined uniquely by
$\c = - |\k|^{\frac{1}{2}} \big ( \mu_0^2 + \frac{1}{\mu_0^2} \big )$;
\begin{align*}
\Psi_{\star} (Y)  = |\k|^{\frac{1}{4}} & \Big[ \Big(
\frac{}{} - \mu_0 + \frac{\mu_0}{4}
\big ( y^2 + z^2 - x^2 \big ) \Big )\partial_x
+ \Big( - \frac{\mu_0}{2} xy + \frac{1}{\mu_0} z \Big) \partial_y \\
& +\Big ( - \frac{\mu_0}{2} xz  - \frac{1}{\mu_0} y \Big) \partial_z
\Big ],
\end{align*}
\item[] Case 1 (b): $\k=0, \c <0, \bar r=1:$
\begin{align*}
\Psi_{\star} (Y)  = |\c|^{\frac{1}{2}} \left (
 z \partial_y - y \partial_z  \right )
\end{align*}
\item[] Case 2 (a): $\k >0$. With $\lambda>0$ defined
uniquely by $\c = |\k|^{\frac{1}{2}} \left ( \lambda^2 - \frac{1}{\lambda^2}
\right )$:
\begin{align*}
\Psi_{\star} (Y)  = (\k)^{\frac{1}{4}} & \Big [
\Big ( \lambda x +  \frac{1}{\lambda} y \Big ) \partial_x
+ \Big ( \lambda y -  \frac{1}{\lambda} x \Big ) \partial_y
+ \lambda z  \partial_z \Big ]
\end{align*}
\item[] Case 2 (b): $\k=0, \c >0$:
\begin{align*}
\Psi_{\star} (Y)  = (\c)^{\frac{1}{2}}  (
x \partial_x + y \partial_y + z \partial_z   ).
\end{align*}
\item[] Case 3 (a): $\k=0, \c<0, \bar r=2$:
\begin{align*}
\Psi_{\star} (Y)  = |\c|^{\frac{1}{2}}  (
\partial_x + z \partial_y -y \partial_z   ).
\end{align*}
\item []Case 3 (b): $\k=0, \c=0$:
\begin{align*}
\Psi_{\star}(Y) =  \partial_x.
\end{align*}
\end{itemize}
\end{enumerate}
\end{theorem}

\begin{figure}[!h]
\label{range}
\begin{center}
\psfrag{k}{{\small $\k$}}
\psfrag{c}{{\small $\c$}}
\psfrag{fk}{{\small $\c^2 + 4 \k =0$}}
\includegraphics[width=6cm]{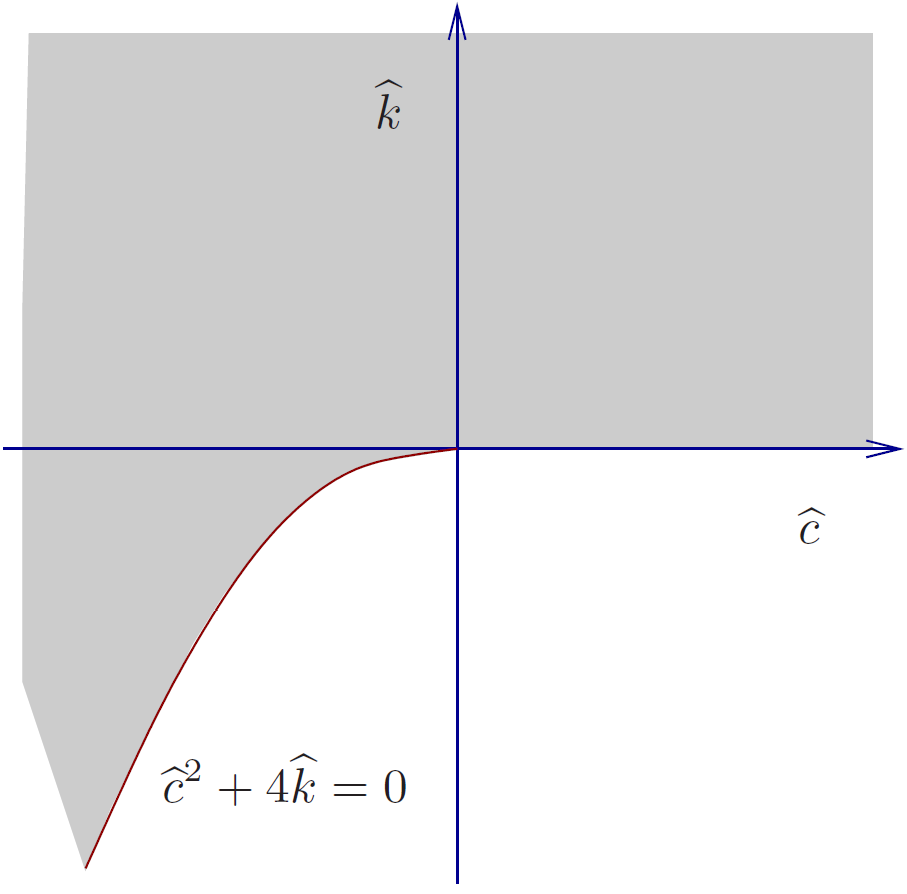}
\caption{Range of the invariants $\k$ and $\c$. The boundary is included
in the set.}
\end{center}
\end{figure}

The subset $\mathcal{A}\subset \mathbb{R}^2$ where the constants $\k$ and
$\c$ can take values is depicted in Figure \ref{range}.
\begin{remark}
\label{barr}
{\rm The rank parameter $\bar r$  in this theorem is used 
to distinguish Cases 1(b) and 3(a). However, this parameter
is defined for all cases (see Appendix \ref{App:2-forms}) and takes
the value $\bar r=2$ for all (a) cases and 
$\bar r=1$ for all (b) cases.}
\end{remark}

\begin{remark}
\label{gauge_freedom_kc}
{\rm The CKVF $Y$ in $(\Sigma,h)$ comes from a Killing vector field $X$ in the
spacetime $(\mcM,g)$  generated by the data
$(\Sigma,h,\Dconst D_Y,Y)$. The freedom in rescaling $X$ translates
into the scaling freedom $Y' = \alpha Y$, with $\alpha \in \mathbb{R}
\setminus \{ 0 \}$. The spacetime will not change if the constant
$\Dconst$ is also transformed to  $\Dconst' = \alpha^3 \Dconst$.
For the quantities $\c(Y)$, $\k(Y)$ this scaling takes the form
(cf. \eq{expression_c}-\eq{simplified_k})
%
\begin{align*}
\k(\alpha Y) = \alpha^4 \k(Y)\;, \quad \quad
\c(\alpha Y) = \alpha^2 \c(Y)\,.
\end{align*}
%

Choosing an appropriate scale  $\alpha$ we   may always achieve
 $\widehat k(Y) \in \{-1,0,+1\}$.
 If $\widehat k(Y)=0$ the scaling freedom remains and
we may achieve $\widehat c(Y) \in \{-1,0,+1\}$. This freedom is used
in Appendix \ref{app_conf_Eucl}
to simplify the expressions. Nevertheless, in the
statement of Theorem \ref{Classification} we have restored
the general values of $\k(Y)$ and $\c(Y)$ for the sake of completeness.}
\end{remark}


Recall that the set of CKVFs in Euclidean 3-space can be parametrized by one constant
$v \in \mathbb{R}$ and three constant vector fields
$\vec{a}, \vec{b}, \vec{w} \in \mathbb{R}^3$ ---this is also shown in the Appendices, see formulas (\ref{CKV}) and (\ref{app_general_CKVF}). The corresponding CKVF $Y$ is given by
(we often identify $Y$ and $\Psi_{\star}(Y)$ under the stereographic
projection $\Psi$ when no confusion arises),
\begin{equation}
Y \,=\, \vec b + \vec w \times \vec x  +v \,\vec x+ (\vec a\cdot \vec x)\vec  x - \frac{1}{2} |\vec x|^2 \vec a \;.
\label{general_CKVF}
\end{equation}
This
yields, as a by-product, explicit expressions for the
constants $\k(Y)$ and $\c(Y)$ in terms of the parametrization
of the CKVFs in $\mathbb{R}^3$ given by (\ref{general_CKVF})
Specifically,
it is proven in (\ref{app_general_k})-(\ref{app_general_c}) that these constants are given by
\begin{eqnarray}
\widehat  c(Y) &=&v^2 -  |\vec w|^2-2\vec b \cdot \vec a
\;,
\label{general_c}
\\
\widehat k(Y) &=& v^2|\vec w|^2 + 2v\,\vec a \cdot (\vec w \times \vec b ) + |\vec a \times\vec b|^2
-2 (\vec a\cdot \vec w)(\vec b \cdot \vec w)
\;.
\label{general_k}
\end{eqnarray}
Obviously, these expressions can also be computed directly from their
definitions (\ref{expression_c})-(\ref{simplified_k}) after using
\begin{eqnarray}
N &=& \mathrm{curl} \,Y \,=\,  2\vec w + 2\vec a\times \vec x
\;,
\label{general_Z}
\\
f &=& \mathrm{div} \,Y \,=\, 3 v+ 3\vec a \cdot \vec x
\;,
\\
Y_kN^k &=& 2\vec b \cdot \vec w +2v\,\vec w \cdot \vec x + 2(\vec b \times \vec a)\cdot\vec x
+|\vec x|^2(\vec w \cdot \vec a)
\;.
\label{general_YZ}
\end{eqnarray}
The computation turns out to be remarkably long, specially for $\k(Y)$,
but serves as a useful check for the validity of the expressions.

The constant $\widehat c$ can be identified with the norm of the CKVF $Y$ w.r.t.\  the ``Killing metric''
$G_{ij} = -\widehat\Gamma^k_{il}\widehat\Gamma^l_{jk}$ considered e.g.\ in \cite{geroch}.

Concerning the rank parameter $\bar r\in\{1,2\}$ that distinguishes between cases $1 (b)$ and $3 (a)$ in Theorem \ref{Classification}, and as shown in Appendix \ref{app_conf_Eucl}, $\bar r=1$ if and only if the following condition holds
\be
v \vec w +\vec b\times \vec a = \vec 0 , \quad \quad \vec w \cdot \vec a = \vec w \cdot \vec b =0 \label{r1} .
\ee
It should be stressed that this condition as well as expressions (\ref{general_c}) and (\ref{general_k}) are independent of the representative of the class $[Y]$. Notice also that (\ref{r1}) implies $Y_k N^k=0$ as follows from \eq{general_YZ}.

\subsection{The classes $[Y]$ according to their fixed points.}
An interesting property of the classification scheme
presented in Theorem \ref{Classification} is that the different
classes can also be (partially) distinguished by the set
of zeros of the CKVF $Y$.

Note first that the
set $\{ q \in \mathbb{S}^3, Y(q) =0\} \subset \mathbb{S}^3$
is transformed diffeomorphically under an element of the conformal group
$\Conf(\mathbb{S}^3)$. Thus, the structure of the zero set
of a CKVF depends only on the equivalence
class, and not on the representative chosen. In order to analyze this
structure we need to find the zero set 
for each one of the canonical representatives in Theorem \ref{Classification}.
Using the stereographic projection of $\mathbb{S}^3$ based at a pole $p$,
the zeros outside $p$ can be directly computed from the explicit expression
(\ref{general_CKVF}). The question about when the pole $p$ can be also a fixed point of $Y$ is discussed and solved in Appendix \ref{newApp}. The Lemma \ref{zero_at_infinity_0} there has an immediate translation to the case of $n=3$ as follows
\begin{lemma}
\label{zero_at_infinity}
Let $Y$ be a CKVF in $(\mathbb{S}^3,\gamSthree)$
and $\Psi: \mathbb{S}^3 \setminus \{p \} \mapsto \mathbb{R}^3$
 the stereographic projection centered at $p$. Define
$v, \vec{a},\vec{b},\vec{w}$ as the constants corresponding to the
CKVF $\Psi_{\star} (Y)$ in Euclidean $3$-space as in (\ref{general_CKVF}).
Then $Y(p)=0$ if and only if $\vec{a}=0$.
\end{lemma}

Therefore, only $\vec{a}$ in (\ref{general_CKVF}) plays a role to determine whether
this vector field vanishes ``at infinity'' or not. Then we have the following interesting result.
\begin{Proposition}
\label{zeroes}
Let $Y$ be a CKVF on $(\mathbb{S}^3, \gamSthree)$. The zeros of $Y$ depend
on the conformal class $[Y]$ as follows:
\begin{itemize}
\item[(i)]  If $[Y]$ belongs to Case 1 (a) then
$Y$ has no zeros.
\item[(ii)]  If $[Y]$ belongs to Case 1 (b)  then
$Y$ vanishes on a closed, embedded curve.
\item[(iii)]  If $[Y]$ belongs to Cases 2
then $Y$ vanishes at precisely two isolated points.
\item[(iv)]  If $[Y]$ belongs to Cases 3 then
$Y$  vanishes at precisely one isolated point.
\end{itemize}
\end{Proposition}

\begin{proof}
As discussed above, the zero set of $Y$ is, up to diffeomorphism,
independent of the representative. We thus use the canonical representative
in each equivalence class as given in Theorem \ref{Classification}.

In Case 1 (a), the CKVF $\Psi_{\star}(Y)$ vanishes nowhere
in $\mathbb{R}^3$. Indeed, setting the
coefficient in $\partial_z$ to zero requires
$y = - \frac{\mu_0^2}{2} xz$ which inserted into the $\partial_y$ component
yields $(1 + (\mu_0^4/4) x^2) z=0$. Hence $\{z=0,y=0\}$ and then the coefficient
in $\partial_x$ vanishes nowhere. Since $\vec{a} \neq 0$ is non-zero
for this CKVF, it follows from Lemma \ref{zero_at_infinity}
that  $\Psi_{\star}(Y)$ is also non-zero
``at infinity'', and claim $(i)$ follows.
In Case 1 (b), it is clear that the CKVF vanishes at the line $\{z=y =0 \}$.
Since $\vec{a}=0$, it also vanishes ``at infinity'', and the canonical
representative vanishes on an $\mathbb{S}^1$ embedded in $\mathbb{S}^3$.
This proves $(ii)$.
In Cases 2, both (a) and (b), $\Psi_{\star}(Y)$ vanishes at
the single point $x=y=z=0$ in $\mathbb{R}^3$. Since in both cases
$\vec{a}=0$, it follows that the canonical representative
vanishes precisely at two isolated points, proving $(iii)$. In Cases 3, either (a) or (b), $\Psi_{\star}(Y)$ does not vanish anywhere on $\mathbb{R}^3$,
but it vanishes ``at infinity'' given that $\vec{a}=0$, and item $(iv)$ follows.
\qed
\end{proof}

\subsection{Existence of a second CKVF $\widehat\varsigma$ and the dimension of $\mathrm{span}\{Y,\widehat\varsigma\}$}
\label{section_dimension}
From the results in \cite{mars-senovilla} we know that any $\Lambda$-vacuum spacetime 
 which admits a Killing vector field $X$ whose associated MST vanishes,  possesses a second Killing vector field  $\varsigma$  (not necessarily, but generally, independent from the original one), which can be expressed in terms of the first one \cite{mars-senovilla} as
\begin{eqnarray}
\varsigma^{\mu} \,=\, \frac{4}{|Q_0\mathcal{F}^2-4\Lambda|^2}X^{\sigma}\ol{\mathcal{F}}_{\sigma}{}^{\rho}\mathcal{F}_{\rho}{}^{\mu} +\mathrm{Re}\Big(\frac{\mathcal{F}^2}{(Q_0\mathcal{F}^2-4\Lambda)^2}\Big) X^{\mu}
\,.
\label{varsigma}
\end{eqnarray}
%
Assuming that the spacetime has a $\scri$, this second KVF induces there a second CKVF $\widehat\varsigma$ \cite{mpss},%
\footnote{
Note that the vector field $\widehat\varsigma$  differs from the corresponding one in \cite{mpss} by a factor of $9$.
The advantage of the current  definition is that \eq{B}-\eq{A}
take a simpler form.
}
%
\begin{eqnarray}
\widehat\varsigma^i (Y)\,:=\, 2\Lambda^2\varsigma^{i}|_{\scri}
&=&
\frac{1}{4}Y_k N^k  N^i
+ Y^i \Big(  Y^kY^l\widehat L_{kl}+
   \frac{1}{3}Y^k\widehat\nabla_k f \Big)
\nonumber
\\
&&
 -|Y|^2 \Big(\widehat L^i{}_k Y^k  + \frac{1}{3}\widehat \nabla^if \Big)
+
 \frac{1}{6} f\widehat\epsilon^{ikl}Y_kN_l
\label{second_Killing_gen}
\;.
\end{eqnarray}
A computer algebra calculation shows that in Euclidean space, where $Y$ is given by \eq{general_CKVF}, we have
%
\begin{eqnarray}
\widehat\varsigma(Y)
&=&
\vec B+ \vec W \times \vec x  +U \,\vec x+ (\vec A\cdot \vec x)\vec  x - \frac{1}{2} |\vec x|^2 \vec A
\;,
\end{eqnarray}
with
\begin{eqnarray}
 \vec B &=&   v \,\vec b\times \vec w  +(\vec b \cdot\vec w) \vec w
+(\vec a\cdot \vec b)\vec b -|\vec b|^2\vec a
\;, \label{B}
\\
\vec W &=& (\vec b\cdot \vec w)\vec a + (\vec a \cdot\vec w)\vec b  -v^2\vec w +v(\vec a\times\vec b)
\;,
\\
U &=& v|w|^2  +\vec a \cdot(\vec w\times\vec b)
\;,\label{V}
\\
\vec A&=& v(\vec w\times\vec a) +  (\vec a\cdot \vec b)\vec a -|\vec a|^2\vec b +(\vec a \cdot\vec w )\vec w\label{A}
\;.
\end{eqnarray}

This second KVF will be associated to a 2-form $\bm{G}$ in flat spacetime $\Minkfour$, in accordance with the discussion in the Appendices, whose decomposition of type (\ref{F=vabw}) is denoted using capitals (and $W^{A}=\epsilon^{ABC}\Omega_{BC}$):
$$
\bm{G} = U dt\wedge dx^1 + dt \wedge (A-B/2) -dx^1 \wedge (A +B/2) -\bm{\Omega} .
$$

Relations (\ref{B})-(\ref{A})
suggest that the endomorphism $G$ is related to $F^3$, and thereby that the second KVF involves $\zeta^{(1)}=\zeta_{F^3}$ in the notation of section \ref{subsec:2nd}.
A direct computation gives
$$
(F^3)^0{}_1 = F^0{}_{\rho}F^{\rho}{}_{\sigma}F^\sigma{}_1 =-v^3 +2v\,  \vec b \cdot \vec a +\vec a \cdot(\vec w\times\vec b)
$$
and using here (\ref{general_c}) and (\ref{V})
$$
(F^3)^0{}_1-\widehat c F^0{}_1 =-v|w|^2  -\vec a \cdot(\vec w\times\vec b) =-U.
$$
This identifies $G$ as being
$$
G=F^3-\widehat c F =F^3 -I_1 F
$$
where $I_{1}=\mbox{tr}(F^{2})/2 = \c$
is the first invariant of $F$ as defined in (\ref{inv}).
Computing the
endomorphism $F^3 - I_1 F$ with $F$ as in (\ref{ExpF}) and reading
off the corresponding coefficients $\{ V,\vec{A},\vec{B},\vec{W} \}$ one
recovers
precisely (\ref{B})-(\ref{A}) which confirms that indeed
$\widehat{\varsigma} = \zeta_{G}$.
Since $G$ obviously commutes with $F$ we find, as a  consequence
of (\ref{commutator}) in Appendix \ref{App:2-forms}, that the
CKVF $\widehat\varsigma$ commutes with $Y$.
This recovers, at $\scri$, the commutation of $Y$ and $\varsigma$
proven in \cite{mars-senovilla} for spacetimes with vanishing MST.

An alternative expression for $G$ can be obtained by
using $I_{1}=\mbox{tr}(F^{2})/2 = - \frac{1}{2} F_{\alpha\beta} F^{\alpha\beta}$,
$$
G_{\mu\nu} =\frac{1}{2} F^{\rho\sigma} (F\wedge F)_{\mu\nu\rho\sigma}=\frac{3}{2} F^{\rho\sigma} F_{\mu[\nu}F_{\rho\sigma]} .
$$

From here one immediately sees that the second CKVF
 is trivial in {\em all} cases with $\bar r=1$ (because then $\bm{F}\wedge \bm{F} =0$). This corresponds --- see Remark \ref{barr}--- to all cases (b) in point 5 of Theorem \ref{Classification}

Concerning the other possibility $\bar r=2$, that is cases (a) in Theorem \ref{Classification}, by computing $\bm{G}$ with the above formula and using the canonical expressions provided, one easily checks that
\begin{itemize}
\item  if $[Y]$ belongs to case 3(a), then the CKVF $\widehat\varsigma$
 belongs to case 1(b), so that dim(span$\{\widehat\varsigma,Y\})=2$.
\item if $[Y]$ belongs to case 2(a) (setting $k=1$), then the CKVF $\widehat\varsigma$
also belongs to the same case 2(a) with $V=1/\lambda$ while $\vec W =-\lambda (0,0,1)$, so that again dim(span$\{\widehat\varsigma,Y\})=2$.
\item finally, if $[Y]$ belongs to case 1(a) (setting $k=-1$), then  the CKVF $\widehat\varsigma$
 belongs to the same case 1(a) with $\mu_0 \rightarrow 1/\mu_0$, so that once more dim(span$\{\widehat\varsigma,Y\})=2$ {\em except} in the case $\mu_0 =1$, where $\bm{G}$ and $\bm{F}$ are 
proportional and thus, $Y$ is parallel to $\widehat\varsigma$. This corresponds to the exceptional case with $k=-1$ and $c=-2$.
\end{itemize}

%
%
%
%

%

We have thus proven 
\begin{lemma}
\label{lemma_dimensions}
If $\scri$ is conformally flat, $\mathrm{dim}(\mathrm{span}\{Y,\widehat\varsigma\})=1$ in Cases (b) of Theorem \ref{Classification} while $\mathrm{dim}(\mathrm{span}\{Y,\widehat\varsigma\})=2$ in Cases (a) of that theorem with the exception of the particular subcase of $1(a)$ with 
$\k + 4 \c^2 =0$ and $\c <0$, where also $\mathrm{dim}(\mathrm{span}\{Y,\widehat\varsigma\})=1$.
\end{lemma}
In other words, the two CKVFs are co-linear in all cases $(b)$ of Theorem \ref{Classification} plus the exceptional case of $1(a)$ with $\k + 4 \c^2 =0$ and $\c <0$ (these correspond to the generalized Kottler spacetimes and to the Taub-NUT-de Sitter spacetime with NUT-parameter $ \ell = \frac{1}{2}\sqrt{\frac{3}{\Lambda}}$, respectively, cf.\ below).

\begin{remark}
{\rm
Apart from the special case just mentioned, that means that $\Lambda>0$-vacuum spacetimes with vanishing MST and conformally flat $\scri$ belonging to Cases $(a)$ ($\bar r=2$) also belong to the class (B.i) in  \cite[Theorem 4]{mars-senovilla}.
}
\end{remark}

%

\section{Classification of Kerr-dS-like spacetimes with conformally flat $\scri$}
\label{section_alternative}

The aim of this section is to provide a complete classification of $\Lambda>0$-vacuum spacetimes with vanishing MST and conformally flat $\scri$
from the point of view of an asymptotic Cauchy problem with data on $\scri$.
In doing so we shall re-derive and extend some of the results in \cite{mars-senovilla}, but
our analysis is to a large extent independent of the results established in \cite{mars-senovilla, mpss}.
We merely need the ``only if''-part of Theorem~\ref{first_main_thm2} that provides necessary conditions at $\scri$ to end up with a Kerr-dS-like spacetime.

Based on that result, the idea to classify Kerr-dS-like spacetimes
 which admit a conformally flat $\scri^-$
now goes as follows (recall that a conformally flat $\scri^-$ corresponds to asymptotic Cauchy data with $\Cconst=0$):
Since $(\scri^-,h)$ is conformally flat and  only the conformal class matters geometrically
(cf.\ Remark~\ref{rem_gauge_freedom} b)),
  $(\scri^-,h)$ can locally be represented by Euclidean 3-space.
Then the Killing vector field $X$ induces on $\scri^-$ a CKVF $Y$ of the form \eq{general_CKVF}.
Theorem~\ref{first_main_thm2} states that
the data for the rescaled Weyl tensor need to be of the form
\begin{eqnarray}
D_{ij} \,=\, \Dconst  |Y|^{-5}(Y_iY_j -\frac{1}{3}|Y|^2 h_{ij})
\label{condition_on_D}
\;,
\end{eqnarray}
  in order to end up with a spacetime with vanishing MST.
The Cauchy data are thus fully determined by the CKVF $Y$ and the parameter $\Dconst$.

It was shown in Section~\ref{section_constant}
that  the spacetime functions $k$ and $c$ defined  by \eq{equation_k} and \eq{equation_c}, respectively,
induce the invariants $\widehat k (Y)$ and $\widehat c (Y)$ on $\scri$ as given by \eq{general_k} and \eq{general_c}, respectively.%
\footnote{For the argument presented here the fact \cite{mars-senovilla} that, in the setting where the MST vanishes, $k$ and $c$ are constants in spacetime is not needed.}
There remains the gauge freedom to perform conformal transformations  (i.e.\ M\"obius transformations)  of Euclidean 3-space.
It follows from Theorem \ref{Classification}
that given $\widehat k $ and $\widehat c$  there exist either none, one, or two equivalence classes of CKVFs which induce these values of $\widehat k$ and $\widehat c$, the latter case happening when $\widehat k=0$ and $\widehat c <0$. In this situation with two classes of CKVFs, one can distinguish between them by checking whether or not the M\"obius invariant condition (\ref{r1}) holds.
Well-possedness of the asymptotic Cauchy problem with data on $\scri^-$, cf.\  Theorem~\ref{thm_well-posedness}, implies
that for given  $\{\widehat k, \widehat c\}$ and $\Dconst\in\mathbb{R}$ there exist, up to isometries, either none, a unique, or precisely  two, $\Lambda>0$-vacuum spacetimes
which induce these values on $\scri$.

Taking further into account  that
a scaling factor in $Y$ can be absorbed by the constant $\Dconst$, the CKVF $Y$ can be assumed to adopt one of the ``standard forms'' presented in Theorem \ref{Classification}
with appropriately normalized values of the constants $\widehat k$ and $\widehat c$ in accordance with Remark \ref{gauge_freedom_kc}.
In the case of a conformally flat $\scri^-$,
Kerr-de Sitter-like spacetimes are therefore generated by asymptotic KID which can be parametrized  by $\widehat k\in \{1,0,-1\}$, $\widehat c\in \mathbb{R}$ and $\bar r\in\{1,2\}$ according to Theorem \ref{Classification}
plus $\Dconst\in \mathbb{R}$.

Subsequently, we shall show that letting aside de Sitter (dS) spacetime 
which, as explained above, corresponds to $\Dconst =0$,
the Kerr-de Sitter family \eq{KdS_metric_old}, the generalized Kottler spacetimes \eq{kottler_metric}, the Taub-NUT-de Sitter spacetime \eq{taub_spec} with $ \ell = \frac{1}{2}\sqrt{\frac{3}{\Lambda}}$, the Wick-rotated Kerr-anti-de Sitter family \eq{wick-KdS_metric_old} and the $a\rightarrow \infty$-KdS-limit-spacetime \eq{KdS_limit} exhaust all possible values of $\widehat k$, $\widehat c$, $\bar r$ and $\Dconst\in\mathbb{R}\setminus\{0\}$.
Since we know that these spacetimes can be generated by a Cauchy problem with asymptotic KID,
we then  conclude that these are the only $\Lambda>0$-vacuum spacetimes with vanishing MST and conformally flat $\scri^-$. We further express their mass and angular momentum in terms of the asymptotic  data.


The following lemma will be useful to classify the possible geometries
of $\scri$ of $\Lambda>0$-vacuum spacetimes with vanishing MST and admitting
a smooth conformal compactification.

\begin{lemma}
\label{hmetric}
Let $(\Sigma,h)$ be a three-dimensional Riemannian manifold with metric
\begin{align*}
h =  \left ( dt + A(\theta)d\phi \right )^2 + H^2(\theta) d\theta^2 + W^2(\theta)
d\phi^2
\end{align*}
where $H(\theta)$ and $W(\theta)$ are positive.
With $Y = \partial_t$, the functions $\widehat{c}(Y)$ and $\widehat{k}(Y)$ defined by \eq{general_c} and   \eq{general_k}, respectively,
 are
\begin{align}
\widehat{c}(Y) & =- \frac{1}{H^2} \frac{W''}{W}  +
\frac{H' W'}{H^3 W} - \frac{3}{2} s^2, \label{exphatc} \\
\widehat{k} (Y) & = \frac{s}{2} \Cot_{t\, t}
+ \frac{1}{4} \Big ( \frac{s'{}^2}{H^2} + s^2 \widehat{c}(Y) + \frac{1}{4} s^4
\Big ),   \label{exphatk}
\end{align}
where $s=s(\theta)$ is defined as $s(\theta):= H^{-1} W^{-1} A'$,
prime denotes derivative with respect to $\theta$
and $\Cot_{ij}$ is the Cotton-York tensor of $(\Sigma,h)$. $(\Sigma,h)$
is locally conformally flat if and only if
the following three conditions hold:
\begin{align*}
& \widehat{c}(Y)=  \mbox{\rm{constant}}, \quad \quad
\frac{s'}{HW}= \mbox{\rm{constant}}, \quad \quad
\frac{1}{2} H^2 s \left ( 2 \widehat{c}(Y) + s^2 \right )
+ \frac{W' s'}{W} =0.
\end{align*}
\end{lemma}

\begin{proof}
By direct computation one checks that the following equalities
hold
\begin{align}
\Cot_{t \, t} & = - \frac{2 \Cot_{\theta \, \theta}}{H^2} -
\frac{W}{2 H} \frac{\partial}{\partial {\theta}} \Big (
\frac{s'}{HW} \Big ), & & &  \Cot_{t \, \theta}  & =0, \label{iden1}\\
\Cot_{t \, \phi }  & = A \Cot_{t \, t} + \frac{W}{2H} \frac{\partial
\Gamma}{\partial \theta},
& & &  \Cot_{\theta \, \phi}  & =0, \label{iden2} \\
\Cot_{\phi \, \phi} & = - \frac{W^2}{H^2} \Cot_{\theta\, \theta}
- (A^2+W^2) \Cot_{t\, t} + 2 A \Cot_{t \phi } ,
 \label{iden3}
\end{align}
where  $\Gamma$ is just a shorthand for the right-hand side of
(\ref{exphatc}) and the last identity corresponds to
the fact that
the Cotton-York tensor is trace-free. From (\ref{iden1})-(\ref{iden3})
it follows that  $\Cot_{ij}=0$  iff $\Cot_{\theta \, \theta}=0$ and
$\Gamma, H^{-1} W^{-1} s'$ are both constant. The explicit form of $\Cot_{\theta\, \theta}$
is
\begin{align*}
\Cot_{\theta\, \theta} =
\frac{1}{4} H^2 s \left ( 2 \Gamma + s^2 \right )
+ \frac{1}{2} \frac{W' s'}{W},
\end{align*}
and the second part of the lemma is proven, as long as we can show that $\Gamma =\widehat{c}(Y)$. Concerning the
expressions for $\c(Y)$ and $\k(Y)$
note that $Y$ is a Killing vector so that
$f := \mbox{div} Y =0$. Moreover, given that $\bm{Y} = dt + A d\phi$,
the curl of $Y$ is (we take the orientation
$\{t,\theta,\phi\}$ positive, but the result is independent of this choice)
\begin{align*}
N^k = \eta_h^{ijk} \nablah_i Y_j = \delta^k_t \frac{A'}{H W}
\quad \quad \Longrightarrow \quad \quad N = s \, Y,
 \end{align*}
where in the second equality we used that $\mbox{det}(h) = HW$
and the definition of $s$ has been inserted
in the last step. For Killing vectors,
the definitions of $\widehat{c}, \widehat{k}$
\eq{expression_c}-\eq{simplified_k}
simplify to
\begin{align*}
\widehat{c} (Y)  = & - \frac{1}{4} |N|^2 - 2 Y^i Y^j \widehat{L}_{ij}, \\
\widehat{k} (Y) = & - \frac{(Y_k N^k)^2}{2 |Y|^2} \widehat{L}_{ij} Y^i Y^j
+ \frac{1}{2} (Y_k N^k) \Cot_{ij} Y^i Y^j
- \frac{1}{8} |Y|^{-2} \nablah_i |Y|^2  \nablah^i ( Y_k N^k)^2  \\
& + \frac{1}{4} \nablah_i ( Y_k N^k) \nablah^i (Y_j N^j).
\end{align*}
The result then follows from $|Y|^2 =1$, $Y_k N^k = s$, $|N|^2 = s^2$
and the fact that the $\{ t \, t\}$ component of the Schouten
tensor of $h$
turns out to be
\begin{align*}
\widehat{L}_{t \, t} = \frac{1}{2 H^2} \frac{W''}{W}
- \frac{1}{2} \frac{H' W'}{H^3 W}
 + \frac{5}{8} s^2.
\end{align*}
\qed
\end{proof}

%
%



We are now ready to start our classification.

\subsection{Cases (b): Generalized Kottler spacetimes}
Consider first the so-called
generalized Kottler metrics (cf.\ e.g.\ \cite{cs}),
%
\begin{equation}
 g \,=\, -\Big(\varepsilon -\frac{2m}{r}-\frac{\Lambda}{3}r^2\Big) \mathrm{d}t^2 + \Big(\varepsilon -\frac{2m}{r}-\frac{\Lambda}{3}r^2\Big)^{-1}\mathrm{d}r^2 + r^2\mathrm{d}\Omega_{\varepsilon}^2
\;,
\label{kottler_metric}
\end{equation}
with
\begin{equation}
 \Lambda >0\;, \quad m \in\mathbb{R}\setminus \{0\}
\;,
\end{equation}
and where $\varepsilon=0,\pm 1$ and $\mathrm{d}\Omega_{\varepsilon}^2$ denotes a metric of constant Gau\ss\enspace  curvature $\varepsilon$.
A direct computation shows that the MST associated to the Killing vector
$X=\partial_t$ vanishes.

In the conformally rescaled spacetime with conformal factor $\Theta=\sqrt{\frac{3}{\Lambda}} r^{-1}$ one obtains as induced metric on $\scri^-$, defined
by $\{ \Theta =0 \}$,\footnote{
Here and in the subsequent sections we stick to the standard notation of Boyer-Lindquist-type coordinates. That means the coordinate normal to $\scri^-$
is denoted by $r$, while $t$ is used as a coordinate to parametrize $\scri^-$.
}
\begin{equation}
 h \,=\, \mathrm{d}t^2 + \frac{3}{\Lambda}\mathrm{d}\Omega_{\varepsilon}^2
\;.
\end{equation}
The CKVF induced on $\scri^{-}$ is $Y= \partial_t$.
Writing
$$d \Omega_{\varepsilon}^2 = d \theta^2 + \Sigma^2 (\theta,\varepsilon) d \phi^2, \hspace{1cm}
\partial^2_{\theta\theta}\Sigma = - \varepsilon \Sigma
$$
we
can apply Lemma \ref{hmetric} with
$H = \sqrt{\frac{3}{\Lambda}}$, $W = H \Sigma$ and
$A = 0$ (so that $s= 0$). It follows that $h$ is locally conformally flat and
 (we drop the reference to $Y$)
%
%
\begin{align*}
 \widehat k&=0\;, \quad \quad \widehat c = \frac{\Lambda}{3} \varepsilon
\;.
\label{Kottler_kc}
\end{align*}
%
%
%
In terms of the rescaled
%
$Y_{\mathrm{norm}} =\sqrt{ \frac{3}{\Lambda}} Y$ we have
\begin{equation}
 \widehat k=0\;, \quad \quad \widehat c =  \varepsilon
\;.
\end{equation}
This corresponds to cases $2 (b)$ and $3 (b)$ whenever $\varepsilon =1$ or $0$, respectively.
Concerning $\varepsilon =-1$, there still are two possibilities depending on the value of $\bar r$.
We can identify the conformal class $[Y]$ in this case by simply checking that (\ref{r1}) holds for $Y_{\mathrm{norm}}$, implying that $\bar r=1$ and thus we are in case $1 (b)$. (An alternative way of seeing this is by noting that this spacetime admits four independent commuting Killing vector fields, so that $\scri^{-}$
must admit four independent commuting CKVFs: Only the class
$\widetilde{k}=0$, $\widetilde{c} <0$ and $\bar r=1$ has this
property ---see the discussion after (\ref{specialcase})).

We further find that
(we always use the
spacetime Killing vector associated to 
$Y_{\mathrm{norm}}$ to compute $b_1$ and $\Dconst$).
\begin{equation}
 b_1
\,=\,  36  \,\mathrm{Im}\Big(\frac{ Q_0 \mathcal{F}^5}{ (Q_0\mathcal{F}^2-4\Lambda)^3} \Big)
\,=\, 2m \Big(\frac{3}{\Lambda}\Big)^{3/2}
\;,
\end{equation}
%
%
hence
\begin{eqnarray}
 \Dconst \,=\,- \frac{1}{6}\Lambda^2\sqrt{\frac{3}{\Lambda}} b_1
 \,=\,- 3m
\;.
\label{kottler_param_m_C}
\end{eqnarray}

In summary, the generalized Kottler spacetimes correspond to the three cases characterized by $\bar r=1$, that is, Cases $1(b), 2 (b)$ and $3 (b)$ in Theorem \ref{Classification} 
(see Remark \ref{barr}).

\subsection{Case $2 (a)$: Kerr-de Sitter spacetimes}
\label{SectKdSmetric}
Consider now the Kerr-de Sitter metric
in Boyer-Lindquist-type coordinates,
cf.\ e.g.\  \cite{ashtekar, oelz},
%
\begin{eqnarray}
g &=& -\frac{\Delta_r}{\rho^2\Xi^2}\Big(\mathrm{d}t - a\sin^2\theta\mathrm{d}\phi\Big)^2 + \frac{\Delta_{\theta}\sin^2\theta}{\rho^2\Xi^2}\Big(a\mathrm{d}t-(r^2+a^2)\mathrm{d}\phi\Big)^2
\nonumber
\\
&&+ \frac{\rho^2}{\Delta_r}\mathrm{d}r^2+\frac{\rho^2}{\Delta_{\theta}}\mathrm{d}\theta^2
\;,
\label{KdS_metric_old}
\end{eqnarray}
where
\begin{eqnarray}
\rho^2 &:=& r^2 +a^2\cos^2\theta
\;,
\\
\Xi &:=& 1 +\frac{\Lambda}{3}a^2
\;,
\\
\Delta_{\theta} &:=& 1 + \frac{\Lambda}{3} a^2\cos^2\theta
\;,
\\
\Delta_r &:=& (a^2+r^2)\Big( 1-\frac{\Lambda}{3}r^2\Big) -2mr
\;,
\end{eqnarray}
and furthermore
\begin{equation}
\Lambda >0\;, \quad  a\in\mathbb{R}\setminus\{0\}\;, \quad m \in\mathbb{R}\setminus\{0\}
\;.
\end{equation}
Redefining $t$ and $\phi$ one can eliminate $\Xi$, whence  \eq{KdS_metric_old} becomes
\begin{eqnarray}
g &=& -\frac{\Delta_r}{\rho^2}\Big(\mathrm{d}t - a\sin^2\theta\mathrm{d}\phi\Big)^2 + \frac{\Delta_{\theta}\sin^2\theta}{\rho^2}\Big(a\mathrm{d}t-(r^2+a^2)\mathrm{d}\phi\Big)^2
\nonumber
\\
&&+ \frac{\rho^2}{\Delta_r}\mathrm{d}r^2+\frac{\rho^2}{\Delta_{\theta}}\mathrm{d}\theta^2
\;.
\label{KdS_metric}
\end{eqnarray}

The solutions with $a<0$ or $m<0$ can be transformed into those with $a>0$ and $m>0$ by appropriate coordinate transformations (cf.\ e.g.\ \cite{oelz}).
Our gauge will be such that $a>0$.
As explained in Remark~\ref{rem_gauge_freedom} point c) and after Theorem~\ref{first_main_thm2}, we have no gauge freedom left to arrange that $m$ is positive.
In the limit where  $m=0$ one obtains the de Sitter metric while  for $a=0$ one ends up with the Schwarzschild-de Sitter metric.

A computer algebra calculation shows that the MST associated to the  Killing vector field
$X=\partial_t$ vanishes.
In the conformally rescaled spacetime with conformal factor $\Theta= \sqrt{\frac{3}{\Lambda}}r^{-1}$ one obtains
for the induced metric on $\scri^-$, compare \cite{ashtekar,oelz}%
\begin{align*}
 h & \,=\,\mathrm{d}t^2 - 2a\sin^2\theta\mathrm{d}t\mathrm{d}\phi
+ \frac{3}{\Lambda}\frac{1}{\Delta_{\theta}} \mathrm{d}\theta^2 +  \frac{3}{\Lambda}\Xi\sin^2\theta \mathrm{d}\phi^2
\;, \\
& \, =
(dt - a \sin^2 \theta d \phi)^2 + \frac{3}{\Lambda} \left (
\frac{ d \theta^2}{\Delta_{\theta}} + \Delta_{\theta} \sin^2 \theta
d \phi^2 \right ).
\end{align*}
The CKVF $Y$ induced by $X$ on $\scri^-$ is $Y =  \partial_t$.
We need to compute $\widehat k(Y)$, $\widehat c(Y)$ and $\Dconst$.
Applying  Lemma (\ref{hmetric}) with $H = \left ( 3/(\Lambda \Delta_{\theta})
\right )^{1/2}$, $W = \Delta_{\theta} H \sin \theta$ and
$A = - a \sin^2 \theta$ (which implies $s= - (2/3) \Lambda a  \cos \theta$)
it follows immediately that $h$ is locally conformally flat and that 
\begin{align}
\widehat{k} & =
\Big(\frac{\Lambda}{3}\Big)^3a^2 \;, \quad \quad
\widehat{c}  =
\, \frac{\Lambda}{3}\Big(1-\frac{\Lambda}{3}a^2\Big)\;.
\label{KdS_c}
\end{align}
With $Y$ appropriately normalized, i.e.\
%
$Y_{\mathrm{norm}} = \Big(\frac{3}{\Lambda}\Big)^{3/4}a^{-1/2} Y$
these constants become
\begin{equation*}
\widehat k
\,=\, 1 \;,\quad \quad
\widehat c \,=\, \Big(\sqrt{\frac{\Lambda}{3}} a\Big)^{-1}-\sqrt{\frac{\Lambda}{3}}a
\;,
\end{equation*}
and thus we are in the Case $2 (a)$ of Theorem \ref{Classification}. Observe that
\begin{equation}
a \,=\, \frac{1}{2}\sqrt{\frac{3}{\Lambda}}(\sqrt{\widehat c^2+4}-\widehat c)
\;,
\label{KdS_param_a}
\end{equation}
%
so that the function $a\mapsto \widehat c$ defines a bijection from $(0,\infty)$ onto $\mathbb{R}$.
It remains to compute $\Dconst$.
%
%
A computer algebra calculation shows that 
\begin{eqnarray}
 b_1
\,=\,   36  \,\mathrm{Im}\Big(\frac{ Q_0 \mathcal{F}^5}{ (Q_0\mathcal{F}^2-4\Lambda)^3} \Big)
\,=\,  2 \Big(\frac{3}{\Lambda}\Big)^{9/4} \frac{m}{a^{3/2}}
\label{KdS_param_m_b1}
\end{eqnarray}
%
and therefore
%
\begin{eqnarray}
 \Dconst \,=\,- \frac{\Lambda^2}{6}\sqrt{\frac{3}{\Lambda}} b_1
\,=\,  - 3\frac{m}{a^{3/2}}\Big(\frac{3}{\Lambda}\Big)^{3/4}
\;.
\label{KdS_param_m_C}
\end{eqnarray}
%

For $\widehat c=0$ we have $a^2 =  \frac{3}{\Lambda}$ which
 is an important value for $a^2$ since the conformal diagrams for  $0< a^2 < \frac{3}{\Lambda}$
 and  $a^2 \geq  \frac{3}{\Lambda}$ look completely different 
due to
the fact that their horizon structure differs significantly (cf.\ e.g.\ \cite{oelz}).
The impact of the parameter $\widehat c$ on the global structure of the associated KdS-spacetime is not visible at all from our analysis at $\scri$.
It would be interesting to know whether it
can be detected by an analysis of the poles where the horizons ``touch'' $\scri$.

\subsection{Case $3 (a)$: $a\rightarrow \infty$-KdS-limit-spacetime}
Case $3(a)$ is defined by the properties that 
$\widehat k =0$, $\widehat c < 0$ and rank parameter 
$\bar r = 2$. Contrarily to all other cases, where we had available
candidates of spacetime metrics admitting a smooth conformal
compactification and potentially fulfilling all our requirements
on $\widehat k$, $\widehat c$ and  $\bar r$, in the present case
no such candidate was known to us. Nevertheless, from
Theorem \ref{Classification}
and the Cauchy well-possedness Theorem \ref{thm_well-posedness} 
one knows a priori that such spacetime must exist. This 
a priori knowledge was in fact crucial
in our efforts to find the appropriate metric.
Observe that
Case $3(a)$ lies in the closure of Case $2(a)$ in the
$\{ \widehat k, \widehat c\} $ parameter space, in the sense
that the former can be viewed as a limit
$\widehat{k} \rightarrow 0$ along a curve with $\widehat{k} >0$ and
$\widehat c <0$ (i.e. a curve fully contained in Case $2(a)$).
Note that the situation is different for all other cases
with $\widehat k =0$, because they all have $\bar r =1$.
Only in Case $3(a)$ the
rank parameter remains unchanged at the limit. These considerations indicate that 
perhaps the spacetime we are seeking may be obtainable as a limit
of the Kerr-de Sitter family. To gain further insight on how this limit might go,
consider the rescaled CKVF
$Y' = \frac{3}{\Lambda} \frac{1}{a} Y$ in the Kerr-de Sitter spacetime.
According to the results in Section \ref{SectKdSmetric}, the constants
associated to this CKVF are
\begin{equation*}
\widehat{k} (Y') = \frac{3}{\Lambda} \frac{1}{a^2}, \quad
\quad
\widehat{c} (Y') = \frac{3}{\Lambda} \frac{1}{a^2} -1,
\end{equation*}
which have the property that their limit as $a \rightarrow 
+ \infty$ approach the desired values $\widehat k = 0$,
$\widehat c = -1$. Thus, the possibility arises that
the sought metric  may obtainable as a limit as $a \rightarrow + \infty$
of the Kerr-de Sitter metric. With these hints at hand, we can proceed as
follows.

Let us consider the Kerr-de Sitter spacetime \eq{KdS_metric},
%
and perform the coordinate change 
%
\begin{equation*}
t'=a t\;, \quad r'=a^{-1}  r\;, \quad \phi' = a^{2}\phi\;,
\end{equation*}
which combined with the parameter redefinition
%
$m' \,:=\,  a^{-3} m$
%
brings  the Kerr-dS line element \eq{KdS_metric} into the form
 \begin{eqnarray*}
g = -\frac{\Delta'_r}{\rho'^2}\Big( \mathrm{d}t' -\sin^2\theta\mathrm{d}\phi'\Big)^2 + \frac{\Delta'_{\theta}\sin^2\theta}{\rho'^2}\Big( \mathrm{d}t'-( r'^2+1)   \mathrm{d}\phi'\Big)^2
+ \frac{\rho'^2}{\Delta'_r}\mathrm{d}r'^2+\frac{\rho'^2}{\Delta'_{\theta}}\mathrm{d}\theta^2
\;,
\end{eqnarray*}
with
\begin{eqnarray*}
\rho'^2 &:=&  r'^2 +\cos^2\theta
\;,
\\
\Delta'_{\theta} &:=& \frac{1}{a^2} + \frac{\Lambda}{3} \cos^2\theta
\;,
\\
\Delta'_r &:=& (1+ r'^2)\Big( a^{-2} - \frac{\Lambda}{3} r'^2\Big) -2  m' r'
\;.
\end{eqnarray*}
The limit $a\rightarrow\infty$ in this spacetime leads
to a regular metric ---dropping the primes---
%
 \begin{equation}
g \,=\, \Delta^{\infty}\Big(\mathrm{d}t -  \sin^2\theta\mathrm{d}\phi\Big)^2 + \Sigma^{\infty}\sin^2\theta\Big( \mathrm{d}t-( r^2+1) \mathrm{d}\phi\Big)^2
- \frac{\mathrm{d}r^2}{\Delta^{\infty}}+\frac{\mathrm{d}\theta^2}{\Sigma^{\infty}}
\;,
\label{KdS_limit}
\end{equation}
with
\begin{eqnarray*}
\Sigma^{\infty}&:=& \frac{\frac{\Lambda}{3} \cos^2\theta}{ r^2 +\cos^2\theta}
\;,
\\
\Delta^{\infty} &:=&\frac{ \frac{\Lambda}{3} r^2(1+ r^2)+ 2 m r}{r^2 +\cos^2\theta}
\;.
\end{eqnarray*}
We call the metric \eq{KdS_limit}, the \emph{$a\rightarrow \infty$-KdS-limit-spacetime}.  As
far as we are aware, this metric has not been discussed before in the
literature, 
and we emphasize that the process of finding it involved
in a crucial way the general classification scheme developed in this paper. 
Observe that the subcase with $m=0$ is locally de Sitter spacetime ---in a peculiar coordinate system---, and that the spacetime possesses a group $G_2$ of motions.
Since furthermore this spacetime appears as a limit of the KdS-family, it may  represent a model of physical relevance.
It is therefore of interest to study  its properties, an issue which is to be accomplished  elsewhere.

By direct computation
one shows that \eq{KdS_limit} solves the $\Lambda>0$-vacuum Einstein equations (\ref{efe}), and that the MST associated to the Killing vector field
$X=\partial_t$ vanishes.

Concerning  $\scri^{-}$, we use the rescaled
 metric with the conformal factor $\Theta=\sqrt{\frac{3}{\Lambda}}  r^{-1}$
and send
$r \rightarrow \infty$ to obtain the following
induced metric  on $\scri^-$,
\begin{align*}
 h & =\mathrm{d}t^2- 2\sin^2\theta\mathrm{d}t\mathrm{d}\phi
+\Big(\frac{3} {\Lambda}\Big)^2 \frac{1}{\cos^2\theta}\mathrm{d}\theta^2 + \sin^2\theta \mathrm{d}\phi^2 \\
& = \left ( dt - \sin^2 \theta d \phi \right )^2
+ \left ( \frac{3}{\Lambda} \right )^2 \frac{1}{\cos^2 \theta}
d \theta^2 + \sin^2 \theta \cos^2 \theta \, d \phi^2\;,
\end{align*}
The induced CKVF is $Y = \partial_t$. Applying Lemma (\ref{hmetric})
with $H = \frac{3}{\Lambda \cos \theta}$, $W = \sin \theta \cos \theta$
and $A= - \sin^2 \theta$ (from which $s = - (2/3) \Lambda \sin \theta$)
it follows immediately that $h$ is
locally conformally flat and
%
%
%
%
\begin{align*}
 \widehat k&= 0, \quad \quad
 \widehat c =
-\Big( \frac{\Lambda}{3}\Big)^2
\; .
\end{align*}
%
In terms of
%
$Y_{\mathrm{norm}} = \frac{3}{\Lambda} Y$ the constants
are
\begin{equation*}
 \widehat k=0\;, \quad \quad \widehat c =  -1
\;.
\end{equation*}
%
One can now easily see that (\ref{r1}) does not hold for $Y_{\mathrm{norm}}$ (for instance by noting that $Y_k N^k \neq 0$ as $s\neq 0$), and thus the conformal class $[Y]$ has $\bar r=2$ and this corresponds to Case $3 (a)$ in Theorem \ref{Classification}. (Another way of seeing this is to realize that the $a\rightarrow \infty$-KdS-limit-spacetime has precisely two independent Killing vector fields, whence it is generated by the CKVF $Y^{(1)}$ of  (\ref{specialcase})).

As before, we compute $b_1$ and $\Dconst$ using the spacetime expressions
(\ref{equation_b1b2}) and (\ref{expression_b1})
to get
\begin{align}
 b_1
\,=\,  2m \Big(\frac{3}{\Lambda}\Big)^{3}, \quad \quad
%
%
\Dconst \,=
 -3    m \Big(\frac{3}{\Lambda}\Big)^{3/2}
\;.
\label{limit_KdS_param_m_C}
\end{align}

\subsection{Case $1 (a)$: Wick-rotated Kerr-AdS and Taub-NUT-dS with $ \ell = \frac{1}{2}\sqrt{\frac{3}{\Lambda}}$ spacetimes}
It remains to identify spacetimes in the parameter range where $\widehat k=-1$ and $\widehat c
\leq -2$. We are going to see that the subcase $\widehat{c}=-2$ is somehow special, but that anyway it can be included in a general expression for this set of spacetimes as a limit. We start with the generic case, leading to the Wick-rotated Kerr-anti-de Sitter (AdS) spacetime

\subsubsection{Wick-rotated Kerr-anti-de Sitter spacetime}
In \cite{kmv} a Wick rotation
\begin{eqnarray}
&t\mapsto it\;, \quad r\mapsto  i r\;, \quad
\theta\mapsto i\theta\;, \quad \phi\mapsto \phi\;, &
\\
&\Lambda \mapsto -\Lambda \;, \quad m\mapsto -im \;, \quad a\mapsto ia
\;,&
\end{eqnarray}
 has been performed  to the Kerr-de Sitter spacetime to construct  black-hole solutions with negative cosmological constant.
In fact, the whole procedure is independent of the sign of the cosmological constant (as long as one makes sure that the Lorentzian signature of the metric
is preserved), and one is led to a 2-parameter family of solutions with positive
cosmological constant by a Wick rotation of the Kerr-AdS metric.
In Boyer-Lindquist-type coordinates it reads
\begin{eqnarray}
g &=& -\frac{\widehat \Delta_r}{\widehat \rho^2\widehat \Xi^2}\Big(\mathrm{d}t + a\sinh^2\theta\mathrm{d}\phi\Big)^2 + \frac{\widehat \Delta_{\theta}\sinh^2\theta}{\widehat \rho^2\widehat \Xi^2}\Big(a\mathrm{d}t-(r^2+a^2)\mathrm{d}\phi\Big)^2
\nonumber
\\
&&+ \frac{\widehat \rho^2}{\widehat \Delta_r}\mathrm{d}r^2+\frac{\widehat \rho^2}{\widehat \Delta_{\theta}}\mathrm{d}\theta^2
\;,
\label{wick-KdS_metric_old}
\end{eqnarray}
where
\begin{eqnarray}
\widehat \rho^2 &:=& r^2 +a^2\cosh^2\theta
\;,
\\
\widehat \Xi &:=& 1 -\frac{\Lambda}{3}a^2
\;,
\\
\widehat \Delta_{\theta} &:=& 1 - \frac{\Lambda}{3} a^2\cosh^2\theta
\;,
\\
\widehat \Delta_r &:=& -(a^2+r^2)\Big( 1+\frac{\Lambda}{3}r^2\Big) -2mr
\;, \label{Deltar}
\end{eqnarray}
and with
\begin{equation}
\Lambda >0\;, \quad  a \in\mathbb{R}\setminus\{0\}\;, \quad m \in\mathbb{R}\setminus\{0\}
\;.
\end{equation}

The Wick-rotated Kerr-AdS spacetime is a generalization of the Kottler spacetime with $\varepsilon =-1$ just as the Kerr-de Sitter spacetime
generalizes the  Kottler spacetime with $\varepsilon =+1$, which is the Schwarzschild-de Sitter metric.
However,  \eq{wick-KdS_metric_old} defines a Lorentzian metric 
of signature $\{ -,+,+,+\}$ 
if  and only if
%
\begin{equation}
a^2< \frac{3}{\Lambda},
\end{equation}
namely for
(note that $\det g|_{\theta=0}=0$)
%
\begin{equation}
\label{rangeWick}
0< \theta < \mathrm{arcosh}\Big(\sqrt{\frac{3}{\Lambda}} a^{-1}\Big)
\;.
\end{equation}
One checks that this is indeed a solution of Einstein's field equations (\ref{efe}).
As in the Kerr-dS-case the sign of $a$ and $m$ is a matter of gauge. In our gauge $a$ will be positive while $m$ will have both signs.

As in the Kerr-de Sitter case, we can get rid of $\widehat \Xi$ by a simple
rescaling of $t$ and $\phi$, whence  \eq{wick-KdS_metric_old} becomes
\begin{eqnarray}
g &=& -\frac{\widehat \Delta_r}{\widehat \rho^2}\Big(\mathrm{d}t + a\sinh^2\theta\mathrm{d}\phi\Big)^2 + \frac{\widehat \Delta_{\theta}\sinh^2\theta}{\widehat \rho^2}\Big(a\mathrm{d}t-(r^2+a^2)\mathrm{d}\phi\Big)^2
\nonumber
\\
&&+ \frac{\widehat \rho^2}{\widehat \Delta_r}\mathrm{d}r^2+\frac{\widehat \rho^2}{\widehat \Delta_{\theta}}\mathrm{d}\theta^2
\;.
\label{wick-KdS_metric}
\end{eqnarray}
This metric has vanishing MST
with respect to the Killing vector field  $X = \partial_t$. Performing the
conformal rescaling with conformal factor
$\Theta= \sqrt{\frac{3}{\Lambda}}r^{-1}$ and sending $r \rightarrow \infty$,
the induced metric on $\scri^-$ turns out to be
%
\begin{align*}
 h \,& =\,\mathrm{d}t^2 + 2a\sinh^2\theta\mathrm{d}t\mathrm{d}\phi
+ \frac{3}{\Lambda}\frac{1}{\widehat \Delta_{\theta}} \mathrm{d}\theta^2 +  \frac{3}{\Lambda}\widehat \Xi\sinh^2\theta \mathrm{d}\phi^2 \\
& = \Big ( \mathrm{d}t + a \sinh^2 \theta \mathrm{d}\phi\Big )^2 + \frac{3}{\Lambda}
\Big ( \frac{1}{\widehat \Delta_{\theta}} d \theta^2 + \widehat \Delta_{\theta} \sinh^2 \theta
\mathrm{d} \phi^2 \Big )
\;,
\end{align*}
and the CKVF associated to $X$ is $Y= \partial_t$.
With $H = (3/(\Lambda \widehat \Delta_{\theta}))^{1/2}$, $W=
\widehat \Delta_{\theta} H \sinh \theta$
and $A = a \sinh^2 \theta$ (which gives $s = (2/3) \Lambda a \cosh \theta$),
Lemma \ref{hmetric} implies that $h$ is locally conformally flat and that
\begin{align*}
\widehat k
&=  - \Big(\frac{\Lambda}{3}\Big)^3a^2 \;, \quad \quad
 \widehat c = -\frac{\Lambda}{3}\Big(1+\frac{\Lambda}{3}a^2\Big)
\;.
\label{WRKAdS_c}
\end{align*}
Normalizing $Y_{\mathrm{norm}} = \Big(\frac{3}{\Lambda}\Big)^{3/4} a^{-1/2}Y$
we are led to
%
\begin{equation}
\label{ckWick}
\widehat k
\,=\,- 1 \;,\quad \quad
\widehat c \,=\, - \Big(\sqrt{\frac{\Lambda}{3}} a\Big)^{-1}-\sqrt{\frac{\Lambda}{3}}a
\;.
\end{equation}
Recall that $0<\sqrt{\frac{\Lambda}{3}} \, a<1$.
The function $\sqrt{\frac{\Lambda}{3}} \, a \mapsto \widehat c$ defines a bijection from $(0,1)$ onto $(-\infty,-2)$ and we have
\begin{equation}
a^2 = \frac{3}{4\Lambda} (\sqrt{\widehat c^2-4}+\widehat c)^2
\;.
\label{wick_param_a}
\end{equation}
Therefore, this metric corresponds to Case $1 (a)$ in Theorem \ref{Classification} but with $\widehat{c}<-2$, with the strict inequality. The value $\widehat{c}=-2$ is not included here (in other words, the constant $\mu_0$ in Theorem \ref{Classification} is such that $\mu_0\in (0,1)$, excluding the right end of this interval).

For the constants $b_1$ and $\Dconst$ we find
%
\begin{align}
b_1
\,=\,   2 \Big(\frac{3}{\Lambda}\Big)^{9/4} \frac{m}{a^{3/2}}
\;, \quad \quad
 \Dconst
\,=\, -  3\frac{m}{a^{3/2}}\Big(\frac{3}{\Lambda}\Big)^{3/4}
\;.
\label{wick_param_m_C}
\end{align}
%
%

%

\subsubsection{Taub-NUT-dS with $ \ell = \frac{1}{2}\sqrt{\frac{3}{\Lambda}}$}
\label{TaubNUTSec}

Consider now the Taub-NUT-de Sitter metric (cf.\ e.g.\ \cite{griffiths}),
\begin{equation}
g =- V_r \Big( \mathrm{d}t - 4\ell\sin^2\frac{\theta}{2}\mathrm{d}\phi\Big)^2 + \frac{1}{V_r}\mathrm{d}r^2
 + (r^2 + \ell^2)
(\mathrm{d}\theta^2  + \sin^2\theta\mathrm{d}\phi^2)
\;,
\label{taub}
\end{equation}
where
\begin{equation}
\label{Vr}
V_r = \frac{1}{r^2 + \ell^2}\Big(r^2 -\ell^2-2mr - \frac{\Lambda}{3}( r^4 + 6\ell^2r^2-3\ell^4  )\Big)
\;,
\end{equation}
and with
\begin{equation}
 \Lambda>0\;, \quad m\in\mathbb{R}\setminus \{0\}\;, \quad \ell \in\mathbb{R}\setminus\{0\}\;.
\end{equation}
This spacetime also has vanishing MST with respect to the Killing vector field
$X=\partial_t$ .
With the conformal factor
$\Theta= \sqrt{\frac{3}{\Lambda}}r^{-1}$ the induced metric at
$\scri^{-}$ (i.e. in the limit $r \rightarrow \infty$) is
%
%
\begin{align*}
 h & = \mathrm{d}t^2 - 8\ell \sin^2\frac{\theta}{2}\mathrm{d}t\mathrm{d}\phi   + \frac{3}{\Lambda} \mathrm{d}\theta^2
+ \Big(16\ell^2\sin^4\frac{\theta}{2}
 + \frac{3}{\Lambda}\sin^2\theta\Big) \mathrm{d}\phi^2 \\
& = \Big( dt - 4 \ell \sin^2 \frac{\theta}{2} d \phi \Big)^2
+ \frac{3}{\Lambda} \big( d \theta^2 + \sin^2 \theta d \phi^2 \big)\,,
\end{align*}
and the induced CKVF  is $Y = \partial_t$. Lemma (\ref{hmetric})
with $H = (3/\Lambda)^{1/2}$, $W = H \sin \theta$ and
$A = - 4 \ell \sin^2 (\theta/2)$ (so that $s = - (2/3) \Lambda \ell$) gives
\begin{align*}
\Cot_{\theta \, \theta} = \frac{\ell \Lambda}{3} \Big( \frac{4}{3} \Lambda \ell^2
-1 \Big).
\end{align*}
Thus, $h$ can be locally conformally flat only if either $\ell=0$, which
is the Schwarzschild-de Sitter case already discussed, or
\begin{equation}
 \ell \,=\, \frac{1}{2}\sqrt{\frac{3}{\Lambda}}
\;,
\label{ell0}
\end{equation}
which we assume henceforth. For later reference we quote the
spacetime metric in this case
\begin{equation}
g  =- V_r \Big( \mathrm{d}t - 2\sqrt{\frac{3}{\Lambda}}\sin^2\frac{\theta}{2}
\mathrm{d}\phi\Big)^2 + \frac{1}{V_r}\mathrm{d}r^2
 + \Big(r^2 + \frac{3}{4\Lambda}\Big)
(\mathrm{d}\theta^2  + \sin^2\theta\mathrm{d}\phi^2)
\;,  \label{taub_spec}
\end{equation}
where
\begin{equation}
V_r  = -\frac{\Lambda}{3} \Big( r^2  + \frac{3}{4\Lambda}\Big) - 2mr \Big(r^2+
\frac{3}{4\Lambda}\Big)^{-1}
\;. 
\end{equation}
The expressions of $\widehat{k}$ and
$\widehat{c}$ as given in (\ref{exphatc})-(\ref{exphatk}) yield
\begin{align*}
\widehat k \,=\, -\frac{\Lambda^2}{144}
\;, \quad \quad \widehat c \,=\,  -\frac{\Lambda}{6}.
\end{align*}
After rescaling
$Y_{\mathrm{norm}} =2  \sqrt{\frac{3}{\Lambda}} Y$, the constants become
\begin{align*}
\widehat k
\,=\, -1 \;,\quad \quad
\widehat c \,=\, -2
\;.
\end{align*}
Hence, this is the special Case $1 (a)$ of Theorem \ref{Classification} left over before (corresponding to $\mu_0 =1$ in that theorem). Observe that this is also the exceptional case identified in Lemma \ref{lemma_dimensions}.

The constants $b_1$, $\Dconst$ can be computed from the spacetime metric to be
\begin{align}
 b_1 & =
16m \Big(\frac{3}{\Lambda}\Big)^{3/2}
\;, \quad \quad
 \Dconst
 \,=\,- 24 m
\;.
\label{taub_param_m_C}
\end{align}

\subsubsection{Taub-NUT-dS as a limit of the Wick-rotated Kerr-AdS}
\label{NUT_Wick_relation}

Theorem \ref{Classification} and the two previous subsections clearly indicate that
the Taub-NUT-dS
spacetime with $\ell = \frac{1}{2} \sqrt{\frac{\Lambda}{3}}$
and the  Wick-rotated Kerr-AdS spacetimes belong to the same single class, Case $1 (a)$,
characterized by the property $\hat{k} <0$ ---within the allowed set $\mathcal A\subset \mathbb{R}^2$ of that theorem.
The values of $\hat{c}$ in Taub-NUT-dS ($\hat{c} = - 2 \sqrt{|\hat{k}|}$)
and in the Wick rotated spacetime (\ref{ckWick})
suggest that the former
should be obtainable as the limit $\frac{\Lambda}{3} a^2
\rightarrow 1$ of the latter. In this section we show that this is indeed the case.

Given that the value $\frac{\Lambda}{3} a^2 =1$ is not allowed by the metric
(\ref{wick-KdS_metric_old})
we need to perform a coordinate transformation which is singular
when $\widehat \Xi = 1  -\frac{\Lambda}{3} a^2 \rightarrow 0$ and take the limit
of the resulting metric  as $\widehat \Xi \rightarrow 0$. The appropriate coordinate
change is suggested by the range (\ref{rangeWick}) of the coordinate $\theta$.
Assume $\widehat \Xi > 0$ and define a coordinate $\hat{\theta}$ by
\begin{align*}
\theta = \beta (\widehat \Xi) \sin \frac{\hat{\theta}}{2}
\end{align*}
where
\begin{align*}
\beta(\widehat \Xi) & := \mathrm{arcosh} \left ( \frac{1}{\sqrt{1- \widehat \Xi}} \right ).
\end{align*}
Note that the range of $\theta$ corresponds to the range $\hat{\theta} \in [0,
\pi)$. Expanding $\beta(\widehat \Xi)$ near $\widehat \Xi=0$, a function $U(\widehat \Xi)$,
vanishing at $\widehat \Xi=0$ and smooth in a neighbourhood of this point, can
 be defined by $\beta(\widehat \Xi) = \sqrt{\widehat \Xi} ( 1+ U(\widehat \Xi))$. We can also
expand $\widehat \Delta_{\theta}(\theta) = 1 - ( 1 - \widehat \Xi) \cosh^2 \theta$
in terms of the new coordinate, to obtain
\begin{align*}
\widehat \Delta_{\theta} =
\widehat \Xi \Big ( 1 - ( 1- \widehat \Xi) (1 + U(\widehat \Xi))^2
\sin^2 \frac{\hat{\theta}}{2} + O(\widehat \Xi) \Big ),
\end{align*}
from which the term $\frac{d \theta^2}{\widehat \Delta_{\theta}}$ in the metric is
immediately seen to become, in terms of the new variable,
\begin{align*}
\frac{d \theta^2}{\widehat \Delta_{\theta}} & =
\frac{ \widehat \Xi \left (
( 1 + U(\widehat \Xi))^2 \cos^2 \frac{\hat{\theta}}{2}
\right ) d \hat{\theta}^2}{
4 \widehat \Xi \left ( 1 - ( 1- \widehat \Xi) (1 + U(\widehat \Xi))^2
\sin^2 \frac{\hat{\theta}}{2} + O(\widehat \Xi) \right )}
= \frac{\cos^2 \frac{\hat{\theta}}{2} \, d \hat{\theta}^2}{
4 \big( 1 - \sin^2 \frac{\hat{\theta}}{2} \big) } + O (\widehat \Xi) \\
& = \frac{d \hat{\theta}^2}{4} + O(\widehat \Xi).
\end{align*}
Note also that
\begin{align*}
\frac{\sin^2 \theta}{\widehat \Xi} = \sin^2 \frac{\hat{\theta}}{2} +
O(\widehat \Xi)
\,,
\end{align*}
and that both $\widehat \rho^2$ and $\widehat \Delta_r$ have smooth limits
at $\widehat \Xi=0$, which we denote respectively as
$\widehat \rho^2_0$ and $\widehat \Delta_r^0$. Observe in particular that
$\widehat \rho_0^2 = r^2 + 3 / \Lambda$. With the coordinate transformation
$t = \frac{\widehat \Xi}{2} \hat{t}$, the Wick rotated metric
(\ref{wick-KdS_metric_old}) takes the form
\begin{align*}
g_{\widehat \Xi} & =
- \frac{\widehat \Delta_r^0}{\widehat \rho^2_0}
\Big ( \frac{1}{2} d \hat{t} + \sqrt{\frac{3}{\Lambda}}
\sin^2 \frac{\hat{\theta}}{2} d \phi
\Big )^2 +
\widehat \rho^2_0 \sin^2 \frac{\hat{\theta}}{2}
\cos^2 \frac{\hat{\theta}}{2} d \phi^2 + \frac{\widehat \rho^2_0}{\widehat \Delta_r^0} dr^2
+ \frac{1}{4} \widehat \rho^2_0 d \hat{\theta}^2 + O(\widehat \Xi) \\
& = - \frac{\widehat \Delta_r^0}{4 \widehat \rho^2_0} \left ( d \hat{t} -
4 \ell_0  \sin^2 \frac{\hat{\theta}}{2} d \hat{\phi} \right )^2
+ \frac{1}{4} \widehat \rho^2_0 \left ( d \hat{\theta}^2 + \sin^2 \hat \theta
d \hat{\phi}^2 \right )
+ \frac{4 \widehat \rho^2_0}{\widehat \Delta_r^0} d\hat{r}^2 + O (\widehat \Xi)
\,,
\end{align*}
where in the second equality we have defined new coordinates
$r = 2 \hat{r}$, $\phi = - \hat{\phi}$ and introduced the quantity
$\ell_0 = \frac{1}{2} \sqrt{\frac{3}{\Lambda}}$.
It is immediate to check that
\begin{align*}
\left . \frac{\widehat \Delta_r^0}{4 \widehat \rho^2_0}
\right |_{r = 2 \hat{r}}
\frac{- \hat{r}^4 \ell_0^{-2} - 2 \hat{r}^2 - m \hat{r} - \ell_0^2}{4 (\hat{r}^2
+ \ell_0^2)} := V_{\hat{r}}.
\end{align*}
Note that $V_{\hat{r}}$ is defined exactly as in (\ref{Vr})
with the value of $m$ replaced by $\frac{m}{8}$. So, we conclude
\begin{align*}
g_{\widehat \Xi} & =
- V_{\hat{r}} \Big ( d\hat{t} - 4 \ell_0 \sin^2 \frac{\hat{\theta}}{2} d \hat{\phi}
\Big )^2
+ \frac{1}{V_{\hat{r}}} d \hat{r}^2
+ (\hat{r}^2 + \ell_0^2 ) \left ( d \hat{\theta}^2 +
\sin^2 \hat{\theta} d \hat{\phi}^2 \right ) + O(\widehat \Xi)
\,,
\end{align*}
which indeed shows that the Taub-NUT-dS spacetime with $2 \ell
= \sqrt{\frac{3}{\Lambda}}$  and the
Wick-rotated Kerr-AdS spacetimes belong to a single class with parameter
range $1 - \frac{\Lambda}{3} a^2 \geq 0$, with the equality case
corresponding to the Taub-NUT-dS spacetime with $2 \ell
= \sqrt{\frac{3}{\Lambda}}$.

\section{Main results}
\label{sec_main_result}

We are now able to state our  main results.
 As indicated at the beginning of Section~\ref{section_alternative} they combine a reconsideration of \cite[Theorem 4 \& 6]{mars-senovilla} with some important refinements, under the additional assumptions that the cosmological constant is positive and that the  spacetime admits a conformally flat $\scri$. In addition, our results provide a clear identification of the KID needed at $\scri$ for the different spacetimes, and inform us of the underlying topology of the connected components of $\scri^-$.
In summary, it becomes possible to give a clearer statement on which spacetimes
emerge under which conditions, and, moreover, to achieve a classification in terms of the initial data of an asymptotic Cauchy problem.

The various possibilities appearing in  \cite[Theorem 4]{mars-senovilla}, cf.\ also Section~6.1 of that reference, are restricted in such a way that only the spacetimes in (i), (ii), (iv) and (v) of Theorem~\ref{second_main_thm} below remain.
\begin{theorem}
\label{second_main_thm}
Consider a spacetime $(\mcM,g)$, solution of Einstein's vacuum field equations (\ref{efe}) with $\Lambda>0$, which admits a smooth conformally flat $\scri^-$
and which contains a Killing vector field $X$
w.r.t.\ which the MST vanishes.
Then the restrictions \eq{expression_b1}, \eq{general_k} and \eq{general_c} of the functions   $b_1$, $k$ and $c$, as given by \eq{equation_b1b2}-\eq{equation_k}, on $\scri^-$ are constant.%
\footnote{Recall that
$\widehat c= (\Lambda / 3) c|_{\scri^-}$ and $\widehat k=(\Lambda/3)^3 k|_{\scri^-}$.
}

Assuming
 that $b_1\ne 0$ (equivalently $\Dconst\ne 0$; if $b_1=0$,  $(\mcM,g)$ is locally given by the de Sitter spacetime
 in the DoD of $\scri^-$),
 we have furthermore:
\begin{enumerate}
\item[(i)] If $\widehat k>0$, $(\mcM,g)$ belongs locally to the  Kerr-de Sitter family \eq{KdS_metric_old} in the DoD of $\scri^-$.
For $\widehat k$ normalized to $1$, the parameters $m\ne 0$ and $a$ are given by \eq{KdS_param_a} and \eq{KdS_param_m_C}, respectively. This corresponds to Case $2 (a)$ in Theorem \ref{Classification}.
\item[(ii.1)] If $\widehat k=0$ and $\widehat c\geq 0 $, $(\mcM,g)$ is locally
a generalized Kottler spacetime \eq{kottler_metric} with  $\varepsilon=\mathrm{sign}(\widehat c)$. For $\widehat c$ normalized to $1$ or $0$, respectively,  its mass is given by
\eq{kottler_param_m_C}. These correspond to Cases $2 (b)$ and $3 (b)$ in Theorem \ref{Classification}, respectively.
\item[(ii.2)] If $\widehat k=0$ and $\widehat c< 0 $ there are two possibilities:
\begin{enumerate}
\item[(a)] Either $(\scri^-,h)$ admits four CKVFs, then  $(\mcM,g)$ is locally given by the
 generalized Kottler spacetime \eq{kottler_metric} with  $\varepsilon=-1$, its mass (in $\widehat c=-1$-gauge) is given by
\eq{kottler_param_m_C}. This corresponds to Case $1 (b)$ in Theorem \ref{Classification}.;
\item[(b)]
or, $(\scri^-,h)$ admits merely two CKVFs, then
$(\mcM,g)$ is locally given by the
$a\rightarrow \infty$-KdS-limit-spacetime \eq{KdS_limit}; the parameter $m\ne 0$
is given, again for $\widehat c=-1$, by \eq{limit_KdS_param_m_C}. This corresponds to Case $3 (a)$ in Theorem \ref{Classification}.
\end{enumerate}
\item[(iii)] A spacetime which satisfies the above assumptions with $\widehat k <0 $ and $\widehat c> -2\sqrt{|\widehat k|}$ does not exist.
\item[(iv)] If $\widehat k< 0 $ and $\widehat c=-2\sqrt{|\widehat k|}$, $(\mcM,g)$  is locally the  Taub-NUT-de Sitter spacetime \eq{taub_spec}
 with NUT-parameter $ \ell = \frac{1}{2}\sqrt{\frac{3}{\Lambda}}$
 in the DoD of $\scri^-$.
For $\widehat k$ normalized to $-1$ its mass is given by \eq{taub_param_m_C}. This corresponds to Case $1 (a)$ in Theorem \ref{Classification} with $\mu_0 =1$, and coincides with the exceptional case identified in Lemma \ref{lemma_dimensions}.
\item[(v)] If $\widehat k< 0$ and $\widehat c<-2\sqrt{|\widehat k|}$,  $(\mcM,g)$ belongs locally to the Wick-rotated   Kerr-anti-de Sitter family \eq{wick-KdS_metric_old}
 in the DoD of $\scri^-$.
 For $\widehat k$ normalized to $-1$ the parameters $m\ne 0$ and $a$ are given by \eq{wick_param_a} and \eq{wick_param_m_C}, respectively. This corresponds to Case $1 (a)$ in Theorem \ref{Classification} with $\mu_0 \in (0,1)$.
\end{enumerate}
\end{theorem}

\begin{remark}
{\rm
In fact, it follows from the results in \cite{mars-senovilla}
that $b_1$, $k$ and $c$ are constant in the whole spacetime, whence
their values  can be computed at any point, not necessarily on $\scri^-$.
}
\end{remark}

\begin{remark}
{\rm
Recall that the $\varepsilon=+1$-Kottler spacetime and the $a\rightarrow\infty$-KdS-limit-spacetime
are limits of the Kerr-de Sitter spacetime, while the   $\varepsilon=-1$-Kottler spacetime and the
$\ell=\frac{1}{2}\sqrt{\frac{3}{\Lambda}}$-Taub-NUT-de Sitter spacetime  are limits
of the Wick-rotated Kerr-anti-de Sitter spacetime.
The  $\varepsilon=0$-Kottler spacetime, in turn, can be obtained as limit  of the $\varepsilon\ne 0$-Kottler spacetimes if
the Gau\ss\enspace  curvature is not normalized to  $\pm 1$ and one  passes to the limit $\varepsilon\rightarrow 0$.
Theorem~\ref{second_main_thm} therefore tells us that a $\Lambda>0$-vacuum spacetime which admits
a  conformally flat $\scri^-$ and a Killing vector field w.r.t.\ which  the MST vanishes is locally given by
the Kerr-de Sitter spacetime, or
the  Wick-rotated Kerr-anti-de Sitter spacetime, or certain limits thereof. This is summarized in 
Table \ref{table}.
}
\end{remark}


\begin{table}
\caption{All Kerr-dS-like spacetimes with conformally flat $\scri^-$ according to the values of the constants $\widehat{k}$ and $\widehat{c}$}
\vspace{2mm}

\hspace{-2em}


\begin{tabular}{c||c|c|c|c|c|}
& $\widehat c<-2$ & $\widehat c=-2$ & $-2<\widehat c<0$ & $ \widehat c=0$ & $\widehat c>0$
\\ \hline\hline
\multirow{2}{*}{$\widehat k=+1$} & \multicolumn{3}{c|}{KdS, $a>\sqrt{\frac{3}{\Lambda}}$ 
}     &    \multicolumn{2}{c|}{KdS, $a\leq \sqrt{\frac{3}{\Lambda}}$ }
\\
& \multicolumn{3}{c|}{Case 2(a)} & 
\multicolumn{2}{c|}{Case 2(a)} \\
 \hline
\multirow{2}{*}{$\widehat k =0$}  &  \multicolumn{3}{c|}{Kottler,  $\varepsilon=-1$ \quad or \quad $a\rightarrow \infty$-KdS-limit}        & Kottler,  $\varepsilon=0$    & SdS  \\
& \multicolumn{3}{c|}{Case 1(b) \quad \quad \quad \quad \quad \quad Case 3(a)}
& Case 3(b) & Case 2(b) 
\\ \hline
\multirow{2}{*}{$\widehat k =-1$} & Wick-KAdS  &  T-NUT-dS,$\,\ell = \sqrt{\frac{3}{4\Lambda}}$  &   \multicolumn{3}{c|}{no solution} \\
& Case 1(a) & Case 1(a) & \multicolumn{3}{c|}{} 
\\
\hline
\end{tabular}
\label{table}
\end{table}

\subsection{Global considerations: topology of $\scri^-$}
\label{topology_sec}

Let us complement our classification results based at $\scri$ with some global considerations of the Kerr-dS-like spacetimes with a locally conformally flat $\scri$, i.e.\ those which appear in Theorem~\ref{second_main_thm}, concerning  in  particular the topology of $\scri$.

A convenient way to set up the asymptotic Cauchy problem for spacetimes with vanishing MST is to take as initial Riemannian manifold the
standard 3-sphere $(\mathbb{S}^3,\gamSthree)$, which takes cares of the assumption that $\scri$ is locally conformally flat. The asymptotic Cauchy KID are completed by the choice of a CKVF $Y$ on $\mathbb{S}^3$ and some constant $\Dconst$.
In general, though, $Y$ will have zeros on  $\mathbb{S}^3$ and, letting aside the de Sitter spacetime, these points do not belong to the null infinity of the so-emerging $\Lambda>0$-vacuum spacetime
\cite{mpss}.
Its topology will therefore not be $\mathbb{S}^3$ but  rather
$$\scri^-\cong \mathbb{S}^3\setminus\{ p\in \mathbb{S}^3 | |Y(p)|^2=0\}$$
(or a connected component thereof if the zero set
of $Y$ separates the 3-sphere).
The zeros of $Y$ were presented in Proposition~\ref{zeroes} for the different possibilities, and thus we have all we need to perform the analysis of the $\scri^-$-topology. In the following we provide such a global structure for each Kerr-dS-like spacetime
and discuss briefly their 4-dimensional global structure.

A particularly interesting situation arises if $Y$ has no zeros on $\mathbb{S}^3$. If then the  emerging spacetime
is asymptotically simple 
(note that a compact $\scri^-$ is a prerequisite for that \cite{ag}),
an argument analogous to the one which was used to prove the non-linear stability of the de Sitter spacetime   \cite{F2, F2B} applies
 and shows that the corresponding spacetime is stable as well.

\paragraph*{De Sitter spacetime} The de Sitter spacetime is
 asymptotically simple with $\scri^-\cong \mathbb{S}^3$,
$\mcM\cong \mathbb{R}\times \mathbb{S}^3$ and $\widetilde{\mcM\enspace}\hspace{-0.5em} \cong [0,1]\times \mathbb{S}^3$.  Its global structure is discussed e.g.\ in \cite{gh}.

\paragraph*{Kerr-de Sitter spacetime and Schwarzschild-de Sitter spacetime} From Theorem \ref{second_main_thm} these are Cases $2 (a)$ and $2 (b)$ in Theorem \ref{Classification} and then Proposition~\ref{zeroes} tells us that the CKVF $Y$ has two zeros $p_1$ and $p_2$, representing the poles where the event horizon and the cosmological horizon ``touch'' $\scri^-$. Therefore, $\scri^-\cong \mathbb{S}^3\setminus \{p_1,p_2\} \cong \mathbb{S}^2\times \mathbb{R}$.
None of its members is asymptotically simple.
Their  global structure is discussed e.g.\ in \cite{gh, oelz}.

\paragraph*{Kottler spacetime with  $\varepsilon=0,-1$}
The topology of past null infinity is known to be $\scri^-\cong\mathbb{R}^3$ \cite{MNT}. This can also be seen  from the fact that $Y$ vanishes on a closed embedded curve in Case $1 (b)$ of Theorem \ref{Classification}, or at one isolated point in Case $3 (b)$ of that Theorem, and as then follows from Proposition~\ref{zeroes}. Observe that in the first case the relevant part of the initial manifold $\mathbb{S}^3$ is merely a hemisphere.
The spacetimes are not asymptotically simple.


\paragraph*{$a\rightarrow \infty$-KdS-limit-spacetime}
This is Case $3 (a)$ in Theorem \ref{Classification}, so that from Proposition~\ref{zeroes} the CKVF $Y$ vanishes at precisely one isolated point whence the topology of past null infinity is again $\scri^-\cong\mathbb{R}^3$ and the spacetime is not asymptotically simple.

\paragraph*{Wick-rotated Kerr-AdS and Taub-NUT-dS spacetimes} These correspond to Case $1 (a)$ in Theorem \ref{Classification}. 
From Proposition \ref{zeroes} all its members admit a $\scri^-$ which is topologically $\mathbb{S}^3$. Actually, this is true also for the general Taub-NUT-dS metric, independently of having a conformally flat $\scri$ or not. 
Its global structure is determined by the function $V_r$ given in (\ref{Vr}) and appearing in \eq{taub}. More precisely, by the
(real) zeros of the function $f(r)= -(r^2 + \ell^2)V_r$ \cite{beyer0, beyer}.
 If  $f$  has no zeros the spacetime is asymptotically simple with $\mcM\cong \mathbb{R}\times \mathbb{S}^3$ and $\widetilde{\mcM\enspace}\hspace{-0.5em} \cong [0,1]\times \mathbb{S}^3$.
If the function  $f$ has zeros the spacetime will be extendible and causally geodesically incomplete (with an extension which is not globally hyperbolic).

\begin{itemize}
\item
Let us start then with the exceptional situation in Case $1 (a)$, given by Taub-NUT-dS spacetime with $\ell=\frac{1}{2}\sqrt{\frac{3}{\Lambda}}$, so that $\scri$ is conformally flat.
To analyze the zeros of $f(r)$ observe that in the case with $\ell=\frac{1}{2}\sqrt{\frac{3}{\Lambda}}$
%
\begin{equation}
f(r) \,=\, \frac{\Lambda}{3}\Big(r^2 + \frac{3}{4\Lambda}\Big)^2 + 2mr 
\;.
\label{eqn_f}
\end{equation}
Hence, if $m$ is positive, the roots must be negative and vice versa.
Without loss of information, we can assume that $m \geq 0$.
Equation \eq{eqn_f}   can be written as
\begin{equation}
f(r) \,=\,  \Big( 2r + \frac{1}{\sqrt{\Lambda}}\Big)^2\underbrace{\Big( \frac{\Lambda}{12} r^2  - \frac{\sqrt{\Lambda}}{12} r + \frac{3}{16} \Big)}_{>0}
 + 2 \Big( m-\frac{1}{3\sqrt{\Lambda}}\Big) r
\;.
\end{equation}
We deduce that $f$ cannot have zeros for $0\leq m <\frac{1}{3\sqrt{\Lambda}}$. For $m=\frac{1}{3\sqrt{\Lambda}}$ we have a double root at $r=-\frac{1}{2\sqrt{\Lambda}}$.
For $m>\frac{1}{3\sqrt{\Lambda}}$ we have $f(-\frac{1}{2\sqrt{\Lambda}})<0$ and therefore $f$ must have zeros.

Consequently, the  $\ell=\frac{1}{2}\sqrt{\frac{3}{\Lambda}}$-Taub-NUT-de Sitter spacetime is asymptotically simple if and only if
\begin{equation}
|m| <\frac{1}{3\sqrt{\Lambda}}
\;.
\label{m_range_nut}
\end{equation}
This is in accordance with the stability of de Sitter.
 In particular,  the Taub-NUT-dS spacetime with $\ell=\frac{1}{2}\sqrt{\frac{3}{\Lambda}}$  is stable in this parameter range.

\item
Consider now the generic situation in Case $1 (a)$, the Wick-rotated Kerr-anti-de Sitter spacetime.
Again, the corresponding CKVFs do not have zeros whence $\scri^- \cong \mathbb{S}^3$. The metric is given by (\ref{wick-KdS_metric}) 
and, by analogy with the previous case,
one might expect that the global structure of the spacetime
(in particular whether it is asymptotically simple) may depend
on the zeroes of the function $\widehat\Delta_r$ in (\ref{Deltar}).
With this hypothesis, it makes sense to 
determine under which conditions on $m$ and $a$ this function has no zeros on $\mathbb{R}$, which we do next. Recall that
$$
\widehat\Delta_r =-(a^2+r^2)\Big( 1+\frac{\Lambda}{3}r^2\Big) - 2mr
\;, 
\quad 0<a<\sqrt{\frac{3}{\Lambda}}
\;.
$$
Without restriction we can assume $m\geq 0$ when we analyze the roots of this function. 
Note that, as before, $m\geq 0$ implies that the roots (if any)
must be negative.
It is convenient to set
\begin{eqnarray*}
\hat f(\hat r) &:=&  (\hat a^2+\hat r^2)( 1+\hat r^2) + 2\hat m\hat r
=  \hat r^4 + (\hat a^2 +1)\hat r^2 + 2\hat m \hat r + \hat a^2
\;,
\end{eqnarray*}
%
%
\begin{eqnarray*}
 \hat r := \sqrt{\frac{\Lambda}{3}} r
\;, \quad
\hat a := \sqrt{\frac{\Lambda}{3}} a  \quad (0<\hat a<1)
\;, \quad
\hat m := \sqrt{\frac{\Lambda}{3}} m \quad (\hat m \geq 0).
\end{eqnarray*}
Note that $\hat f(\hat r) = - \frac{\Lambda}{3} \widehat\Delta_r(r(\hat r))$
so that its zeroes determine those of $\widehat\Delta_r$.
The function $\hat f$ has a double root at
\begin{equation}
 \hat r_0\,=\,-\sqrt{\frac{-1-\hat a^2+\sqrt{1+14 \hat a^2 + \hat a^4}}{6}} <0
\;,
\end{equation}
%
for the following specific value of $\hat m$
\begin{equation}
 \hat m \,=\,\hat m_0 \,:=\, \frac{-1-\hat a^4 +10\hat a^2 +(1+ \hat a^2 ) \sqrt{1+14\hat a^2 + \hat a^4}}{3\sqrt{6}\sqrt{-1-\hat a^2 + \sqrt{1+14 \hat a^2 + \hat a^4}}}
\in (0, 3^{-3/2} 8)
\;.
\end{equation}
We  write $\hat f$ as
\begin{eqnarray*}
\hat f(\hat r)
&=&  \hat r^4 + (\hat a^2 +1)\hat r^2  + \hat a^2 +2\hat m_0\hat r+ 2(\hat m-\hat m_0) \hat r
\\
&=&(\hat r -\hat r_0)^2 \underbrace{(\hat r^2 + 2 \hat r_0 \hat r + \hat a^2\hat r_0^{-2})}_{>0}+ 2(\hat m-\hat m_0) \hat r
\;,
\end{eqnarray*}
(the second factor of the first term is positive since it is clearly positive for $\hat r\rightarrow \pm\infty$ and its discriminant is negative for $0<\hat a <1$).

As above, we conclude from this representation that $\hat f$ cannot have zeros for $0 \leq \hat m < \hat m_0$.
For $\hat m =\hat m_0$ it has a double root at $\hat r =\hat r_0$.
For $\hat m >\hat m_0$ we have $\hat f(\hat r_0) <0$ and therefore $\hat f$ must have zeros.
It follows that $\widehat\Delta_r$ 
has no zeros on $\mathbb{R}$  if and only if $|\hat m| < \hat m_0$, or, equivalently,
\begin{equation}
| m| <  \frac{-1-\big(\frac{\Lambda}{3}\big)^2 a^4 +\frac{10\Lambda}{3} a^2 +(1+ \frac{\Lambda}{3}  a^2 ) \sqrt{1+\frac{14\Lambda}{3}  a^2 + \big(\frac{\Lambda}{3}\big)^2 a^4}}{\sqrt{18\Lambda}\sqrt{-1-\frac{\Lambda}{3}  a^2 + \sqrt{1+ \frac{14\Lambda}{3}  a^2 + \big(\frac{\Lambda}{3}\big)^2 a^4}}}
\;.
\label{bound}
\end{equation}
%
It is an interesting problem to determine for which values of 
$m$ and $a$, the spacetime (\ref{wick-KdS_metric_old}) is asymptotically simple. 
From the preliminary analysis above, and by comparison with 
the simpler Taub-NUT-dS case, we put forward the following
conjecture.

\begin{conjecture}
The Wick-rotated Kerr-anti-de Sitter spacetime
(\ref{wick-KdS_metric_old})-(\ref{Deltar}) is
asymptotically simple 
and
$\mcM\cong \mathbb{R}\times \mathbb{S}^3$, $\widetilde{\mcM\enspace}\hspace{-0.5em} \cong [0,1]\times \mathbb{S}^3$
if and only if $m$ satisfies the bound
(\ref{bound}).
In particular the spacetime is stable under the same bound.
\end{conjecture}

Note that for $a\rightarrow \sqrt{3/\Lambda}$ (\ref{bound})
becomes $|m|< \frac{8}{3\sqrt{\Lambda}}$, which recovers \eq{m_range_nut} when taking into account that the ``$m$''-parameters differ by a factor
of $8$, cf. Section~\ref{NUT_Wick_relation}.

%
%
%

%
%

\end{itemize}

\section{A characterization result for   Kerr-de Sitter}
\label{section_global}

Let us consider one more time  the Kerr-de Sitter family of spacetimes, which we want to characterize asymptotically at null infinity. However, in this section we pay attention to a full connected component of the Kerr-dS $\scri$ and not just some local subset thereof as in the previous parts.
Our goal is to  weaken the asymptotic conditions on a given $\Lambda>0$-vacuum spacetime to represent a Kerr-de Sitter spacetime as much as possible.

Let us recall once again that, as proven in \cite{mpss}, for spacetimes which admit a smooth $\scri^-$ the requirement
on the MST to vanish can be replaced by the conditions  (i) and (ii) of Theorem~\ref{first_main_thm2}.
If, in addition, we assume that $\scri^-$ is conformally flat (ergo $\Cconst =0$), there only remains  condition (ii) relating the $\scri^-$-trace $D$ of the rescaled Weyl tensor with the CKVF $Y$ induced on $\scri^-$ by the spacetime KVF according to (\ref{condition_on_D}).
%

However, $D$ and $Y$ are not independent of each other. They satisfy the reduced KID equations \eq{reduced_KID}.
%
%
Given $Y$, these equations provide transport equations for the TT-tensor $D$.
One therefore merely needs to impose certain conditions on some appropriate hypersurface in $\scri^-$ to make sure that  \eq{condition_on_D} holds.
That is what we intend to work out in detail  to establish a characterization result for the Kerr-de Sitter spacetime, or rather for the domain of dependence of its $\scri^-$.
In fact,  as ``degenerate hypersurface'' we shall use one of the poles in $\mathbb{S}^3$ not belonging to $\scri^-$ where the Kerr-dS-metric becomes singular.

The key to the following discussion is the next result, which is valid
for any symmetric and trace-free tensor satisfying the transport
equation \eq{reduced_KID}, independently of whether
it is transverse or not.

\begin{Proposition}
\label{StructureD}
Let $(\Sigma,h)$ be a
 Riemannian $3$-dimensional manifold
admitting a conformal Killing vector field $Y$ and assume
the existence of a point $p \in \Sigma$ which is a
limit point for  all the integral lines of $Y$.

Let $D$ be a symmetric, trace-free $C^1$ tensor field defined on the
open set $U:= \{ p \in \Sigma, Y(p) \neq 0\}$ and satisfying the transport
equation (\ref{reduced_KID}). Assume that the functions $|Y|D(Y,Y)$ and
$|Y|^6 |D|^2$ have finite limits at $p$ and satisfy
\begin{align}
\lim_{q \rightarrow p} |Y| D(Y,Y) := \frac{2}{3} \alpha, \quad \quad
\lim_{q \rightarrow p} |Y|^6 |D|^2 = \frac{2}{3} \alpha^2
\label{limits}
\end{align}
Then
$$D = \frac{\alpha}{|Y|^5} (\bm{Y} \otimes \bm{Y}   - \frac{|Y|^2}{3} h )$$
on all of $U$.
\end{Proposition}

\begin{proof}
We work on $U$.
We start by showing that $|Y| D(Y,Y), |Y|^6 |D|^2$ are constant
along the integral lines of $Y$.  The conformal Killing equations
$\mcL_{Y} h_{ij} = \frac{2}{3} (\nablah_l Y^l) h_{ij}$ imply
$
  \mcL_Y |Y| = \frac{1}{3} |Y|  \nablah_l Y^l,
$
from which the constancy of $|Y| D(Y,Y)$
follows as a consequence of (\ref{reduced_KID}).
The constancy of $|Y|^6 |D|^2$ follows from
$\mcL_{Y} |D|^2 = - 2 (\nablah_l Y^l) |D|^2$, which in turn
is a direct consequence of (\ref{reduced_KID})  and
$\mcL_{Y} h^{ij} = - \frac{2}{3}(\nablah_l Y^l) h^{ij}$.

Recalling the definition \eq{DY} of $D_Y$ set
\begin{align*}
W  := |Y|^6 \left (  D \otimes D_Y - D_Y \otimes D \right )
\end{align*}
which, except for a global factor $|Y|^6$, 
is the wedge product of the symmetric tensors $D$ and $D_Y$ (that is, $W$ is symmetric in the first and last pair of indices and antisymmetric under interchange of pairs). Given
that $D_Y$ is nowhere zero on $U$, $W$ vanishes if
and only if $D = f D_Y$ for some $C^1$ function. Now, an immediate
computation shows
\begin{align*}
|W|^2 = \frac{4}{3}
|Y|^6 |D|^2  - 2 |Y|^2 D(Y,Y)^2
\end{align*}
which implies that $|W|^2$ is constant along the integral lines of $Y$. Given
that $p$ lies in the closure of all integral lines of $Y$
and that $\lim_{q \rightarrow p} |W|^2 =0$ from~(\ref{limits}), it follows that
$|W|^2$ vanishes everywhere on $U$. Therefore, $W =0$
and $D = f D_Y$. Thus $|Y| D(Y,Y) = \frac{2}{3} f $ and since this is constant
along the integral lines of $Y$ with limit $\frac{2}{3} \alpha$, we conclude
$f=\alpha$ and the proposition is proven.
\qed
\end{proof}


Now recall that, as discussed in section \ref{topology_sec}, the Kerr-de Sitter $\scri^-$ (including the Schwarzschild-dS case) is topologically a 3-sphere with two poles  removed, see also \cite{ashtekar},
\begin{equation}
\scri^-\cong \mathbb{S}^3\setminus\{p_1,p_2\}\cong \mathbb{S}^2\times\mathbb{R}
\;.
\end{equation}
 The conformally rescaled ``unphysical'' spacetime metric is singular at these two points on the sphere, whereas the  Riemannian metric $h$ which is  induced on $\scri^-$ remains regular there.
Let us therefore consider an asymptotic  initial 3-manifold $\Sigma$ which is topologically $\mathbb{S}^3$, $\Sigma:= \scri^-\cup\{p_1,p_2\}$.
The Riemannian manifold $(\Sigma, h)$ is compact, simply connected and locally conformally flat.
We can thus apply a classical theorem due to Kuipers (cf.\ e.g.\ \cite{kuehnel}) according to which there exists
a conformal diffeomorphism onto the standard sphere $\mathbb{S}^3$.
Since only the conformal class of the initial manifold matters, we may assume $(\Sigma ,h)$ to be the standard 3-sphere from the outset, which we shall do henceforth.

The Kerr-de Sitter family (including the Schwarzschild-dS case) corresponds to Cases 2 in Theorem \ref{Classification}, or in other words to the
choice of CKVF equivalence class $[Y]$ in $(\Sigma,h)$ satisfying $\widehat k >0$ (or
$\k =0$, $\c >0$ for the Schwarzschild-dS case).
The canonical representatives of $Y$ in $\mathbb{E}^3$
according to Theorem \ref{Classification} are given by
\begin{align*}
Y & = \lambda (x \partial_x + y \partial_y + z \partial_z) +
\lambda^{-1} (y \partial_x  - x \partial_y)
& & &  \mbox{if } &\quad \k >0, \\
Y & = x \partial_x + y \partial_y + z \partial_z
& & & \mbox{if } &\quad  \k =0, \, \c > 0. \\
\end{align*}
$\mathbb{E}^3$ and $\mathbb{S}^3$ are conformally related by the stereographic
projection. 
Selecting (in accordance with Appendix \ref{Ster}) the plane $z^1=1$
and the point $p=(-1,0,0,0)$ to define this projection and
choosing standard spherical coordinates $\{ \psi,\theta,\phi\}$
on $\mathbb{S}^3$ centered
at the poles $p$ and $-p$ (with $\psi=0$ at $-p$ and $\psi=\pi$ at $p$),
the CKVF  $Y$ on
$\mathbb{S}^3$ becomes
\begin{align}
Y & = \lambda \sin \psi \partial_{\psi} - \frac{1}{\lambda} \partial_{\phi}
& & &  \mbox{if } &\quad \k >0, \label{KillYKdS1} \\
Y & = \sin \psi \partial_{\psi}
& & & \mbox{if } &\quad  \k =0, \, \c > 0.
\label{KillYKdS2}
\end{align}

Our aim here is to show that,
in order to characterize the Kerr-de Sitter family at infinity, we may
replace the specific form (\ref{condition_on_D}) for $D$ by very mild conditions on one of the excluded
poles, say $p_1$.

\begin{theorem}
Consider a  maximal globally hyperbolic $\Lambda>0$-vacuum spacetime $(\mcM,g)$
which admits a smooth $\scri^-$ and  a Killing vector field $X$. Let the asymptotic
KID be $(\scri^{-},h,D,Y)$ and
suppose that the following conditions are satisfied:
\begin{enumerate}
\item[(i)] $\scri^-\cong {\mathbb S}^3\setminus\{p_1,p_2\}$,
\item[(ii)] $(\scri^-,h)$ is conformally flat,
\item[(iii)]  $\widehat k(Y)$ as given by \eq{simplified_k} is either
positive  (in fact it suffices to be positive at one point) or
vanishes, in which case we assume that $\c(Y)$ (as given by \eq{expression_c} )
is positive.
\item[(iv)] $p_1$ is a fixed point of $Y$ and
the functions $|Y| D(Y,Y)$ and $|Y|^6 |D|^2$
have finite limits at  $p_1$ satisfying
\begin{align*}
\left(\lim_{q \rightarrow p_1} |Y| D(Y,Y) \right)^2 =  \frac{2}{3}  \lim_{q \rightarrow p_1} |Y|^6 |D|^2
\end{align*}
\end{enumerate}
Then $(\mcM,g)$ is isometric to the domain of dependence of
a connected component of past null infinity of a
Kerr-de Sitter spacetime (belonging to the Schwarzschild-de Sitter case if and only if $\widehat k (Y)=0$).
\end{theorem}

\begin{proof}
By uniqueness of the maximal Cauchy development of asymptotic KID data,
it suffices to show that the data $(\scri^{-},h,D,Y)$ are isometric to
the Kerr-dS (or Schwarzschild-dS) data. 
Since $\widehat{k} >0$ (or $\widehat{k} =0$ and $\widehat{c} >0$)
we may assume
(possibly
after applying a conformal diffeomorphism that leaves $p_1$ fixed)
that $Y$ is of the form (\ref{KillYKdS1}) (or (\ref{KillYKdS2}))
in standard spherical
coordinates $\{ \psi,\theta,\phi\}$ with $\psi(p_1) =0$.
All integral lines of $Y$ except the fixed points at the
north pole $p_1$ and the south pole $p_S$ (given by $\psi = \pi$)
have  $p_1$
in their closure. This is because  $\psi$
is a strictly monotonic quantity along $Y$ on $U$ ($\mcL_{Y} (\psi)=
\lambda \sin \psi >0$) and the integral lines of $Y$ are  complete.
If $p_2$ is not the
south pole of $\mathbb{S}^3$, then $Y|_{p_2} \neq 0$ and there exist
precisely one integral line $s(\tau)$
of $Y$ in $\scri^{-}$ connecting $p_2$ and $p_S$. In this case
we define $U := \mathbb{S}^3 \setminus \{ p_1, p_2, s(\tau)\}
\subset \scri^{-}$.
If $p_2$ is the south pole we let $U := \scri^{-} =
\mathbb{S}^3 \setminus \{ p_1, p_2 \}$.  In either case, $U$ is
dense in $\scri^{-}$. Thus,
we can apply Proposition \ref{StructureD} to conclude that $D = \Dconst D_Y$ for some constant $\Dconst :=\lim_{q \rightarrow p_1} |Y| D(Y,Y)$
on $U$, and hence also in $\mathbb{S}^3 \setminus \{ p_1, p_S \}$.
Given that $D_Y$ is singular at the south pole, it follows that
the excluded point $p_2$ must in fact be $p_S$ (otherwise
$p_S$ would be an interior point of
$\scri^{-}$ and $D$ would not be smooth everywhere at past null infinity).
Thus~$(\Sigma,h,D,Y)$ agrees with the data
for Kerr-dS (or Schwarzschild-dS)  and the theorem is proven.
\qed
\end{proof}


\section{Comparison with known results
and existence of conformally
flat $\scri$}
\label{sec_comparison}

A complete classification of $\Lambda$-vacuum spacetimes satisfying the alignment condition
(\ref{alig}) has been obtained in \cite{mars-senovilla} ($\F^2\not\equiv 0)$ and \cite{mars-senovilla-null} ($\F^2 =0$). In the latter case $\Lambda \leq 0$ necessarily and thus this is of no relevance here. Concerning \cite{mars-senovilla}, and as already emphasized, the results in this paper are largely independent of those in \cite{mars-senovilla}. However, having accomplished here a complete classification
of $\Lambda>0$-vacuum spacetimes satisfying the
alignment condition (\ref{alig}) and admitting a conformally flat (or
conformally spherical)
$\scri$, we can now combine both sets of results and find, as
an interesting by-product,  which specific spacetimes
in the general class do admit a conformally flat
 $\scri$ and which do not. Even more, the analysis reveals 
which subset of spacetimes satisfying $b_2 = 0$
(defined in (\ref{equation_b1b2})) {\it do not}
admit a smooth conformal compactification at all.

The starting point is Theorem 4 in 
\cite{mars-senovilla},
where it is  shown that any $\Lambda$-vacuum spacetime
satisfying (\ref{alig}) and $Q\F^2$ not identically zero belongs to one
of three explicit classes, labeled (A), (B.i) and (B.ii) in
\cite{mars-senovilla}. Class (A) corresponds to the situation with
$Q \F^2 - 4 \Lambda=0$ at some point (and hence everywhere
\cite{mars-senovilla}), while
classes (B) correspond to $Q \F^2 - 4 \Lambda \neq 0$ everywhere.
In the latter case, the metric depends on a number of parameters
in addition to $\Lambda$: 
$b_1, b_2, c, k \in \mathbb{R}$ for class (B.i) and
$\beta, n \in \mathbb{R}$, $\epsilon \in \{ -1,1\}$, 
$\kappa \in \{ -1,0,1\}$ for class (B.ii). All these constants
are defined geometrically in terms of the choice of Killing vector
$\xi$ under which (\ref{alig}) holds.  It is clear from
the results in \cite{mars-senovilla} that any two spacetimes
either belonging to different classes, or to the same class but
with different parameters, cannot be locally isometrically
transformed to each other while at the same time transforming the
chosen Killing vectors $\xi$ into one another. However,
in order to make use of Theorem 4 in \cite{mars-senovilla} to draw
general conclusions, we need to know 
under which conditions two given spacetimes in the family are
locally isometric to each other independently of whether the
Killing vectors satisfying (\ref{alig}) get transformed into one another.
One way of addressing this issue is to fix the 
spacetime metric under consideration and ask what is the most
general Killing vector $\xi$ for which (\ref{alig}) holds.
This question is analyzed in the following Lemma.
\begin{lemma}
Let $(\mcM,g)$ be a $\Lambda$-vacuum spacetime satisfying the alignment
condition (\ref{alig})
with $Q\F^2$ not identically zero for a choice of
Killing vector $\xi$. If $Q \F^2 - 4 \Lambda =0$ for $\xi$,
then $Q\F^2 - 4 \Lambda=0$ for any other Killing vector $\xi^{\prime}$
for which (\ref{alig}) holds. On the other hand, if $Q \F^2 - 4 \Lambda \neq 0$
then any other Killing vector satisfying \eq{alig} must be of the form $\xi^{\prime} = \alpha \xi$ where $\alpha$ is a non-zero constant.
\end{lemma}

\begin{proof}
Let $p \in \mcM$ be chosen so that $Q|_p \neq 0$ and $\F^2 |_p \neq 0$.
We work on an open connected neighbourhood $U_p$ of $p$ where
$Q \F^2$ vanishes nowhere. From the assumption
\begin{align*}
\C_{\alpha\beta\mu\nu} = 
Q \left (\F_{\alpha\beta} \F_{\mu\nu}
- \frac{1}{3} \F^2 \I_{\alpha\beta\mu\nu} \right )
= Q^{\prime} \left (\F^{\prime}_{\alpha\beta} \F^{\prime}_{\mu\nu}
- \frac{1}{3} \F^{\prime}{}^2 \I_{\alpha\beta\mu\nu} \right )
\end{align*}
where $Q^{\prime}$, $\F^{\prime}_{\alpha\beta}$ are the quantities
associated to $\xi^{\prime}$,
it follows that $Q^{\prime}$ and $\F^{\prime}_{\alpha\beta}$ are
nowhere zero on $U_p$. Moreover, since $\I$
is of rank 3 (in the complex three-dimensional vector
space of self-dual two-forms at a point), while
$\mbox{span} \{ \F \otimes\F, \F^{\prime}\otimes\F^{\prime}\}$
is at most of rank two, we conclude that
$Q \F^2 = Q^{\prime} \F^{\prime}{}^2$ and 
$\F^{\prime}_{\alpha\beta} = u \F_{\alpha\beta}$ for some
non-zero smooth complex function $u$. The first
equality proves the first claim in the Lemma.

For the second statement we assume $Q \F^2 - 4 \Lambda \neq 0$
and show first that
$u$ is in fact constant. Indeed, expression (\ref{equation_b1b2}) can be written in the
form
\begin{align*}
(\F^2)^3 = \frac{ ( b_2 - i b_1)^2 (Q \F^2 - 4 \Lambda )^6}{36^2 (Q \F^2)^2}.
\end{align*}
Denoting by $b_1^{\prime}, b_2^{\prime}$ the constants corresponding
to $\xi^{\prime}$
we find, using the invariance $Q \F^2 = Q^{\prime} \F^{\prime}{}^2$,
\begin{align*}
u^6 =  \left ( \frac{\F^{\prime}{}^2}{\F^2} \right )^3
= \frac{(b_2^{\prime} - i b_1^{\prime} )^2}{
(b_2 - i b_1)^2}
\end{align*}
and $u$ is constant as claimed. We now recall the definition
of the Ernst-one
form $\sigma_{\alpha} := 2 \xi^{\beta} \F_{\beta\alpha}$
and use the fact \cite[Theorem 4]{mars-senovilla} that
\begin{align*}
\sigma_{\alpha} = 6 \nabla_{\alpha} \left ( 
\F^2 \frac{ Q \F^2 + 2 \Lambda}{(Q \F^2 - 4 \Lambda)^2} \right )
\end{align*}
which readily implies $\sigma_{\alpha}^{\prime} = u^2 \sigma_{\alpha}$. The
Killing vector can be reconstructed out of $\sigma_{\alpha}$ by
means of the general identity
\begin{align}
\sigma_{\beta} \F_{\mu}^{\phantom{\mu}\beta}
= 2 \xi^{\alpha} \F_{\alpha\beta} \F_{\mu}^{\phantom{\mu}\beta}
= \frac{1}{2} \F^2 \xi_{\mu},
\label{idenchi}
\end{align}
the last equality following from the standard algebraic
property of self-dual two forms $\F_{\alpha\beta}\F_{\mu}^{\phantom{\mu}\beta}
= \frac{1}{4} \F^2 g_{\alpha\mu}$. Inserting the transformation law for
$\F_{\alpha\beta}$ and $\sigma_{\alpha}$ it follows
\begin{align*}
\xi_{\mu}^{\prime} = u \xi_{\mu}
\end{align*} 
which automatically implies that $u$ is in fact a real constant
$\alpha$.
\qed
\end{proof}

As an immediate Corollary, two given metrics
in Class (B), either belonging to different subclasses (i) and (ii)
or to the same subclass but with constants not related to each other
by a rescaling of $\xi$, are automatically not locally isometric to each 
other. When dealing with spacetimes admitting a smooth $\scri$, this
normalization freedom will be fixed as described in 
Remark \ref{gauge_freedom_kc}.

We can now proceed with the  identifications of the spacetimes within classes
(A) and (B) admitting a conformally flat smooth $\scri$.
Class (A) is very simple because the corresponding metric
is the so-called Nariai metric, which is known not to admit
a smooth $\scri$ \cite{beyerNariai}. This result also follows
from the considerations in \cite{mpss}, where we showed that
any $\Lambda>0$-vacuum spacetime satisfying (\ref{alig})
and admitting a smooth $\scri$ necessarily satisfies that
both $\F^2$ and $Q \F^2 - 4 \Lambda$ are not identically zero 
near $\scri$. 

In order to deal with class (B) we recall that the splitting 
into  (B.i) and (B.ii) depends on whether
$\mathrm{dim}(\mathrm{span}\{\xi,\varsigma\})$ 
is $2$ or $1$ respectively,
where 
$\varsigma$ 
is the KVF (\ref{varsigma}) commuting with $\xi$.
In \cite{mpss} we have
shown that, whenever the spacetime admits a smooth $\scri$,
one has $\mathrm{dim}(\mathrm{span}\{\xi,\varsigma\})
=\mathrm{dim}(\mathrm{span}\{Y,\widehat\varsigma\})$. 
As a consequence of Lemma \ref{lemma_dimensions}, class (B.ii) admitting
a conformally flat $\scri$ corresponds to all Cases (b)
(which have $\widehat k=0$ and 4 CKVFs) plus the exceptional Case $1 (a)$ with
the specific values $(\widehat k, \widehat c)=(-1,-2)$.

We have proven in Section \ref{section_alternative} that Cases (b) correspond
to  the generalized Kottler spacetimes and that Case 1(a)
with $(\widehat k, \widehat c)=(-1,-2)$ corresponds to
the Taub-NUT-dS spacetime. On the other hand, the metrics in case (B.ii) are 
\cite[Theorem4]{mars-senovilla}
\begin{align}
g & = \left \{ \begin{array}{lr}
       - V \left ( dv - \hat{\bm{w}} \right )^2 + 2 d\y \left ( dv - 
\bm{\hat{w}} \right ) + (\v^2+ \y^2) h_{+} & \mbox{if } \hspace{2mm} 
\epsilon =1  \\
     V^{-1}  d\y^2+ V \left ( dv - \hat{\bm{w}} \right )^2 
+ (\v^2+ \y^2) h_{-} & \mbox{if } \hspace{2mm}\epsilon = -1
 \end{array} \right . 
\label{metricBii} \\
V & = (\v^2 + \y^2)^{-1}   \left ( 
- \frac{\Lambda}{3} \left ( \y^4  + 6 \v^2 \y^2 - 3 \v^4 \right ) - \kappa
\left ( \v^2 - \y^2 \right ) + n \y  \right ), 
 \nonumber \\
\xi & = \partial_v, \quad \quad
\hat{d} \hat{\bm{w}} = 2 \v \bm{\eta_{\epsilon}} \nonumber
\end{align}
where $\beta,n \in \mathbb{R}$, $\kappa\in \{ -1,0,1\}$
and $h_{\epsilon}$ is a metric of constant curvature
$\kappa$, signature $\{\epsilon,1\}$ and volume form
$\bm{\eta_{\epsilon}}$. It follows from the results in \cite{mars-senovilla}
(also by direct computation) that the constant $b_2$ 
for the metric (\ref{metricBii}) reads
$b_2 =2\beta  ( \kappa - \frac{4}{3} \Lambda \beta^2 )$  when $\epsilon = +1$
and 
$b_2=n$ when $\epsilon=-1$. 

Restricting ourselves to 
$\epsilon = +1$, the condition $b_2=0$ (necessary for
the metric to admit a conformally flat $\scri$, cf.\ \eq{expression_b1}) is equivalent to
either $\beta=0$ and $\kappa$ unrestricted,
or $\beta = \frac{1}{2} \sqrt{\frac{3}{\Lambda}}$ and $\kappa =1$. 
The metric (\ref{metricBii}) with $\epsilon = +1$, $\beta=0$ is isometric
to the Kottler spacetime. Indeed, the equation $\hat{d} \hat{\bm{w}} =0$
shows that $\hat{\bm{w}}$ is locally exact. The coordinate change
$dv = dt + \frac{dy}{V} + \hat{\bm{w}}$ brings the metric into 
the generalized Kottler metric 
(\ref{kottler_metric})
after identifying $n = -2 m$ and $\kappa = \varepsilon$.
When $\epsilon = +1$ and $\kappa = 1$, the metric (\ref{metricBii}) is locally
isometric to Taub-NUT-dS \cite{mars-senovilla} with the
identification $\beta = \ell$, and we recover the fact that
this metric admits a conformally flat $\scri$ if and only if $\beta =
\ell =0$ or $\beta = \ell = \frac{1}{2} \sqrt{\frac{3}{\Lambda}}$.
Since we have exhausted all spacetimes within Cases (b) or
Case 1(a) with $(\widehat{k},\widehat{c}) = (-1,-2)$, the following
Proposition holds
\begin{Proposition}
Consider the class of $\Lambda>0$-vacuum spacetimes 
$(\mcM,g)$ satisfying the alignment
(\ref{alig}) with $Q \Fsq|_p \neq 0$ at some point $p \in M$
and $Q \F^2 - 4 \Lambda$ not identically zero. The subclass
(B.ii) defined
by
$\mathrm{dim}(\mathrm{span}\{\xi,\varsigma\})=1$,
and given explicitly in (\ref{metricBii}),
admits a conformally flat $\scri$ if and only if
$\epsilon = 1$, and either $\beta=0$ or $(\kappa = 1,
\beta = \frac{1}{2} \sqrt{\frac{3}{\Lambda}})$. Moreover, the metrics
with $\epsilon = -1$ and $n=0$ do not admit a smooth conformal
compactification.
\end{Proposition}
Note that the last statement of the theorem holds 
because $\epsilon =-1$, $n=0$ has $b_2=0$. Hence, if
the spacetime admitted a smooth conformal compactification, $\scri$ would
have to be locally conformally flat, but this is excluded by our results and the fact that all cases with $\mathrm{dim}(\mathrm{span}\{\xi,\varsigma\})=1$ have been exhausted.

One also knows that the metrics (\ref{metricBii})
with $\epsilon =1$,
$\kappa=1$ (being the Taub-NUT-dS metric with arbitrary NUT parameter)
admit a  smooth $\scri$, so it is tempting to conjecture
that the class with $\epsilon =1$ admits a smooth $\scri$ for all values
of $\kappa$ and $\beta$, while the class with $\epsilon = -1$ does not
admit a smooth conformal compactification irrespective
of the value of the parameters. One way to address this question 
would be to classify the asymptotic Killing initial data arising
in  Theorem \ref{first_main_thm2} with $\Cconst \neq 0$, similarly
as we have done here in the case $\Cconst =0$.

It remains to understand how Class (B.i) fits into the present
framework. This corresponds to the situation when
$\mathrm{dim}(\mathrm{span}\{\xi,\varsigma\})=2$ and the metric
can be written locally as \cite{mars-senovilla}
\begin{align}
g = & - N \left ( dv - Z^2 dx \right )^2
+ 2 \left ( dy + V dx \right ) \left ( dv - Z^2 dx \right ) \nonumber \\
 & + (y^2 + Z^2) \left ( \frac{dZ^2}{V} + V dx^2 \right ) \label{metricBi}\\
  \xi  = &  \partial_v, \quad N := c - \frac{\Lambda}{3} \left (y^2 - Z^2\right )
- \frac{b_1 y + b_2 Z}{y^2 + Z^2}, 
\quad V :=k +b_2 Z - c Z^2 - \frac{\Lambda}{3} Z^4. \nonumber
\end{align}
or, away from the hypersurfaces $\{y=y_0\}$ where $W(y) := k - b_1 y + c y^2
- \frac{\Lambda}{3} y^4$ has zeroes, in the alternative Pleba\'nski form
\begin{align}
g & = \frac{1}{y^2+Z^2} \left ( V(Z) \left ( d \tau + y^2 
d \sigma \right )^2 - W(y) \left ( d \tau - Z^2 d \sigma \right )^2 \right )
\nonumber \\
& + (y^2 + Z^2) \left ( \frac{dZ^2}{V(Z)} + \frac{dy^2}{W(y)}\right ).
\label{PlebanskiForm}
\end{align}
It is convenient to make the redefinitions
$W=\Big(\frac{3}{\Lambda}\Big)^3 \widehat W$, $V=\Big(\frac{3}{\Lambda}\Big)^3 \widehat V$, $Z=\frac{3}{\Lambda}\widehat Z$, $y=\frac{3}{\Lambda}\widehat y$, $\sigma=\Big(\frac{\Lambda}{3}\Big)^2\widehat \sigma$,
$b_1=\Big(\frac{3}{\Lambda}\Big)^2\widehat b_1$ 
and
$b_2=\Big(\frac{3}{\Lambda}\Big)^2\widehat b_2$, which 
brings (\ref{PlebanskiForm}) into
\begin{eqnarray}
 g &=&  \frac{3}{\Lambda}\frac{1}{\widehat y^2 + \widehat Z^2}\Big( \widehat V(\widehat Z)(\mathrm{d}\tau + \widehat y^2\mathrm{d}\widehat \sigma)^2 -\widehat W(\widehat y)(\mathrm{d}\tau- \widehat Z^2\mathrm{d}\widehat \sigma)^2\Big)
\nonumber
\\
&&
+ \frac{3}{\Lambda} (\widehat y^2 +\widehat Z^2)\Big(\frac{\mathrm{d}\widehat Z^2}{\widehat V(\widehat Z)} + \frac{\mathrm{d}\widehat y^2}{\widehat W(\widehat y)}\Big)
\;,
\label{PD_metric} \\
 \widehat W(\widehat y)
 & =&  \widehat k - \widehat b_1 \widehat y + \widehat c\widehat y^2 -\widehat y^4
\;, \quad \quad
\widehat V(\widehat Z)
= \widehat k + \widehat b_2 \widehat Z -\widehat c\widehat Z^2 -\widehat Z^4
\; . \nonumber
\end{eqnarray}
We emphasize that the constants $\widehat k$ and $\widehat c$  here
agree with the constants $\widehat k$ and $\widehat c$ used elsewhere in the
paper (more precisely, when the spacetime admits a smooth
$\scri$, the traces of these constants on $\scri$ agree with those
in the rest of the paper).

Given that our classification has been restricted to the
conformally flat case, we need to set $b_2 = 0$ (or
$\widehat b_2 =0$), which we assume henceforth.
We have proven in Section \ref{section_alternative} that the values of $(\widehat k,\widehat c)$
compatible with a smooth conformal compactification
are (using the normalization in Remark \ref{gauge_freedom_kc}):
Case $1(a)$ with $\widehat k = -1$, $\widehat c < -2$, Case $2(a)$
(defined by $\widehat k >0$), and Case $3 (a)$ where $\widehat k=0$ 
and (\ref{r1}) fails to hold so that there are just two CKVFs.

By simply combining all the results, the following
Proposition holds.
\begin{Proposition}
Consider the class of $\Lambda>0$-vacuum spacetimes 
$(\mcM,g)$ satisfying the alignment
(\ref{alig}) with $Q \Fsq|_p \neq 0$ at some point $p \in M$
and $Q \F^2 - 4 \Lambda$ not identically zero. The subclass
(B.i) defined
by
$\mathrm{dim}(\mathrm{span}\{\xi,\varsigma\})=2$ 
and given explicitly by (\ref{metricBi}) (or (\ref{PlebanskiForm}))
admits a conformally
flat $\scri$ if and only if
$b_2 =0$ and $(k, c)$ lie in the subset
\begin{align*}
\{ k <0, c < - 2 \sqrt{|k|}\} \cup \{ k =0, c <0 \} \cup \{ k >0 \}.
\end{align*} 
Moreover, the metrics with $b_2 =0$ and 
$(k,c)$ not lying in this set do not admit
a smooth conformal compactification.
\end{Proposition}
Nothing can be said from our results here
concerning the general case $b_2 \neq 0$. As already indicated, 
a complete classification of asymptotic KIDs satisfying (i)
and (ii) in Theorem \ref{first_main_thm2} would determine, as a Corollary,
which spacetimes in Class (B.i) do admit a smooth conformal
compactification.

It is however, of interest to see how exactly the metric
(\ref{PlebanskiForm}) with $b_2=0$
fits with the results in Section
\ref{section_alternative} for each allowed value of $(\widehat k, \widehat c )$. We devote the following subsections to address this.

\subsection{Wick-rotated Kerr-anti-de Sitter spacetime: Case $1 (a)$}

We restrict ourselves to $\widehat k = -1$,
$\widehat c < -2$, so that
\begin{eqnarray}
 \widehat W(\widehat y)
 & =&  -1 - \widehat b_1 \widehat y + \widehat c\widehat y^2 -\widehat y^4\;,
\quad \quad
\widehat V(\widehat Z) =  -1 -\widehat c\widehat Z^2 -\widehat Z^4
\;,
\end{eqnarray}
The metric (\ref{PD_metric}) is Lorentzian iff $\widehat{V} > 0$, which will
be the case if and only if
\begin{equation}
 \widehat a \,<\, \widehat Z^2 \,<\,\widehat a + \sqrt{\widehat c^2 -4} = \widehat a^{-1}
\;, \quad
\widehat a:= -\frac{1}{2}(\sqrt{\widehat c^2 -4}+\widehat c)>0
\;.
\label{range_Z}
\end{equation}
Since the line element  \eq{PD_metric} is invariant under the transformation $\widehat Z\mapsto -\widehat Z$ we may assume that $\widehat Z>0$.
We define new coordinates (compare \cite{kmv}),
\begin{eqnarray}
\widehat  Z  =\sqrt{ \widehat a }\cosh \theta
\;,
\quad
\widehat y =\sqrt{\frac{\Lambda}{3\widehat a}}\, r
\;,
\quad
\widehat \sigma = -\sqrt{\widehat a}\, \phi\;,
\quad
 \tau = \sqrt{\frac{\Lambda\widehat a}{3}}\, t - \widehat a^{3/2} \phi
\;,
\end{eqnarray}
which transform \eq{PD_metric} to the line element \eq{wick-KdS_metric_old} of the Wick-rotated Kerr-anti-de Sitter metric
in Boyer-Lindquist-type coordinates.
We note that the coordinate transformation  reveals that
%
\begin{eqnarray}
a\,=\,\sqrt{\frac{3}{\Lambda}}\widehat a\;, \quad
\Big(\frac{\Lambda}{3}\Big)^2b_1\,=\,  
\widehat b_1  \,=\, 2\Big(\frac{3}{\Lambda}\Big)^{1/4} m  a^{-3/2}
\;,
\end{eqnarray}
which recovers  \eq{wick_param_m_C}.

\subsection{$a\rightarrow \infty$-KdS-limit-spacetime: Case $3 (a)$}

Consider next the case $\widehat k =0$, $\widehat c=-1$, so that now
\begin{eqnarray}
\widehat W(\widehat y)&=&-\widehat b_1 \widehat y - \widehat y^2- \widehat y^4
\;, \quad \quad
\widehat V(\widehat Z) =\widehat Z^2- \widehat Z^4
\; . \label{Case3a}
\end{eqnarray}
%
The condition of $\widehat V>0$ holds iff
%
%
%
\begin{equation}
\widehat Z^2\,\in\, (0,1)
\;.
\end{equation}
As before, the invariance $\widehat Z\mapsto -\widehat Z$ of the metric
allows us to assume that $\widehat Z>0$, i.e.\  $\widehat Z\,\in\, (0,1)$.
We define new coordinates $(t,r,\theta,\phi)$,
\begin{equation}
\widehat Z \,=\,\cos\theta\;, \quad \widehat y \,=\,  r\;, \quad \widehat \sigma \,=\, -\frac{\Lambda}{3} \phi\;, \quad \tau \,=\,\frac{\Lambda}{3}( t - \phi)
\end{equation}
and together with
%
\begin{equation}
  m\,=\, \frac{1}{2}\frac{\Lambda}{3} \widehat b_1 \,=\, \frac{1}{2}\Big(\frac{\Lambda}{3}\Big)^3  b_1
\;,
\end{equation}
which recovers \eq{limit_KdS_param_m_C},
the line element  \eq{PD_metric} 
indeed takes the form \eq{KdS_limit}.
%

\subsection{The Kerr-de Sitter family: Case $2 (a)$}
\label{section_KdSI}
It only remains the Case $2 (a)$, corresponding to the Kerr-dS family. Given the importance of this family we make a deeper analysis in this section, in particular comparing with \cite[Theorem\,1]{mars-senovilla} and \cite[Theorem\,6]{mars-senovilla}.
Our first goal is to analyze under which conditions
all the hypotheses for the application of  \cite[Theorem\,1]{mars-senovilla} are fulfilled in our setting,
namely when attention is restricted to Kerr-dS-like spacetimes.
Apart from those metrics belonging to case $2 (a)$ this analysis will also include the limiting case $2(b)$, i.e.\ the Schwarzschild-de Sitter family.
In particular, we are going to show that, assuming a conformally flat $\scri$, the positivity of $\widehat k$ ensures that all the assumptions of
\cite[Theorem 6]{mars-senovilla} are  satisfied, where it is shown how the Kerr-dS metric can be brought to the Pleba\'nski form given in  (B.i) of \cite[Theorem 4]{mars-senovilla}. This will then complete our comparison with \cite[Theorem 4]{mars-senovilla}.

It is shown in \cite{mars-senovilla} that,
among spacetimes with vanishing MST and satisfying $Q\F^2$ and
$Q\F^2 - 4 \Lambda$ both not identically zero, 
the Kerr-de Sitter family (including Schwarzschild-de Sitter) corresponds
exactly to the case when the polynomial
\begin{equation}
V(\zeta) \,=\, k +b_2\zeta-c\zeta^2-\frac{\Lambda}{3}\zeta^4
\end{equation}
can be factored as
\begin{equation}
V(\zeta) \,=\, \check V(\zeta)(\zeta_1^2-\zeta^2)  
\;,
\end{equation}
with $\zeta_1 \geq 0$, $\hat{V}(\zeta)$ strictly positive in
$[-\zeta_1,\zeta_1]$ and, in addition, 
Image$(Z)\in [-\zeta_1,\zeta_1]$. 
These conditions require necessarily that $b_2 =0$
%
which recovers the fact that $\scri$ needs to be conformally flat, 
cf. (\ref{expression_b1}).

For $b_2=0$ the polynomial $V(\zeta)$ happens to be closely related to the characteristic polynomial $\mathcal{P}$ used in (\ref{carpol}) of Appendix \ref{app_conf_Eucl} to classify the different equivalence classes $[Y]$ of CKVFs in Euclidean space up to M\"obius transformations, the relation being given simply by
$$
V(\zeta) \,=\, k -c\zeta^2-\frac{\Lambda}{3}\zeta^4 =\frac{i}{\zeta} \left(\frac{3}{\Lambda}\right)^4 \mathcal{P}\left(\frac{\Lambda}{3}i\zeta \right).
$$
%
%
Therefore, the restrictions posed by the requirement that $V(\zeta)$ be factored as
\begin{eqnarray}
 k-c\zeta^2 - \frac{\Lambda}{3}\zeta^4 \,=\, \check V(\zeta)(\zeta_1^2-\zeta^2)
\end{eqnarray}
with $\check V(\zeta)>0$ on $[-\zeta_1,\zeta_1]$ easily follows from the analysis in Appendix \ref{app_conf_Eucl}, as ${\mathcal P}$ must have a couple of complex conjugate eigenvalues $\pm i \mu_1 =\pm i (\Lambda/3) \zeta_1$
and a couple of {\em non-vanishing} opposite real eigenvalues $\pm \lambda$. This can only happen in Cases 2 of Theorem \ref{Classification}
that is to say, whenever
%
\begin{eqnarray}
 && \widehat k> 0 \quad \text{or}
\label{condition_alter1}
\\
&&   \widehat k=0\;, \enspace \widehat c >0
\label{condition_alter2}
\end{eqnarray}
where the following relations hold
$$
\lambda^2 =\frac{1}{2}\left( \sqrt{\widehat c^2 +4\widehat k} + \widehat c\right), \hspace{1cm} \mu_1^2=\frac{\widehat k}{\lambda^2}, \hspace{1cm}
\mu_1^2 =\frac{1}{2}\left( \sqrt{\widehat c^2 +4\widehat k} - \widehat c\right)
$$
and furthermore the factor $\check V(\zeta)$ takes the simple form
\begin{equation}
 \check V(\zeta)\,=\, \frac{\Lambda}{3}\zeta^2 + \frac{3}{\Lambda} \lambda^2
\;,
\end{equation}
which is obviously positive on $[-\zeta_1,\zeta_1]$, and in fact everywhere.
Hence the inequality $V(Z) \geq 0$ is equivalent to 
$Z \in [-\zeta_1, \zeta_1]$. One of the consequences of
the analysis in \cite{mars-senovilla} is that $V(Z) \geq 0$
everywhere on any $\Lambda$-vacuum spacetime satisfying the
alignment condition (\ref{alig}) with $Q\Fsq$
and $Q \Fsq - 4 \Lambda$ not identically zero. As a consequence,
the condition Image$(Z) \in [-\zeta_1,\zeta_1]$ becomes
redundant when $b_2 =0$.

It is, however, an interesting consistency check to analyze the
range of $Z$ at $\scri$ and show that indeed satisfies
Image$(Z) \in [-\zeta_1,\zeta_1]$.
Let us therefore analyze whether
\begin{eqnarray}
Z^2 \leq \zeta_1^2 =\Big(\frac{3}{\Lambda}\Big)^2\mu_1^2 =\Big(\frac{3}{\Lambda}\Big)^2 \frac{1}{2}\Big(\sqrt{\widehat c^2 +4 \widehat k}-\widehat c \Big)
\;.
\label{cond_imag_Z}
\end{eqnarray}
holds. It has been shown in \cite{mpss} that the function $Z$ satisfies
\begin{equation}
Z|_{\scri} \,=\, -\frac{1}{2} \frac{3}{\Lambda} |Y|^{-1}Y_kN^k
\;.
\end{equation}
Thus, \eq{cond_imag_Z} becomes
\be
 (Y_kN^k)^2 \leq 2 |Y|^{2} \Big(  \sqrt{\widehat c^2  + 4\widehat   k} - \widehat c \Big)=4\widehat{k}|Y|^{2} \lambda^{-2}.
 \label{im_iequality2}
\ee
%


To check whether the inequality \eq{im_iequality2}  is satisfied 
we exploit again
the conformal symmetries of Euclidean 3-space,
as well as the gauge scaling freedom explained in Remark \ref{gauge_freedom_kc}.
From Theorem \ref{Classification} we know that  $Y$ can be represented by  ``standard forms'', with $\widehat k$ and $\widehat c$ being appropriately normalized, as follows
\begin{eqnarray}
Y &=&\lambda \begin{pmatrix} x  \\  y \\ z
\end{pmatrix}
+ \lambda^{-1}\begin{pmatrix} y \\  -x \\ 0
\end{pmatrix}
\;, \quad \lambda>0\;,  \quad \text{for $\widehat k=+1$}
\;,
\label{special_CKVF1}
\\
Y &=& \begin{pmatrix} x  \\  y \\ z
\end{pmatrix}
\;,  \quad \text{for $\widehat k=0$ and $\widehat c=+1$}
\;.
\label{special_CKVF2}
\end{eqnarray}
%

Let us first study the case $\widehat k=+1$ (i.e. $\zeta_1 >0$). It is easily found from the expression \eq{special_CKVF1} that
%
\begin{equation}
Y_kN^k \,=\, 2 z, \hspace{1cm}
|Y|^2 \,=\, \lambda^2z^2 +(\lambda^2 + \lambda^{-2}) (y^2+x^2)
\;,
\end{equation}
and \eq{im_iequality2} becomes
$$
(1+\lambda^{-4})(x^2+y^2)\geq 0
$$
which is obviously true.
%
%
%
We therefore have Image$(Z^2)\in [0,\zeta_1^2]$ on $\scri^-$,
as required.

Knowing  that \eq{special_CKVF1} generates a KdS-spacetime we compute its mass $m$ and its angular momentum $a$ in terms of
the parameters $\widehat c$ and $\Dconst$ from the results in \cite{mars-senovilla}. From (\ref{cond_imag_Z}) together with
%
\begin{eqnarray*}
b_1 &=& -6\Dconst\Lambda^{-2}\sqrt{\frac{\Lambda}{3}}
\;,
\\
\check V (0) &=&  \frac{3}{\Lambda}\lambda^2
\,=\, \frac{1}{2}\frac{3}{\Lambda}(\sqrt{\widehat c^2 + 4} + \widehat c)
\;,
\end{eqnarray*}
we deduce that
\begin{align*}
a^2 & = \frac{\zeta_1^2}{\check V(0)} \,=\,  \frac{3}{4\Lambda}(\sqrt{\widehat c^2 + 4}-\widehat c)^2
\\
m & = \frac{b_1}{2(\check V(0))^{3/2}}\,=\, \frac{1}{2}\Big(\frac{\Lambda}{3}\Big)^{9/4}a^{3/2} b_1
\,=\,-\frac{1}{3} \Big(\frac{\Lambda}{3}\Big)^{3/4}a^{3/2} \Dconst
 \,=\,-\frac{\Dconst}{3} \lambda^{-3}
\;.\phantom{xx}
\end{align*}
in agreement with  \eq{KdS_param_a} and \eq{KdS_param_m_C}.
%
%


Let us pass to the case $\widehat k=0$  and $\widehat c=1$, which
is equivalent  to $\zeta_1=0$. Now the image of $Z$ is simply 
$\{0\}$ everywhere, in particular at $\scri$. We again check the latter
for consistency using 
our results in this paper. Indeed, $Z=0$
 at $\scri$ will hold 
if and only if $Y_kN^k=0$, which is automatically satisfied by
the CKVF \eq{special_CKVF2} because $ Y$ is a gradient
and hence its $\mbox{curl}$, $N^k$, vanishes.


When $Z=0$, the results in \cite{mars-senovilla} show that 
the spacetime is locally isometric to the
Schwarzschild-de-Sitter case with mass
$$
 m\,=\,\frac{b_1}{2(\check V(0))^{3/2}} \,=\, \Big(\frac{\Lambda}{3}\Big)^{3/2}\frac{b_1}{2}\,=\,  -\frac{\Dconst}{3}
$$
in full agreement with \eq{kottler_param_m_C} (note that $\check V(0)=c=(3/\Lambda)\widehat c = 3/\Lambda$).

\begin{remark}
{\rm
Due to the relation  $Y_kN^k=0$ the restriction of $\widetilde X$ to
$\scri^-$ is  hypersurface orthogonal.
In Schwarzschild-dS this is true also off $\scri$, i.e.\ the KVF whose associated MST vanishes is hypersurface orthogonal in the physical spacetime $(\mcM,g)$.
Indeed, this follows from   Proposition~\ref{prop_hyp_orth}, as  in Kerr-dS-like spacetimes the hypothesis  (ii) imposed there   holds.
}
\end{remark}

\section*{Acknowledgments}
TTP is grateful to Bobby Beig for pointing out reference \cite{geroch} and to  Piotr Chru\'sciel for useful discussions.  
MM acknowledges financial support under the project
FIS2015-65140-P (Spanish MINECO/FEDER).
TTP acknowledges financial support  by the Austrian Science Fund (FWF): P~23719-N16 and P~28495-N27.
JMMS is supported under grants FIS2014-57956-P (Spanish MINECO-fondos FEDER),  UFI 11/55 (UPV/EHU), and EU COST action CA15117 ``CANTATA". 

\section*{Appendices}

\appendix

\section{Relation between CKVFs in $\mathbb{S}^n$ (and $\mathbb{E}^{n}$) with 2-forms in flat spacetime $\mathbb{M}^{1,n+1}$}
\label{app_SHM}

For our analysis, we need to classify the conformal Killing vector fields (CKVF) in
the $3$-dimensional round sphere $\mathbb{S}^3$ up to conformal transformations.
Thus, we devote this Appendix to that end, without restricting the dimension of the sphere, by presenting a classification of the CKVF in
the $n$-dimensional sphere $\mathbb{S}^n$ ---up to conformal transformations--- via its direct relation with 2-forms in Minkowski spacetime. The companion Appendix \ref{App:2-forms} deals with the algebraic classification of these 2-forms and its implications concerning canonical forms of the CKVF, which are then fully detailed in Appendix \ref{app_conf_Eucl}.

Let $\CSn$ denote the conformal group of $\mathbb{S}^n$
with the standard round metric $\gSn$, that is to say,  $\CSn$ is the set of diffeomorphisms
$\Phi: \mathbb{S}^n \mapsto \mathbb{S}^n$ satisfying
$$\Phi^{\star} (\gSn) = \Omega^2 \gSn$$
for some positive $\Omega \in {\cal F} (\mathbb{S}^n)$. Let $\KSn$ be the conformal
Killing algebra, i.e. the set of (smooth) vector fields $\xi$ satisfying
$$\mcL_{\xi} \gSn = 2 \Psi \gSn$$
for some
$\Psi \in {\cal F} (\mathbb{S}^n)$. It is well-known that $\Phi_{\star} (\xi)
\in \KSn$ for $\Phi \in \CSn$ and $\xi \in \KSn$. This defines an equivalence
relation $\sim$  in $\KSn$: $\xi_1  \sim \xi_2$ if there exists $\Phi \in \CSn$
such that $\xi_1 = \Phi_{\star} (\xi_2)$. We wish to classify the corresponding equivalence classes.

The conformal group of $\mathbb{S}^n$ is also called the M\"obius group and
is well-known to be isomorphic to the isometry group $\IsoHn$ of the
hyperbolic space $\mathbb{H}^{n+1}$. Moreover, $\mathbb{H}^{n+1}$
can be viewed as the
hyperboloid
$$\H := \{ \eta_{\alpha\beta} x^{\alpha} x^{\beta} = -1, \, \, \,
t :=x^0 > 0 \}$$
of the Minkowski spacetime $\mathbb{M}^{1,n+1}$,
where $\{x^{\alpha}\}$ is a fixed choice of Minkowskian coordinates.
The embedding of the hyperboloid $\H$ leading to the disk model representation of hyperbolic space
is given by
\begin{align*}
\varphi: B_1 := \{ |y|< 1 \} \subset \mathbb{R}^{n+1}  & \longrightarrow
\Mink \\
y   & \longrightarrow \varphi(y) :=  \left \{ \begin{array}{cc}
t & =  \frac{1 + |y|^2}{1- |y|^2} \\
x & = \frac{2 y}{1 - |y|^2} \\
\end{array}
\right .
\end{align*}
where $| \cdot |$ denotes the Euclidean norm in $\mathbb{R}^{n+1}$
and we have written $x^{\alpha}  = (t,x)$, $x \in \mathbb{R}^{n+1}$.
The induced metric is immediately found to be
\begin{align*}
\gHn = \frac{ 4 |dy|^2}{(1- |y|^2)^2}
\end{align*}
which is  indeed the hyperbolic metric in the disk model representation.

The isometry group of $\mathbb{H}^{n+1}$ is thus isomorphic to the orthochronous Lorentz
group $\Lor^{+}:= O^+(1,n+1)$, i.e. the subset
of the Lorentz group preserving the time orientation of causal vector fields.
It follows that there exists a Lie-algebra isomorphism between the set of conformal Killing vectors
$\KSn$ and the set of Killing vectors $\KillM$ generating the connected
component of the identity of $\Lor^+$. The
latter are those Killing vectors of $\mathbb{M}^{1,n+1}$ leaving the origin
$o = \{x^{\alpha} =0 \}$ invariant. Elements in  $\KillM$ are thefore
characterized by a 2-form $\bm{F}$ over $T_{o} \Mink$
via the expression
\be
\zeta_F := F^\mu{}_\nu x^\nu \partial_\mu \label{kil}.
\ee
These Killing vectors are
obviously tangent to $\H$ (because the hyperboloid remains invariant under
the isometries generated by $\zeta_{F}$).
Hence, there exists a vector field
$\hat{\zeta}_{F} \in \X (\mathbb{H}^{n+1})$
 such that $\varphi_{\star}
(\hat{\zeta}_{F} |_p ) = \zeta_{F} |_{\varphi(p)}$ for all $p \in
\mathbb{H}^{n+1}$, and the following must hold
\begin{align*}
\hat{\zeta}_{F}^a |_{y} \frac{\partial \varphi^{\alpha}}{\partial y^a} =
\zeta^{\alpha}_F |_{\varphi(y)} =
F^{\alpha}_{\phantom{\alpha}\beta}\, \varphi^{\beta}(y) =
\frac{1 + |y|^2}{1- |y|^2} F^{\alpha}_{\phantom{\alpha}t}
+ \frac{2 y^a}{1 - |y|^2} F^{\alpha}_{\phantom{\alpha}a}
\end{align*}
where Latin indices $a,b,\cdots$ take values in $\{1,\cdots, n+1\}$. Computing the
Jabobian of the embedding $\varphi$, it is easy to check that $\hat{\zeta}_{F}$ is given by
\begin{align}
\hat{\zeta}_{F} =
\Big ( \frac{1}{2} F^b{}_{t} (1 + |y|^2)
+ F^b{}_{a} y^a - F^t{}_{a} y^a y^b
\Big ) \frac{\partial}{\partial y^b}.
\label{zetaomega}
\end{align}

The conformal metric $g_E: =\frac{(1 - |y|^2)^2}{4} \gHn$ is  the
flat metric in $B_1$ and can be smoothly extended to the boundary
$\partial B_1$ in an obvious way. The space $\partial B_1$ with the metric induced
from $g_E$ is the $n$-sphere $\mathbb{S}^{n}$. Any vector field $\hat{\zeta}_{F}$
is a Killing vector in $(B_1,\gHn)$ and hence a conformal Killing vector in
$(B_1,g_E)$. It also extends smoothly to $\partial B_1$
as a tangential vector. Indeed, $|y|^2=1$ defines the
boundary $\partial B_1$ and using (\ref{zetaomega}) (Latin indices are raised and lowered with the
Euclidean metric $g_E$) one has 
\begin{align*}
\hat\zeta_{F} (|y|^2 ) = 
F_{bt} (1+|y|^2)  y^b + 2 F_{ba} y^b y^a - 2 F^{t}{}_{a}y^a
|y|^2 =  \left ( 1 - |y|^2 \right ) F_{a t} y^a
\stackrel{|y|^2 =1}{=} 0
\end{align*}
where the antisymmetry of $F_{\alpha\beta}$ has been used.

Thus, the restriction of $\hat\zeta_{F}$ to $\partial B_1$ is a conformal Killing
vector of $\mathbb{S}^n$ which we denote by
$\bar{\zeta}_{F}$. The M\"obius group of $\mathbb{S}^n$
is precisely the extension to the boundary $\partial B_1$
 of the isometry group $\IsoHn$. At the level of generators, the
isomorphism between the M\"obius group and $\IsoHn$ is translated into the
map
\begin{align*}
\Tau: \KillM & \rightarrow \CSn \\
\zeta_F &\mapsto \Tau(\zeta_{F}) = \bar{\zeta}_{F}
\end{align*}
defined by using the procedure just explained. It follows that $\Tau$ is
necessarily a Lie-algebra isomorphism (this can also be
checked directly).
In summary, classifying the set of conformal Killing vectors $\CSn$ up to
the M\"obius group amounts to classifying the vectors fields
$\zeta_{F}$ up to the orthochronous Lorentz group. For this reason, we devote next section to the action of the Lorentz group on $\zeta_{F}$ and on the underlying 2-form $\bm{F}$.

\subsection{Lorentz transformations}
Recall that a Lorentz transformations $L^\mu{}_\alpha $ (at any point $x\in \mcM$) is defined by
$$
g_{\mu\nu} L^\mu{}_\alpha L^\nu{}_\beta  = g_{\alpha \beta}
$$
so that, as usual,
$$
L_{\nu\alpha} L^\nu{}_\beta = g_{\alpha\beta} , \hspace{1cm} L_\nu{}^\alpha L^\nu{}_\beta = \delta^\alpha_\beta
$$
which provides the formula for the inverse Lorentz transformation
$$
(L^{-1})^\alpha{}_\nu = L_{\nu}{}^\alpha \, .
$$
Recall also that, as is obvious
$$
\det (L) =\pm 1, \hspace{1cm} g_{\mu\nu} (L^{-1})^\mu{}_\alpha (L^{-1})^\nu{}_\beta  = L_{\alpha}{}^{\mu} L_{\beta}{}^{\nu}g_{\mu\nu}=g_{\alpha \beta}.
$$

Under Lorentz transformations, the different tensorial objects transform as follows:
$$
\omega'_\mu = L_\mu{}^\nu\omega_\nu, \hspace{3mm} v'^\mu =L^\mu{}_\nu v^\nu, \hspace{3mm} F'_{\alpha\beta} =L_\alpha{}^\mu L_\beta{}^\nu F_{\mu\nu}, \hspace{3mm}  F'^\alpha {}_\beta =L^\alpha{}_\nu F^\nu{}_\mu L_\beta{}^\mu \, .
$$
Therefore, under a Lorentz transformation the vector fields $\zeta_F$ in (\ref{kil}) transform according to the previous rules as
$$
\zeta'_F{} ^\alpha = L^\alpha{}_\mu \zeta^\mu = L^\alpha{}_\mu F^\mu{}_\nu x^\nu =
L^\alpha{}_\mu F^\mu{}_\nu (L_\rho{}^\nu L^\rho{}_\sigma) x^\sigma =
F'^\alpha{}_\rho (L^\rho{}_\sigma x^\sigma) =\zeta_{F'}{}^\alpha
$$
and therefore, in order to classify the set $\{\zeta_F\}$ up to the orthochronous Lorentz group Lor$^+$,
it is enough to classify 2-forms over $T_o\Mink$ up to Lor$^+$.

Denoting simply by $F$ the endomorphism $F^{\mu}{}_{\nu}$ associated to $F_{\mu\nu}$, one can define the following set of invariants associated to $\bm{F}$
\be
I_{s}:= \frac{1}{2} \mbox{tr} (F^{2s}), \hspace{1cm} s=1,\dots , [n/2+1] \label{inv0} .
\ee
Notice that the odd powers of $F$ have identically zero traces, thus zero invariants.

One immediately notices that all 2-forms $\bm{F}'$ which are Lorentz transformed of $\bm{F}$ have exactly the same set of invariants $I_a$. This follows from
\begin{align*}
2I_1(F') & = F'^\alpha {}_\beta  F'^\beta {}_\alpha = L^\alpha{}_\nu F^\nu{}_\mu L_\beta{}^\mu L^\beta{}_\rho F^\rho{}_\sigma L_\alpha{}^\sigma =L^\alpha{}_\nu F^\nu{}_\mu F^\mu{}_\sigma L_\alpha{}^\sigma =
F^\nu{}_\mu F^\mu{}_\nu \\
& = 2 I_1(F)
\end{align*}
and so on.
However, as discussed in detail in the companion Appendix \ref{App:2-forms}, the converse does not hold, and the set of invariants $I_a (F)$ does not determine the algebraic type of $F$, and thus it does not permit to characterize the required classes of 2-forms. The full classification requires, in addition to the invariants, the {\em rank} $2\bar r$ of the 2-form $\bm{F}$, see Section~\ref{subsec:2-formClassification}

\subsection{CKVF in $\mathbb{S}^n$ versus $\mathbb{E}^n$}
\label{Ster}
For many purposes, it is often useful to consider the set of conformal transformations
of Euclidean space $\mathbb{E}^n$ instead of the
conformal group of the sphere $\mathbb{S}^n$. This requires some care because
$\mathbb{E}^n$ is conformal to $\mathbb{S}^n \setminus \{q\}$ where
$q\in \mathbb{S}^n$ is a point on the sphere. Thus, the conformal group of $(\mathbb{E}^n,g_E)$
is, strictly speaking, the subset of the conformal group of $\mathbb{S}^n$
leaving $q$ invariant, which is a proper subset of the M\"obius group.

For this reason, one considers maps
 $$\Phi : \mathbb{E}^n \setminus \{q_1\} \rightarrow
\mathbb{E}^n \setminus \{q_2\}$$
satisfying  $\Phi^{\star} (g_E) = \Omega^2 g_E$ where $\Omega \in {\cal F}(\mathbb{E}^n
\setminus \{q_1 \})$ is positive and  the points $q_1, q_2 \in \mathbb{E}^n$
may depend on $\Phi$. In this extended sense, the set of conformal maps
of $\mathbb{E}^n$ is the same as the M\"obius group. It follows that the
conformal Killing algebra of $\mathbb{E}^n$ is isomorphic to the
conformal Killing algebra of $\mathbb{S}^n$. Hence, the arguments
above allow one to classify the conformal Killing vectors of Euclidean
space up to the M\"obius group (understood in the sense above). We are going
to do this explicitly in what follows by constructing the map that sends
any conformal Killing vector $Y \in \KEn$ to the corresponding 2-form
$\bm{F}\in T_o\Mink$.

To that aim, it is convenient to use the half-space representation of
hyperbolic space $\mathbb{H}^{n+1}$
because then the boundary is indeed diffeomorphic to $\mathbb{R}^{n}$.
The relationship betweem the disk and the half-space models of $\mathbb{H}^{n+1}$
is as follows. Fix a point
$p \in \partial B_1 \subset \mathbb{R}^{n+1}$.
The circle inversion of radius $2$
centered at this point is the map
\begin{align}
\chi: \mathbb{R}^{n+1} \setminus \{ p \}& \rightarrow \mathbb{R}^{n+1}
\setminus \{p \} \nonumber \\
 y & \mapsto \chi(y) \defi z  = p + \frac{4}{|y-p|^2} (y-p)
\label{mapchi}
\end{align}
which has inverse
\be
y = \chi^{-1}(z) = p + \frac{4}{|z-p|^2}(z-p).\label{chiinv}
\ee
Observe that $|z-p|^2 =16 |y-p|^{-2}$.
The $n$-sphere outside $p$, $\partial B_1 \setminus \{ p \}$
is mapped onto the hyperplane $ \la z-p,p \ra = -2 $ or, equivalently, given that $|p|=1$
\be
\chi (\partial B_1 \setminus \{ p \}):=
{\mathcal P} = \{ \la z,p \ra = -1 \} \label{boundary}
\ee
and
the (open) ball $B_1$ is mapped diffeomorphically to the half-space
$\{ \la z , p \ra < -1\} $. The pull-back of the hyperbolic metric
$\gHn$ reads
\begin{align}
(\chi^{-1})^{\star}(\gHn)
= \frac{|dz|^2}{\left ( \la z,p \ra + 1 \right )^2}. \label{half}
\end{align}

We have to compute the vector fields $\hat{\zeta}_{F}$ as defined in (\ref{zetaomega}) and extend them to the boundary (\ref{boundary}). As before, this defines a Lie algebra
isomorphism between the Killing vectors of $\mathbb{H}^{n+1}$ and the conformal Killing vectors
of Euclidean space $\mathbb{E}^n$.
We first transform the vector field
(\ref{zetaomega}) into the $z$ coordinates or, more precisely,
compute $\chi_{\star}
(\hat{\zeta}_{F})$. This is given by
\begin{align}
\chi_{\star} (\hat{\zeta}_{F}) |_z =
\left .
\hat{\zeta}_F^b(y) \frac{\partial \chi^{c}}{\partial y^b}
\right |_{y=\chi^{-1}(z)} \partial_{z^c} \label{vectortrans} .
\end{align}
The Jacobian is
\begin{align}
\left .
\frac{\partial \chi^c}{\partial y^b}
\right |_{y = \chi^{-1}(z)}
=
\frac{|z-p|^2}{4} \delta^c_b - \frac{1}{2} (z-p)^c (z-p)_b
\label{Jacob}
\end{align}
which implies, taking (\ref{chiinv}) into account,
\begin{align}
\left. \frac{\partial \chi^c}{\partial y^b}
\right |_{y = \chi^{-1}(z)}   p^b & = \frac{|z-p|^2}{4} p^c - \frac{1}{2} \la z-p,p \ra (z-p)^c , \nonumber \\
\left . \frac{\partial \chi(y)^c}{\partial y^b} y^b
\right |_{y = \chi^{-1}(z)}
& = - \frac{1}{2} (z-p)^c \left ( 1 + \la z,p \ra \right )
+ \frac{|z-p|^2}{4} p^c.
\label{Jacoby}
\end{align}
Moreover, using $|p|^2=1$ and the explicit form of $\chi^{-1}$ given in (\ref{chiinv}) it follows
\begin{align}
|\chi^{-1}(z)|^2 =  1 +\frac{8}{|z-p|^2} \left ( 1 + \la z,p \ra \right ).
\label{y2}
\end{align}
Let us define 
$\bm{F^t}$ as the constant
covector in $\mathbb{R}^{n+1}$
with components
$F^t_{\phantom{t}a}$. Inserting (\ref{Jacob}),(\ref{Jacoby}) and (\ref{y2})
into (\ref{vectortrans}), a straightforward calculation yields
\begin{align}
\tilde{\zeta}^c_{F} \defi \chi_{\star}(\hat{\zeta}_{F})^c & =
\left ( \frac{|z-p|^2}{4} + 1+ \la z,p \ra
\right ) F^c_{\phantom{c}t} +
F^c_{\phantom{c}a}  \left ( p^a \frac{|z-p|^2}{4} + (z-p)^a \right )
\nonumber \\
& +\frac{1}{2}
(z-p)^c \left [ -  F_{b \, a} z^b p^a
-  \la \bm{F^t}, z-p \ra + \la \bm{F^t},p \ra
\left ( 1+ \la z,p \ra \right ) \right ] \nonumber \\
&- p^c \left ( \frac{1}{4} |z-p|^2 \la \bm{F^t},p \ra
+ \la \bm{F^t}, z-p \ra \right ). \label{vectildezeta}
\end{align}
This vector field extends regularly at the boundary
hyperplane  ${\mathcal P}$.
Moreover, the normal vector to ${\mathcal P}$ is $p$ and hence the
normal component of $\tilde{\zeta}_{F}$ is 
\begin{align*}
\la \tilde{\zeta}_{F}, p \ra =
\left( 1 + \frac{1}{2} \la z-p,p \ra \right )
\left ( \frac{}{} \la \bm{F^t},p \ra \left ( 2 + \la z-p,p\ra \right )
+ F_{ba} p^b (z-p)^a - \la \bm{F^t},z-p \ra \right )
\end{align*}
which vanishes at ${\mathcal P}$. Thus
$\tilde{\zeta}_{F}$ is tangent to this hyperplane.

If we let  $p$ be the point $p= (-1,0, \cdots 0 )$ and
define Cartesian coordinates $x \defi z + p $, the domain of $x$ is simply $x^1 > 0$
and the metric (\ref{half}) takes the form
\begin{align*}
(\chi^{-1})^{\star}(\gHn) = \frac{|dx|^2}{(x^1)^2}
\end{align*}
which is indeed the half-space model of hyperbolic space.
 ${\mathcal P}$ is now $\{x^1 =0\}$,
so its inclusion map $i$ is simply $i(x^A) = (x^1 =0,x^A)$, where
indices $A,B,\cdots$ take values in $\{2,\cdots,n+1\}$.
From (\ref{boundary}), there exists a map $\Psi : \mathbb{S}^{n} \setminus \{p\} 
\mapsto \mathbb{R}^{n}  $ such that 
the restriction of $\chi$ to  ${\mathbb{S}^{n} \setminus \{p\}}$ satisfies 
$\chi |_{\mathbb{S}^{n} \setminus \{p\}} = i \circ \Psi$ (after identifying 
$S^{n}$ with $\partial B_1 \subset \mathbb{R}^{n+1}$).
$\Psi$ is the standard stereographic projection centered at $p$.
The tangency of $\tilde{\zeta}_{F}$ to ${\mathcal P}$ implies that
for any CKVF $Y$ in $\mathbb{S}^n$ 
arising as the restriction of $\hat{\zeta}_F$  
to $\mathbb{S}^n$, the vector field $\Psi_{\star}(Y) \in
\X(\mathbb{R}^n)$ satisfies $i_{\star}( \Psi_{\star} (Y |_x)) =
\chi_{\star} (Y|_x) = \tilde{\zeta}_{F}|_{\chi(x)}$.
To find the expression of $\Psi_{\star}(Y)$ we simply need
to evaluate  the $A$-components
of $\tilde{\zeta}_F$ (\ref{vectildezeta}) at $x^1 =0$.
From $(z-p)^A = (x -2 p)^A =x^A$, $(z- p)^1 = (x - 2p)^1 \stackrel{x^1=0}{=} 2$
and
\begin{align*}
\frac{1}{4} |z- p |^2 & = \frac{1}{4} |x - 2 p|^2 = \frac{1}{4} |x|^2 + 1 -  \la x, p \ra \stackrel{x^1 =0}{=}
\frac{1}{4} \delta_{AB} x^A x^B + 1 := \frac{1}{4}|x|^2_{\delta} + 1 \\
\la \bm{F^t},z -p \ra & \stackrel{x^1=0}{=} 2 F^t{}_{1}  + F^t{}_{B} x^B
\end{align*}
we  find
\begin{align}
\Psi_{\star}(Y)^A =
F^A{}_{t} + F^A{}_{1}
+ \frac{|x|_{\delta}^2}{4} \left ( F^A{}_{t}-
 F^A{}_{1} \right )
+ F^A{}_{B} x^B - F^t{}_{1} x^A
+ \frac{1}{2} x^A \left ( F^C{}_{1} - F^C{}_{t}
\right ) x_C. \label{vec}
\end{align}
Defining a (constant)
scalar $v$, vectors $a,b$ and
two-form $\bm{\omega}$ in $\mathbb{R}^n$ by means of
\begin{align}
v&:= -F^t{}_1 =F_{t1}, & & &  2a_B & := - F_{Bt} + F_{B1}=-F^t{}_B -F^1{}_B,
\label{va}\\
b_B &:= F_{Bt}+F_{B1}=F^t{}_B -F^1{}_B, & & &  \omega_{AB} & := - F_{AB}
\nonumber
\end{align}
the conformal Killing vector $\Psi_{\star}(Y)$ on Euclidean space associated to $\zeta_F$ is written as
\be
\Psi_{\star}(Y) =\left(b^B +v x^B +(a_A x^A) x^B-\frac{1}{2} (x_Ax^A) a^B -\omega^B{}_A x^A\right)\partial_B
\label{CKV}
\ee
and we recover the well-known expression for conformal Killing vectors in $\mathbb{R}^n$. In terms of $v,a,b,\bm{\omega}$ the 2-form $\bm{F}$ over $T_o(\Mink)$ that generates $Y$ is given, in matrix form, by
\begin{equation}
\bm{F} = (F_{\mu\nu}) =\left(
\begin{array}{ccc}
0 & v & a-b/2 \\
-v & 0 & -a-b/2 \\
-a+b/2 & a+b/2 & -\bm{\omega}
\end{array}
\right)
\label{ExpF}
\end{equation}
or equivalently by
\be
\bm{F} = v dt\wedge dx^1 + dt \wedge (a-b/2) -dx^1 \wedge (a +b/2) -\bm{\omega}
\label{F=vabw} .
\ee


\subsection{Fixed points of the CKVFs in $\mathbb{S}^n$}
\label{newApp}
In the main text we will need to know the fixed points of the CKVFs in $\mathbb{S}^n$. Since these CKVFs are given explicitly in (\ref{CKV}) in terms of Cartesian coordinates in the stereographic projection of $\mathbb{S}^n$, the zeros outside
the pole $p$ can be identified directly. However, some argument
is needed to determine whether $Y$ vanishes at the pole $p$. There
are several ways of doing this. Here we exploit the above discussion
to find under which conditions $Y$ vanishes at $p$.

\begin{lemma}
\label{zero_at_infinity_0}
Let $Y$ be a CKVF in $(\mathbb{S}^n,\gamma^{\mathbb{S}^n})$
and $\Psi: \mathbb{S}^n \setminus \{p \} \mapsto \mathbb{R}^n$
the stereographic projection centered at $p$.
Then $Y(p)=0$ if and only if $a_B$ in \eq{CKV} vanishes, with $a_B$ given as in \eq{va}.
\end{lemma}
\begin{proof}
\label{proof simplified, given the previous rewordings}
Following the previous section \ref{Ster}, 
$Y$ is the restriction to $\mathbb{S}^n=\partial \overline B_1$ of 
a vector field $\hat{\zeta}_F$ in \eq{zetaomega}. As just described above,
the CKVF $\Psi_{\star}(Y)$ is given by (\ref{CKV})
provided the two sets of constants in the respective
expressions are related by (\ref{va}).
Since $\hat{\zeta}_F$ is tangential to the sphere
$\mathbb{S}^n$, $Y$ vanishes at $p$ if and only
if $\hat{\zeta}_F |_{y=p}=0$. Evaluating \eq{zetaomega} at $p = (-1,0,\dots,0)$ yields
\begin{align*}
\hat{\zeta}_F |_p =
\left ( F^b{}_{t}
- F^b{}_{1}  - F^t{}_{1} \delta^b_1
\right ) \frac{\partial}{\partial y^b}
= \left ( F^B{}_t - F^{B}{}_1 \right ) \partial_{y^B}= -a^B\partial_{y^B},
\end{align*}
where in the last equality we have used 
the second in (\ref{va}). Thus this vector vanishes at $p$ if and only $a_B=0$, as claimed.
\qed
\end{proof}

From the considerations in this Appendix, in order to classify the classes of CKVFs on $\mathbb{E}^n$ (or $\mathbb{S}^n$), we need to classify the 2-forms on $\Mink$ up to Lor$^+$, and thus we need results from the algebraic classification of such 2-forms. This is performed in the next Appendix \ref{App:2-forms}.


\section{Algebraic classification of 2-forms in Lorentzian manifolds}
\label{App:2-forms}
Let $(M,g)$ be a causally oriented $d$-dimensional Lorentzian manifold with metric tensor $g$ (for the case of our interest, $d=n+2$). Let $F_{\mu\nu}=F_{[\mu\nu]}$ be a 2-form at any point $x\in \varietat$. In order to classify the set $\Lambda^{2}_{x}$ of 2-forms {\em up to orthochronous Lorentz transformations}, we need to describe the algebraic classification of $F_{\mu\nu}$, that is to say, the possible Segre types, eigenvalues and eigenvectors with respect to the metric $g_{\mu\nu}$. This is, of course, equivalent to the algebraic classification of the endomorphism $F^{\mu}{}_{\nu}$ acting on $T_{x}M$. Possible references are \cite{BS,M,HOW}.

First of all, from the anti-symmetry of $F_{\mu\nu}$ it is easily checked \cite{BS} that any possible (real) eigenvector $v^{\nu}$
$$
F_{\mu\nu}v^{\nu}= \lambda v_{\mu}
$$
either has a vanishing eigenvalue $\lambda =0$, or is null. In other words, all spacelike or timelike eigenvectors have necessarily a vanishing eigenvalue.

From a classical result, one also knows that any 2-form admits a decomposition of the form
\be
\bm{F} = \bm{\omega}_{1}\wedge \bm{\omega}_{2}+\bm{\omega}_{3}\wedge \bm{\omega}_{4}+\dots + \bm{\omega}_{2r-1}\wedge \bm{\omega}_{2\bar r}, \label{rank}
\ee
where $\bm{\omega}_{1},\dots ,\bm{\omega}_{2\bar r}$ are a set of linearly independent 1-forms. This decomposition is non-unique, but the number $2\bar r\leq d$ depends only on $\bm{F}$ and is called its {\em rank}. The rank is invariantly defined by
$$\underbrace{\bm{F}\wedge \dots \wedge\bm{F}}_{\bar r}\neq 0, \hspace{1cm}
\underbrace{\bm{F}\wedge \dots \wedge\bm{F}}_{\bar r+1}= 0.
$$


Therefore, any 2-form $\bm{F}$ falls into one of the following three algebraic classes:
\begin{enumerate}
\item There is a timelike eigenvector. In this case, {\em all} real eigenvalues vanish and {\em all} the one-forms in (\ref{rank}) can be chosen to belong to an orthogonal co-basis and be spacelike. The Segre type is
$$
\left\{(1,1\dots1)z\bar z\dots z\bar z \right\}
$$
where round brackets (here and below) indicate the degeneracy of the zero eigenvalue. A canonical form for this type of 2-form
is
$$
\bm{F} =\sum_{j=0}^{\bar r-1}\mu_{j}\,  \bm{\theta}^{2j+1}\wedge \bm{\theta}^{2j+2}, \hspace{1cm} \mu_{j} \neq 0,
$$
where $\{\bm{\theta}^{\alpha}\}$ ($\alpha =0,1,\dots ,d-1$) is an orthonormal (ON) co-basis in $(M,g)$. Observe that $\lambda_{j}=\pm i \mu_{j}$ are a pair of complex conjugate eigenvalues for each $j$, with eigen-directions given by $\bm{\theta}^{2j+1}\pm i \bm{\theta}^{2j+2}$. Of course, there can also be degeneracies in these complex eigenvalues.
\item There is a real non-zero eigenvalue $\lambda \neq 0$. In this case, $-\lambda$ is also an eigenvalue and the two corresponding eigenvectors are null. The Segre type is
$$
\left\{11(1\dots1)z\bar z\dots z\bar z \right\}
$$
In the ON co-basis the canonical form is
$$
\bm{F} =\lambda \bm{\theta}^{0}\wedge \bm{\theta}^{1}+\sum_{j=1}^{\bar r-1}\mu_{j}\,  \bm{\theta}^{2j}\wedge \bm{\theta}^{2j+1}, \hspace{1cm} \mu_{j} \neq 0\neq \lambda
$$
so that $\bm{k}^{\pm}=\bm{\theta}^{0}\mp \bm{\theta}^{1}$ are the pair of null eigendirections with eigenvalues $\pm \lambda$, respectively. The same comments as above apply here concerning $\mu_{j}$.
\item There is no timelike eigenvector and no non-zero eigenvalue. Equivalently, there is a triple vanishing eigenvalue with corresponding null eigenvector, say $\bm{k}^{-}$ in the notation above. The Segre type is now
$$
\left\{(31\dots1)z\bar z\dots z\bar z \right\}
$$
and the canonical form reads
$$
\bm{F} = - \frac{1}{2} \bm{k}^{-}\wedge \bm{\theta}^{2}+\sum_{j=1}^{\bar r-1}\mu_{j}\,  \bm{\theta}^{2j+1}\wedge \bm{\theta}^{2j+2}, \hspace{1cm} \mu_{j} \neq 0
$$
Again, the same comments as above apply here. The {\em null} 2-forms are included here (for $\bar r=1$, that is to say, all possible eigenvalues vanish but $\bm{F}\neq 0$). The Segre type for null 2-forms is $\{(31\dots 1)\}$ and they are invariantly characterized by the existence of a null vector $\vec k$ such that
$$
\bm{k} \wedge \bm{F}=0, \hspace{1cm} i_{\vec k} \bm{F} = 0
$$
where $i_{\vec k}$ denotes interior product with $\vec k$.
\end{enumerate}


The characteristic polynomial of $F$, whose roots are the eigenvalues of $F$, is
$$
{\cal P} (\lambda) = \det(\lambda \bm{1} - F)
$$
and, in terms of the invariants introduced in (\ref{inv0}), explicitly reads
\bean
{\cal P} (\lambda) &=& \lambda^d - I_1 \lambda^{d-2} +\frac{1}{2}\left(I_1^2 -I_2\right)\lambda^{d-4}+\dots +c_{[d/2]} \lambda^{d-2[d/2]}\\
&=& \lambda^{d} +\sum_{b=1}^{[d/2]}c_{b}\lambda^{d-2b}
\eean
where only odd (even) powers of $\lambda$ appear for $d$ odd (even), and the different coefficients $c_b$ ($b=1,\dots , [d/2]$) are known polynomials of the invariants $I_a$: each $c_{b}$ is linear in $I_{b}$ and only involves the $I_a$ with $a\leq b$ at the appropriate power.  The explicit form of these coefficients $c_b$ is  complicated for large $d$, but a convenient way to have them all explicitly in a nutshell is to recall the Cayley-Hamilton theorem, that is to say,
$$
{\cal P} (F) =0
$$
which can always be expressed in the following succinct form
$$
F^{\mu_{1}}{}_{[\mu_{1}}\dots F^{\mu_{d}}{}_{\mu_{d}}\delta^{\tau}_{\rho]}=0
$$
providing explicit expressions for $c_b$. There is also a recursive method, known as the Le Verrier-Faddeev algorithm, to compute all the $c_b$ \cite{Gant}.

One sees that the invariants $I_a$ determine fully the characteristic polynomial and thereby the eigenvalues (with all degeneracies) of the 2-form $\bm{F}$.
However, they do not determine its algebraic type. To prove this, we now provide an explicit counterexample.
Consider the following two 2-forms:
\be
\bm{F}_{1}=\mu \bm{\theta}^{3}\wedge \bm{\theta}^{4}, \hspace{1cm} \bm{F}_{2}=\bm{k^{+}}\wedge \bm{\theta}^{2}+\mu \bm{\theta}^{3}\wedge \bm{\theta}^{4}
\label{example}
\ee
where $\bm{k}^{+}$ is null and orthogonal to $\bm{\theta^{2},\theta^{3},\theta^{4}}$ as above. They clearly have the same invariants ($I_{a}=(-1)^{a} \mu^{2a}$), but
they have different ranks.
The first has rank 2, and the second has rank 4. In general, if a 2-form $\bm{F}$ is {\em spacelike} and has rank $2\bar r<d-2$, then one can always add a null 2-form of the above type while keeping the same set of invariants.  However, the rank is a property invariant
under Lorentz transformations.
Indeed, recall that the rank $2\bar r$ of any 2-form $\bm{F}$ is precisely the rank of the matrix $(F_{\mu\nu})$. But given that
$$
F'_{\alpha\beta} =L_\alpha{}^\mu L_\beta{}^\nu F_{\mu\nu} \, \Longrightarrow \det \bm{F}' = (\det L)^2 \det \bm{F} = \det \bm{F}
$$
and similarly for the minors (or directly from the defining formula (\ref{rank})),
one deduces that
the rank remains unchanged under Lorentz transformations.
It follows then that the 2-forms in
 (\ref{example}) cannot be related by {\em any} Lorentz transformation at all, and they must belong
to different algebraic types, despite the fact that they have the same set of invariants.
Thus, the algebraic type of a 2-form
depends not only on the invariants, but also on the number $\bar r$ which
determines its rank.

We want to show that the knowledge of $\bar r$ and of the invariants
is sufficient to determine the algebraic type of $F$. To that aim,
we
start by providing an explicit formula for the invariants $I_a$ in terms of the canonical forms presented above. To do that, let $\{\vec e_\alpha\}$ be the basis dual to $\{\bm{\theta}^\alpha\}$ and observe that, for each pair $\{\bm\theta^{j},\bm\theta^{j+1}\}$ and letting $\bm G_j=\bm\theta^{j}\wedge \bm\theta^{j+1}$ one has $G_{j}^2=-\Pi_j$, where $\Pi_j$ is the projector to the 2-plane spanned by $\left<\vec e_{j},\vec e_{j+1}\right>$. Then, it is easy to derive:
\be
I_{a} =(-1)^{a}\sum_{j=\epsilon }^{\bar r-1}\mu^{2a}_{j}+ \epsilon'\lambda^{2a}, \hspace{1cm} \forall a=1,\dots , [d/2] \label{inv}
\ee
where $\epsilon =0$ (respectively $\epsilon =1$) in case 1 (resp.\ cases 2 and 3) and $\epsilon'=0$ (resp.\ $\epsilon'=1$) in cases 1 and 3, (resp.\ case 2). Notice the signs, as $(-1)^{a}I_{a}>0$ in cases 1 and 3, and $I_a>0$ for even $a$ in all cases.

Given a fixed set of invariants $I_{a}$, formulas (\ref{inv}) allow us to find the $\mu_{j}^{2}$ (and $\lambda^{2}$ in case it exists) and to determine the $\epsilon$'s \underline{{\em as long as we know the rank $2\bar r$}}.



\subsection{How to characterize 2-forms up to Lor$^{+}$}\label{subsec:2-formClassification}

As proven after the formula (\ref{inv0}) defining the invariants, we know that all 2-forms $\bm{F}'$ which are Lorentz transformed of $\bm{F}$ have exactly the same set of invariants $\{I_a\}$. We have also shown that
the converse does not hold, and the set of invariants $\{I_a (F)\}$ is not
enough to characterize the required classes of 2-forms. In order to classify 
non-trivial 2-forms up to Lor$^+$ one must provide the following set of numbers: 
\be
\{\bar{r},I_a\} , \, \, \,  a, \bar r \in\{ 1,
\dots [d/2]\},\hspace{4mm} I_{a}\in \mathbb{R}, 
\label{numbers}
\ee
where $I_a$ is the set of invariants (\ref{inv0}) and $2\bar r$ is the rank. Equivalently, we can provide the set
\be
\{\bar r,c_b\} , \, \, \,  b, \bar r \in \{1, \dots [d/2]\},\hspace{4mm} c_{b}\in \mathbb{R}, 
\label{numbers2}
\ee
where $c_{b}$ are the coefficients of the characteristic polynomials:
$$
c_1 =-I_1, \hspace{1cm} c_2 =\frac{1}{2} (I_1^2-I_2),\hspace{1cm} c_3= -\frac{1}{6} (2I_3-3I_1 I_2 +I_1^3), \hspace{1cm} \dots
$$
This relies on the fact that one can then obtain the canonical expression of the 2-form by solving the system of equations (\ref{inv}). To perform this, we need to know whether or not $F$ has a non-zero real eigenvalue, which is equivalent to saying that $\bm{F}$ `contains a timelike plane'. But this is provided by the sign of $c_{\bar r}$, according to
$$
c_{\bar r}=\left\{
\begin{array}{ccc}
>0 & \mbox{Case 1} & \epsilon =\epsilon'=0 \\
<0 & \mbox{Case 2} & \epsilon =\epsilon'=1 \\
=0 & \mbox{Case 3} & \epsilon =1, \epsilon'=0
\end{array}
 \right.
$$

Hence, once we are given the numbers (\ref{numbers2}), which are Lorentz invariant, the classes of 2-forms inequivalent under Lor$^+$ are fully determined by those with a different set of such numbers. This follows because, if this set is the same for two different 2-forms, we can always write them in their canonical form by solving (\ref{inv}), which is necessarily one of the cases 1, 2 or 3 listed at the beginning, with a fixed $\bar r$ and the same degeneracies in the complex eigenvalues --- and up to the signs of the $\mu_{j}$ and $\lambda$ involved. But one can always connect any such two 2-forms, that belong to a fixed algebraic class, by means of the {\em orthochronous} Lorentz transformation which sends the orthonormal co-basis used for any one of the 2-forms to the orthonormal co-basis of the other one ---composed with the necessary space inversions 
to fit the signs.

We conclude that the equivalence classes of conformal Killing vectors
of $\mathbb{S}^n$ is uniquely characterized by the M\"obius-invariant
constants $\{ c_b \}$ {\em together} with the number $\bar r$.

\subsection{Relation between different Killings $\zeta_F$}\label{subsec:2nd}
Let $\zeta_F$ and $\zeta_G$ be any two Killing vectors of the form (\ref{kil}) for given two-forms $\bm{F}$ and $\bm{G}$. One can easily compute its commutator
\begin{equation}
\label{commutator}
\left[\zeta_F,\zeta_G\right]= \zeta_{[G,F]}
\end{equation}
where $[G,F]$ denotes the commutator of endomorphisms, that is to say
$$
[G,F]^\mu{}_\nu =G^\mu{}_\rho F^\rho{}_\nu - F^\mu{}_\rho G^\rho{}_\nu
$$
which represents indeed a new 2-form (indices down). In particular, one can compute different odd powers $F^{2j+1}$ of $F$ for $j=1,\dots ,\bar r-1$, all of them being 2-forms (when indices are down), and the corresponding Killing vectors are all mutually commuting
$$
\zeta^{(j)} := \zeta_{F^{2j+1}}, \hspace{1cm} \left[\zeta^{(j)},\zeta^{(j')}\right] =0   .
$$
This is used in section \ref{section_dimension}.

\section{CKVFs in conformal Euclidean 3-space}
\label{app_conf_Eucl}

In this Appendix we provide the classification of conformal Killing vector fields (CKVFs) in conformal Euclidean 3-space up to M\"obius transformations.
This corrresponds to the case $d=5$, $n=3$, of the previous Appendices: $\mathbb{S}^3$ or $\mathbb{E}^3$, $\mathbb{H}^4$ and $\mathbb{M}^{1,4}$.
From the discussion above, in this situation $\bar r\in\{1, 2\}$ and, to achieve a complete classification of the CKVF on $\mathbb{E}^3$, one only has to provide three numbers $\{\bar r,\widehat c,\widehat k\}$ according to (\ref{numbers2}), that is to say:
\begin{itemize}
\item $2\bar r$ is the rank of the underlying 2-form $\bm{F}\neq 0$, and $\bar r\in\{1,2\}$
\item $\widehat c \defi I_1 =-c_1 =\mbox{tr}(F^2)/2$ is the first invariant
\item and $\widehat k \defi - c_2 =(I_2-I_1^2)/2$ is the remaining coefficient of the characteristic polynomial of $F$, given now by
\be
{\cal P}(\lambda)=\lambda(\lambda^4+c_1 \lambda^2+c_2)=\lambda(\lambda^4-\widehat c \lambda^2- \widehat k).\label{carpol}
\ee
\end{itemize}

Using classical vector notation, the CKVF found in (\ref{CKV}) can be written as
\begin{equation}
Y = \vec b+ \vec w \times \vec x  +v \,\vec x+ (\vec a\cdot \vec x)\vec  x - \frac{1}{2} |\vec x|^2 \vec a
\label{app_general_CKVF}
\end{equation}
where we have defined a third vector $\vec w$ by means of
$$
\vec w \defi \star \bm{\omega}, \hspace{1cm} w^C \defi \frac{1}{2} \epsilon^{ABC} \omega_{AB}
$$
and $\epsilon^{ABC}$ is the Levi-Civita skew-symmetric symbol.
Then, by using (\ref{F=vabw}) it is straightforward to find, for each $Y$ in (\ref{app_general_CKVF}), explicit expressions for the constants $\widehat c, \widehat k$:
\begin{eqnarray}
\widehat k(Y) &=& v^2|\vec w|^2 + 2\,v\,\vec a \cdot (\vec w \times \vec b ) + |\vec a \times\vec b|^2
-2 (\vec a\cdot \vec w)(\vec b \cdot \vec w)
\;,
\label{app_general_k}
\\
\widehat c(Y) &=&v^2 -  |\vec w|^2-2\,\vec a \cdot \vec b
\;.
\label{app_general_c}
\end{eqnarray}
From the construction in the previous Appendices we know that
\begin{lemma}
The constants  $\widehat k$ and $\widehat c$, as defined in \eq{app_general_k} and \eq{app_general_c}, are invariant under M\"obius transformations.
\end{lemma}

We have shown in remark~\ref{gauge_freedom_kc} that the scalars $\widehat k$ and $\widehat c$ come along with some
gauge freedom:
Rescaling $Y$ by some constant factor $\alpha\in\mathbb{R}\setminus\{0\}$ gives another CKVF.
Under such a rescaling we have
\begin{equation}
\widehat k \mapsto \alpha^4 \widehat k\;, \quad \widehat c \mapsto \alpha^2 \widehat c
\;.
\label{scaling_behavior}
\end{equation}
We thus may always achieve
 $\widehat k\in \{-1,0,+1\}$; if $\widehat k=0$ we may further  assume $\widehat c \in \{-1,0,+1\}$,  which we will do henceforth.
(Indeed, in spacetimes with vanishing MST the CKVF $Y$ enters the asymptotic Cauchy problem only via (\ref{asympt_Cauchy_data_1}),
(\ref{asympt_Cauchy_data_2}),
the factor $\alpha$ can thus be absorbed in the parameters $\Dconst$ and $\Cconst$.)

Concerning $\bar r$, and given that the rank can only be 2 or 4 in our situation, we just need to distinguish the case with $\bar r=1$. But this is invariantly characterized by $\bm F \wedge \bm F =0$, and thus using (\ref{F=vabw}) by
$$
\frac{1}{2} \bm F \wedge \bm F = -dt \wedge dx^1 \wedge \left(v \bm{\omega} +b\wedge a \right) - dt\wedge (a-b/2) \wedge \bm{\omega} + dx^1 \wedge (a+b/2) \wedge \bm{\omega} =0
$$
which immediately leads, in the used vector notation, to
\be
v \vec w +\vec b\times \vec a = \vec 0 , \quad \quad \vec w \cdot \vec a = \vec w \cdot \vec b =0 \label{r=1} .
\ee
Again, this set of conditions is M\"obius invariant. Hence, one only needs to check if (\ref{r=1}) hold: if yes, then $\bar r=1$, otherwise $\bar r=2$. It is easily checked from (\ref{app_general_k}) that (\ref{r=1}) implies $\widehat k (Y) =0$. We are going to see presently that the value of $\bar r$ turns out to be relevant only in the case $\widehat k =0$ with $\widehat c <0$.

Arrived at this point, the classification of the CKVFs is as follows, depending on the different algebraic classes for $\bm{F}$. A canonical form for the CKVF \eq{CKV}
(or \eq{app_general_CKVF}) is provided for each case using the corresponding canonical algebraic form of $\bm{F}$ provided in Appendix \ref{App:2-forms} (and with $\{\bm{\theta}^2,\bm{\theta}^3,\bm{\theta}^4\}=\{dx,dy,dz\}$):
\begin{enumerate}
\item Case 1: The canonical form for the underlying 2-form is $\bm{F} = \mu_0 dx^1\wedge dx +\mu_1 dy\wedge dz$ so that, in this case, using (\ref{inv}) we get $\widehat k=-\mu_0^2 \mu_1^2 \leq 0$ and $\widehat c =-\mu_0^2 -\mu_1^2$. Thus, one has in general $\widehat c<0$,  and $\widehat k<0$ if $\bar r=2$ while $\widehat k=0$ if $\bar r=1$.
\begin{enumerate}
\item If $\bar r=2$ we can set, without loss of generality, $\widehat k=-1$. This means $\mu_1^2=1/\mu_0^2\neq 0$ and thus $\widehat c=-(\mu_0^2+1/\mu_0^2)$. Observe that, by re-ordering the basis, one can assume without loss of generality that $\mu_0\in(0,1]$  and
$\mu_1 = 1/\mu_0$ .
The last expression for $\widehat c$ has a global maximum at $\mu_0^2=1$, and therefore $\widehat c\leq -2$. The particular extreme case with $\widehat c=-2$ is characterized by the degeneracy of the complex eigenvalues $\pm i$.

The decomposition of the canonical form of $\bm{F}$ is
$$
v=0,\hspace{5mm}  \vec a =-\frac{\mu_0}{2} (1,0,0), \hspace{5mm} \vec{b}=-\mu_0 (1,0,0),\hspace{5mm} \vec w=-\frac{1}{\mu_0} (1,0,0)
$$
and thus (\ref{app_general_CKVF}) becomes
\begin{equation}
Y =\left(
\begin{array}{c}
-\mu_0+\frac{\mu_0}{4}(y^2+z^2-x^2)\\
-\frac{\mu_0}{2}xy+\frac{1}{\mu_0} z\\
-\frac{\mu_0}{2}xz -\frac{1}{\mu_0} y
\end{array}
 \right).
\label{CKVF_stand}
\end{equation}
In the generic case with $\mu_0\in (0,1)$, there exists a  2-parameter family  of CKVFs that commutes with $Y$: $X=\lambda_1 (0,-z,y)  +\lambda_2 Y$ with $\lambda_i\in\mathbb{R}$.

The exceptional case with $\widehat c=-2$ is achieved by setting $\mu_0=1$ in (\ref{CKVF_stand}), and then there exists a 4-parameter family  of CKVFs that commutes with $Y$.
This family is characterized by the conditions
$\{ v=0, \vec{b} = 2 \vec{a}, \vec{w}= -2 \vec{a} + \lambda_1 (1,0,0)\}$,
$\vec{a} = \mathbb{R}^3$, $\lambda_1 \in \mathbb{R}$,
in the general CKVF (\ref{app_general_CKVF}), or after
a suitable redefinition of parameters,
%
$$
X = \lambda_1\begin{pmatrix} 0 \\ -z\\y\end{pmatrix}
 + \lambda_2 \begin{pmatrix} 2(xy-2z)\\ 4 -x^2 + y^2-z^2\\ 2(yz+2x)\end{pmatrix}
+ \lambda_3 \begin{pmatrix}2(xz + 2y)\\ 2(yz-2x)\\ 4 -x^2-y^2 + z^2\end{pmatrix}
+ \lambda_4 Y
\;.
$$

\item If $\bar r=1$, so that $\widehat k=0$, one can set $\widehat c=-1$. Without loss of generality we thus choose $\mu_0=0$ and $\mu_1=1$, and one has
$$
v=0,\hspace{5mm}  \vec a =0, \hspace{5mm} \vec b=0,\hspace{5mm} \vec w= (-1,0,0)
$$
so that (\ref{app_general_CKVF}) reads
$$
Y =\left(
\begin{array}{c}
0\\
z\\
- y
\end{array}
 \right).
$$
\end{enumerate}
The most general CKVF commuting with this $Y$ depends on four
parameters and is given by the family
$\{ v= \lambda_1, \vec{a} = \lambda_2 (1,0,0),
\vec{b}= \lambda_3 (1,0,0), \vec{w}= \lambda_4 (1,0,0)\}$,
$\lambda_i\in \mathbb{R}^4$. Explictly
\begin{align}
X = \lambda_1\begin{pmatrix} x \\ y\\ z \end{pmatrix}
 + \lambda_2 \begin{pmatrix} \frac{1}{2} (x^2 - y^2 -z^2) \\
xy \\ xz\end{pmatrix}
+ \lambda_3 \begin{pmatrix} 1 \\ 0\\ 0\end{pmatrix}
+ \lambda_4 \begin{pmatrix} 0 \\ -z\\ y\end{pmatrix}
\;. \label{X2}
\end{align}

\item Case 2:  Now the canonical form is $\bm{F} =\lambda dt \wedge dx^1 +\mu_1 dx \wedge dy$, so that (\ref{inv}) implies $\widehat c =\lambda^2-\mu_1^2$ and $\widehat k =\lambda^2\mu_1^2 \geq 0$ with strict inequality if $\bar r=2$ ($\mu_1\neq 0$) and equality if $\bar r=1$ ($\mu_1=0$).
\begin{enumerate}
\item When $\bar r=2$ we set, without loss of generality, $\widehat k=1$. Using (\ref{inv}) this implies $\mu_1^2=1/\lambda^2$ and
\be
\widehat c=\lambda^2-\frac{1}{\lambda^2} \hspace{1cm} (\lambda > 0)
\label{form_check_c>}
\ee
so that $\widehat c$ can have any sign depending on whether $\lambda^2$ is greater, equal or smaller than 1.
 We can assume without loss of generality that
$\mu_1 = 1/\lambda$.

The decomposition of the canonical form of $\bm{F}$ is
$$
v=\lambda ,\hspace{5mm}  \vec a =0, \hspace{5mm} \vec{b}=0,\hspace{5mm}
\vec w =-\frac{1}{\lambda} (0,0,1)
$$
and thus (\ref{app_general_CKVF}) becomes
\be
Y =\left(
\begin{array}{c}
\lambda x +\frac{1}{\lambda} y\\
\lambda y -\frac{1}{\lambda} x\\
\lambda z
\end{array}
 \right). \label{stand_form_neg_k}
\ee
There exists a 2-parameter family of CKVFs that commutes with $Y$: $X=\lambda_1(x,y,z) + \lambda_2 (-y,x,0) $, where $\lambda_i\in\mathbb{R}$.
This family is generated by $\{ v = \lambda_1, \vec{a}=0, \vec{b}=0,
\vec{w} = \lambda_2 (0,0,1) \}$ in (\ref{app_general_CKVF}).

\item When $\bar r=1$ we must set $\mu_1=0$ so that $\widehat k=0$. Without loss of generality we can set $\lambda =1$ ($\widehat c=1$) and then
$$
v=1 ,\hspace{5mm}  \vec a =0, \hspace{5mm} \vec{b}=0,\hspace{5mm} \vec w =0
$$
so that
\be
Y =\left(
\begin{array}{c}
x\\
y\\
z
\end{array}
 \right). \label{stand_form_zero_k}
\ee
There exists a 4-parameter family of CKVFs  that commutes with $Y$: $X= \vec w \times \vec x + v\vec x$, where $\vec w\in\mathbb{R}^3$, $v\in\mathbb{R}$.
\end{enumerate}
\item Case 3: Now the canonical form is $\bm{F} =- \frac{1}{2}
(dt +dx^1)\wedge dx +\mu_1 dy\wedge dz$ so that (\ref{inv}) gives $\widehat k=0$ and $\widehat c =-\mu_1^2 \leq 0$, the strict inequality holds if $\bar r=2$ ($\mu_1\neq 0$), while $\widehat c=0$ if $\bar r=1$ ($\mu_1=0$).
\begin{enumerate}
\item When $\bar r=2$ one can achieve $\widehat c=-1$ by setting $\mu_1 =1$. The decomposition of the canonical form of $\bm{F}$ is
$$
v=0 ,\hspace{5mm}  \vec a =0, \hspace{5mm} \vec{b}=(1,0,0) ,\hspace{5mm} \vec w = (-1,0,0)
$$
and then
$$
Y=\left(
\begin{array}{c}
1\\
z\\
- y
\end{array}
 \right).
$$
The CKVFs that commute with this $Y$ are given by thw two-dimensional family
generated by $\{ v=0, \vec{a}=0, \vec{b} = \lambda_1 (1,0,0), \vec{w}=
\lambda_2 (1,0,0) \}$ in (\ref{app_general_CKVF}), i.e.
$X = \lambda_1 (1,0,0) + \lambda_2 (0,-z,y)$.

\item If $\bar r=1$, so that $\mu_1=0$ and $\widehat k=\widehat c=0$,
$$
v=0 ,\hspace{5mm}  \vec a =0, \hspace{5mm} \vec{b}=(1,0,0) ,\hspace{5mm} \vec w = 0
$$
and then
$$
Y =\left(
\begin{array}{c}
1\\
0\\
0
\end{array}
 \right).
$$
There exists a 4-parameter family of CKVFs that commutes with $Y$: $X=\vec b +  \lambda_1 (0,-z,y)$, where $\vec b\in\mathbb{R}^3$, $\lambda_1\in\mathbb{R}$.
This family is generated by $\{ v=0, \vec{a}=0, \vec{b}, \vec{w} = \lambda_1
(1,0,0)\}$ in (\ref{app_general_CKVF}).
\end{enumerate}
\end{enumerate}

From the explicit solution we see that $\widehat k=0$ and $\widehat c=-1$ is the only case in our setting where $Y$ is \emph{not} uniquely determined up to M\"obius transformations
by the two invariants $\widehat k$ and $\widehat c$, and one needs to know $\bar r$. Note that (\ref{r=1}) is not satisfied in case 3(a). As we have seen,
$Y$ can be brought to one of the following two alternative forms, corresponding to $\bar r=2$ and $\bar r=1$, respectively
%
\begin{equation}
Y^{(1)} = \begin{pmatrix} 1  \\  z \\ -y
\end{pmatrix}
\;,
\quad
Y^{(2)} = \begin{pmatrix} 0  \\  z \\ -y
\end{pmatrix}
\; .\label{specialcase}
\end{equation}
%
The two cases can be distinguised not only by the rank of $F$, but also from
the   fact that the family of CKVF that commutes with
$Y^{(1)}$ depends on two-parameters: $X^{(1)}=\lambda_1 (1,0,0)  +\lambda_2
(0,-z,y)$, $\lambda_i\in\mathbb{R}$,
while there exists a 4-parameter family of CKVFs that commutes with $Y^{(2)}$
given by (\ref{X2}).

Taking the behavior \eq{scaling_behavior} of $\widehat k$ and $\widehat c$ under rescaling of the CKVF into account we have shown:
%

\begin{proposition}
\label{moebius_lemma}
On Euclidean space $\mathbb{E}^{3}$:
\begin{enumerate}
\item There is no CKVF whose invariants (\ref{app_general_k}-\ref{app_general_c}) satisfy $\widehat k <0 $ and $\widehat c > -2\sqrt{|\widehat k|}$.
\item If the invariant (\ref{app_general_k}) vanishes $\widehat k =0$ and the invariant (\ref{app_general_c}) is negative $\widehat c <0$, there exist precisely two CKVFs $Y^{(1)}$ and $Y^{(2)}$ ---cf. (\ref{specialcase})--- up to conformal symmetries (i.e.\ M\"obius transformations). They are distinguished according to whether or not the M\"obius-invariant condition (\ref{r=1}) holds.
\item In all other possibilities for the invariants (\ref{app_general_k}-\ref{app_general_c}),  there exists a unique CKVF up to M\"obius transformations.
\end{enumerate}
\end{proposition}


\section{Hypersurface orthogonal KVFs}

\label{sec_hyp_orth}

%

In this Appendix we will derive conditions which ensure that hypersurface-orthogonality  of a KVF is preserved under evolution from $\scri^-$.
For this purpose we consider the Ernst 1-form
\begin{equation}
\sigma_{\mu} \,:=\, 2X^{\alpha}\mathcal{F}_{\alpha\mu} \;.
\end{equation}
It is well-known that this covector field has an ``Ernst''-potential $\sigma$ (at least locally), which is defined up to some additive complex constant $a$, the so-called ``$\sigma$-constant''.
We decompose $\sigma$ into its real and imaginary part,
\begin{equation}
\sigma \,=\, \lambda + i\omega\;,
\end{equation}
where $\omega$ is called the \emph{twist potential}.
A KVF $X$ is hypersurface-orthogonal provided that
\begin{equation}
\epsilon_{\alpha\beta\mu\nu}X^{\beta}\nabla^{\mu}X^{\nu} \,=\,0 \quad \Longleftrightarrow \quad  \mathrm{Im}(\sigma_{\mu})\,=\,0
 \quad \Longleftrightarrow \quad  \omega \,=\,\mathrm{const.}
\end{equation}
In fact, the ``hypersurface-orthogonal''-property of a KVF is preserved under conformal rescalings of the metric.

We want to derive a homogeneous wave equation for the twist potential $\omega$.
In the physical spacetime  the Ernst 1-form satisfies the following set of  relations (cf.\ \cite{mars-senovilla})
\begin{eqnarray}
 \nabla_{\alpha}\sigma^{\alpha}&=& -\mathcal{F}^2 + 2R_{\alpha\beta}X^{\alpha}X^{\beta}
\,=\,  -\mathcal{F}^2 + 2\Lambda |X|^2
\;,
\\
\sigma_{\alpha}\sigma^{\alpha} &=&  |X|^2\mathcal{F}^2
\;.
\end{eqnarray}
That yields a wave equation for the Ernst potential,
\begin{equation}
\Box_g \sigma
\,=\, -|X|^{-2} \nabla_{\mu}\sigma\nabla^{\mu}\sigma + 2\Lambda |X|^2
\;.
\label{wave_eqn_sigma}
\end{equation}
supposing that
\begin{equation}
|X|^2 \,>\, 0
\;.
\end{equation}
%
In $\Lambda>0$-vacuum spacetimes which admit a smooth $\scri$ and which are not de Sitter
this will be automatically the case near $\scri$ for KVFs whose associated MST vanishes \cite{mpss}, in particular
\begin{equation}
|\widetilde X|^2 |_{\scri} \,=\, |Y|^2 \,>\, 0
\;.
\end{equation}
%
Taking the imaginary part of \eq{wave_eqn_sigma} yields
\begin{equation}
\Box_g \omega
\,=\, -2 |X|^{-2}\nabla_{\alpha}\lambda\nabla^{\alpha}\omega
\;.
\label{wave_eqn_twist}
\end{equation}
We want to express this equation in terms of the unphysical metric.
To do that we need the asymptotic expansion of the Ernst potential, which has been computed in \cite{mpss} (for our purposes here it
suffices to present  the expansion up to and including the next-to-next-to-leading order):
\begin{proposition}
Consider a $\Lambda>0$-vacuum spacetime which admits a KVF and a smooth $\scri$.
Then the Ernst potential $\sigma$ admits the following asymptotic expansion
\begin{eqnarray}
 \sigma
&=&  -|Y|^2 \Theta^{-2}  + i \sqrt{\frac{3}{\Lambda}} Y_iN^i \Theta^{-1}
 +\frac{1}{3\Lambda} f^2  - \frac{3}{\Lambda}Y^{i}\ol{\widetilde\nabla_t\widetilde \nabla_t  \widetilde X_{i}}
\nonumber
\\
&&
+ ip(x^j) -a + O(\Theta)
\;,
\label{expansion_sigma}
\\
\partial_t\sigma
&=&     2\sqrt{\frac{\Lambda}{3}} |Y|^2\Theta^{-3}   - i  Y_i N^i \Theta^{-2}
-  \frac{1}{2}\sqrt{\frac{3}{\Lambda}} \widehat R |Y|^2 \Theta^{-1}
+O(1)
\label{expansion_sigma_t}
\;,
\end{eqnarray}
where $a$ is the complex $\sigma$-constant and $p(x^j)$ is an (at least locally defined) potential of the closed 1-form
\begin{equation}
P_i \,:=\, -6 \sqrt{\frac{\Lambda}{3}} \widehat \epsilon_{ij }{}^{k}  D_{kl} Y^{j}Y^l
\;.
\end{equation}
\end{proposition}

We have
\begin{equation}
\Box_g \omega \,=\,
g^{\alpha\beta}(\partial_{\alpha}\partial_{\beta} \omega - \Gamma^{\mu}_{\alpha\beta}\partial_{\mu}\omega)
\,=\,
\Theta^2 \Box_{\widetilde g} \omega
-2\Theta \widetilde\nabla_{\alpha}\Theta \widetilde\nabla^{\alpha}\omega
\;,
\end{equation}
and thus
\begin{equation}
\Theta  \Box_{\widetilde g} \omega
-2 \widetilde\nabla_{\alpha}\Theta \widetilde\nabla^{\alpha}\omega
\,=\,
 -2\Theta^3 |\widetilde X|^{-2}\widetilde \nabla_{\alpha}\lambda\widetilde\nabla^{\alpha}\omega
\;.
\end{equation}
Equivalently (off $\scri$ where $\Theta>0$),
\begin{eqnarray}
  \Box_{\widetilde g}\widetilde \omega
&=&
-4\Theta^{-2}\widetilde\nabla_{\alpha}\Theta\widetilde\nabla^{\alpha}\Theta ( 1 +\widetilde\lambda |\widetilde X|^{-2})
\widetilde\omega
+ 4\Theta^{-1} ( 1 +\widetilde\lambda |\widetilde X|^{-2})\widetilde\nabla_{\alpha}\Theta\widetilde\nabla^{\alpha}\widetilde\omega
\nonumber
\\
&&+
\Theta^{-1}\Big(\Box_{\widetilde g}\Theta
+ 2|\widetilde X|^{-2} \widetilde \nabla_{\alpha}\widetilde\lambda\widetilde\nabla^{\alpha}\Theta \Big)\widetilde \omega
-  2 |\widetilde X|^{-2}\widetilde \nabla^{\alpha}\widetilde \lambda \widetilde\nabla_{\alpha} \widetilde \omega
\;,
\label{wave_tilde_omega}
\end{eqnarray}
where we have set $\widetilde \omega:=\Theta\omega$ and $\widetilde\lambda := \Theta^2\lambda$.
 It  follows from \eq{expansion_sigma} that these functions are regular  near $\scri$.
We  need to make sure that the equation \eq{wave_tilde_omega} is regular near $\scri$, i.e.\ that the coefficients  in parentheses have an appropriate fall-off behavior at infinity.
For this, we employ the relations at $\scri$ derived in \cite[Section~2]{mpss}
as well as  the asymptotic expansions \eq{expansion_sigma}-\eq{expansion_sigma_t} of the Ernst potential, to conclude that
\begin{eqnarray}
1 +\widetilde\lambda |\widetilde X|^{-2} &=&  O(\Theta^2)
\;,
\\
\Box_{\widetilde g}\Theta
+ 2|\widetilde X|^{-2}\widetilde \nabla_{\alpha}\widetilde\lambda\widetilde\nabla^{\alpha}\Theta
&=& -  4\frac{\Lambda}{3}|Y|^{-2} \lambda \Theta - 2\sqrt{\frac{\Lambda}{3}}|Y|^{-2}\partial_t\lambda \Theta^2 + O(\Theta)
\nonumber
\\
&=&
O(\Theta)
\;.
\end{eqnarray}
%
It follows that the rescaled twist potential $\widetilde \omega$   satisfies a regular, homogeneous wave equation,
\begin{equation}
  \Box_{\widetilde g}\widetilde \omega
\,=\,
O(1) \widetilde \omega
+ O(1) \widetilde\nabla  \widetilde \omega
\;.
\label{wave_tilde_omega2}
\end{equation}
Recall that the Ernst potential is defined only up to the $\sigma$-constant. This is reflected in the wave equation \eq{wave_eqn_twist} for the twist potential:
If $\omega$ is a solution, then $\omega +c$, $c\in \mathbb{R}$, is a solution, as well.
Consequently,   \eq{wave_tilde_omega} has the property that if $\widetilde \omega$ is a solution, then $\widetilde\omega + c\Theta$
is also a solution, which, though, defines the same twist 1-form (indeed, $\widetilde\omega=\Theta$ solves \eq{wave_tilde_omega}).

It follows from \eq{expansion_sigma}-\eq{expansion_sigma_t}
that
\begin{equation}
 \widetilde \omega|_{\scri^-} \,=\, \sqrt{\frac{3}{\Lambda}} Y_iN^i
\;, \quad
\widetilde\nabla_t\widetilde \omega|_{\scri^-}
\,=\, \sqrt{\frac{\Lambda}{3}}\Big(p(x^i) - \mathrm{Im}(a)\Big)
\;.
\end{equation}
The  rescaled twist potential $\widetilde\omega$ will be proportional to $\Theta$ (equivalently, the twist potential $\omega$ will be constant) in the DoD of $\scri^-$ if and only if  $ \widetilde \omega|_{\scri^-} =0$ and $\widetilde\nabla_t\widetilde \omega|_{\scri^-}=\mathrm{const.}$, i.e.\ if and only if
%
\begin{equation}
 Y_iN^i \,=\, 0 \quad \text{and} \quad  \widehat \epsilon_{ij }{}^{k}  D_{kl} Y^{j}Y^l\,=\,0
\;.
\end{equation}
Note that the former condition states that $\widetilde X$ needs to be hypersurface orthogonal on $\scri^-$.
This result shows that hypersurface orthogonality on the initial manifold together with the KID equations are not sufficient to make sure that this
property is preserved under evolution.

\begin{proposition}
\label{prop_hyp_orth}
Let $(\mcM,g)$ be a $\Lambda>0$-vacuum spacetime which admits a smooth $\scri^-$ and a KVF $X$.
Then, the KVF $X$ is  hypersurface orthogonal in the DoD of $\scri^-$ if and only if
\begin{enumerate}
\item[(i)] $\widetilde X=\phi_*X$
 has this property on $\scri^-$, equivalently, if $ Y\cdot \mathrm{curl} \,Y=0$, and
\item[(ii)] $ \widehat \epsilon_{ij }{}^{k}  D_{kl} Y^{j}Y^l=0$.
\end{enumerate}
\end{proposition}

\begin{remark}
{\rm
By definition, condition (ii) is satisfied in the asymptotically Kerr-de Sitter-like spacetimes of \cite{mpss}.
}
\end{remark}

%


\end{document}

%% file: LeeNewcommandv3.tex
\def\nablah{\widehat \nabla}
\def\Cot{\widehat C}

\newcounter{mnotecount}[section]

\renewcommand{\themnotecount}{\thesection.\arabic{mnotecount}}
\newcommand{\mnote}[1]
{\protect{\stepcounter{mnotecount}}$^{\mbox{\footnotesize
$
\bullet$\themnotecount}}$ \marginpar{
\raggedright\tiny\em
$\!\!\!\!\!\!\,\bullet$\themnotecount: #1} }

\newtheorem{theorem}{\sc  Theorem\rm}[section]

\newtheorem{conjecture}[theorem]{\sc  Conjecture\rm}

\newtheorem{definition}[theorem]{\sc  Definition\rm}

\newtheorem{lemma}[theorem]{\sc Lemma\rm}
\newtheorem{Lemma}[theorem]{\sc Lemma\rm}

\newtheorem{proposition}[theorem]{\sc Proposition\rm}
\newtheorem{Proposition}[theorem]{\sc Proposition\rm}
\newtheorem{remark}[theorem]{\sc Remark\rm}

\newcommand{\ol}[1]{\overline{#1}{}}

\newcommand{\jlcax}[1]{}
\newcommand{\eean}{\nonumber\end{eqnarray}}

\newcommand{\F}{{\bf F}}

\newcommand{\kk}[1]{}

\newcommand{\beq}{\begin{equation}}

\newcommand{\ra}{\rightarrow}

\newcommand{\FS}       
                  {F}

\newcommand{\HS} 
       {H_{\mbox{\scriptsize volume}}}

{\ptc{this should be removed in the oberwolfach version}}%

\newcommand{\I}{\sqrt{-1}}%
\newcommand{\eeal}[1]{\label{#1}\end{eqnarray}}
\newcommand{\C}{{\mathbb C}}
\newcommand{\bed}{\begin{deqarr}}
\newcommand{\eed}{\end{deqarr}}
\newcommand{\bedl}[1]{\begin{deqarr}\label{#1}}
\newcommand{\eedl}[2]{\arrlabel{#1}\label{#2}\end{deqarr}}

\newcommand{\bel}[1]{\begin{equation}\label{#1}}
\newcommand{\bea}{\begin{eqnarray}}
\newcommand{\bean}{\begin{eqnarray}\nonumber}
\newcommand{\beal}[1]{\begin{eqnarray}\label{#1}}
\newcommand{\eea}{\end{eqnarray}}

\newcommand{\qed}{\hfill $\Box$ \medskip}

\newcommand{\be}{\begin{equation}}
\newcommand{\eeq}{\end{equation}}
\newcommand{\ee}{\end{equation}}
\newcommand{\beqa}{\begin{eqnarray}}
\newcommand{\eeqa}{\end{eqnarray}}
\newcommand{\beqan}{\begin{eqnarray*}}
\newcommand{\eeqan}{\end{eqnarray*}}
\newcommand{\ba}{\begin{array}}
\newcommand{\ea}{\end{array}}

\newcommand{\mcM}{{\mycal M}}

\newcommand{\scri}{{\mycal I}}%

\newcommand{\warn}[1]
{\protect{\stepcounter{mnotecount}}$^{\mbox{\footnotesize
$
\bullet$\themnotecount}}$ \marginpar{
\raggedright\tiny\em
$\!\!\!\!\!\!\,\bullet$\themnotecount: {\bf Warning:} #1} }

\newcommand{\eq}[1]{(\ref{#1})}

\newcommand{\ptc}[1]{\mnote{{\bf ptc:}#1}}

\newcommand{\mcL}{{\mycal L}}

\newcommand{\beqar}{\begin{deqarr}}
\newcommand{\eeqar}{\end{deqarr}}

\newcommand{\beaa}{\begin{eqnarray*}}
\newcommand{\eeaa}{\end{eqnarray*}}

\def\gSn{\gamma^{\mathbb{S}^n}}
\def\gHn{\gamma^{\mathbb{H}^{n+1}}}
\def\Conf{\mbox{Conf}}
\def\CSn{\Conf(\mathbb{S}^n)}
\def\KSn{\mbox{CKill}(\mathbb{S}^n)}
\def\KEn{\mbox{CKill}(\mathbb{E}^n)}
\def\IsoHn{\mbox{Iso}(\mathbb{H}^{n+1})}
\def\Mink{\mathbb{M}^{1,n+1}}
\def\Minkfour{\mathbb{M}^{1,4}}
\def\KillM{\mbox{Kill}(\Mink)}

\def\defi{:=}
\def\H{\mathcal{H}}
\def\X{\mathfrak{X}}
\def\Lor{\mbox{Lor}}
\def\Tau{\mathcal{T}}

\def\la{\langle}
\def\ra{\rangle}

\newcommand{\bm}[1]{\mbox{\boldmath $#1$}}

\def\varietat{{\cal V}}